\documentclass[a4paper,fleqn]{cas-dc}

\usepackage[numbers]{natbib}

\newcommand{\eg}[0]{\textit{e.g.}}
\newcommand{\ie}[0]{\textit{i.e.}}
\newcommand{\atp}[0]{\textit{atp}}
\newcommand{\nad}[0]{\textit{nad}}
\newcommand{\glu}[0]{\textit{glu}}
\newcommand{\ac}[0]{\textit{ac}}
\newcommand{\ace}[0]{\textit{ace}}
\newcommand{\triop}[0]{\textit{triop}}
\newcommand{\tp}[0]{\textit{tp}}
\newcommand{\fru}[0]{\textit{fru}}
\newcommand{\pyr}[0]{\textit{pyr}}
\begin{document}
\let\WriteBookmarks\relax
\def\floatpagepagefraction{1}
\def\textpagefraction{.001}
\shorttitle{Kinetic Parameters Identification}
\shortauthors{Weglarz-Tomczak et~al.}

\title [mode = title]{Population-based Optimization for Kinetic Parameter Identification in Glycolytic Pathway in \textit{Saccharomyces cerevisiae}}               

\author[1]{Ewelina Weglarz-Tomczak}[orcid=0000-0001-8080-2801]
\ead{ewelina.weglarz.tomczak@gmail.com}
\cormark[1]
\fnmark[1]
\credit{Conceptualization, Validation, Investigation, Writing - Original Draft, Writing - Review \& Editing}
\author[2]{Jakub M. Tomczak}[orcid=0000-0001-8634-694X]
\ead{j.m.tomczak@vu.nl}
\cormark[1]
\fnmark[2]
\credit{Methodology, Software, Investigation, Writing - Original Draft, Writing - Review \& Editing}
\author[2]{Agoston E. Eiben}[orcid=0000-0002-3106-4213]
\credit{Writing - Review \& Editing}
\author[1]{Stanley Brul}[orcid=0000-0001-5706-8768]
\credit{Supervision, Writing - Review \& Editing}

\address[1]{Swammerdam Institute for Life Sciences, Faculty of Science, University of Amsterdam, the Netherlands}

\address[2]{Department of Computer Science, Faculty of Science, Vrije Universiteit Amsterdam, the Netherlands}

\cortext[cor1]{Corresponding author}

\fntext[fn1]{Tel. +31 20 525 8626, Address: Science Park 904, 1098 XH, Amsterdam, the Netherlands}
\fntext[fn2]{Tel. +31 20 598 8206, Address: De Boelelaan 1111, 1081 HV, Amsterdam, the Netherlands}

\begin{abstract}
Models in systems biology are mathematical descriptions of biological  processes that are used to answer questions and gain a better understanding of biological phenomena. Dynamic models represent the network through rates of the production and consumption for the individual species. The ordinary differential equations that describe rates of the reactions in the model include a set of parameters.
The parameters are important quantities to understand and analyze biological systems. Moreover, the perturbation of the kinetic parameters are correlated with upregulation of the system by cell-intrinsic and cell-extrinsic factors, including mutations and the environment changes. 

Here, we aim at using well-established models of biological pathways to identify parameter values and point their potential perturbation/deviation. We present our population-based optimization framework that is able to identify kinetic parameters in the dynamic model based on only input and output data (\ie, timecourses of selected metabolites). Our approach can deal with the identification of the non-measurable parameters as well as with discovering deviation of the parameters. We present our proposed optimization framework on the example of the well-studied glycolytic pathway in \textit{Saccharomyces cerevisiae}.
\end{abstract}


\maketitle

% ============MAIN TEXT=============================
% =====SECTION=====
\section{Introduction}
Mathematical models in systems biology provide a representation of the information obtained from experimental observations about the structure and function of a particular biological network \cite{ingalls2013mathematical, nielsen2017systems}. The models that include dynamics of the network typically consist of systems of ordinary differential equations (ODE) \cite{tsiantis2018optimality, wolkenhauer2004modeling}. We call them dynamic or kinetic models, and their crucial element are parameters. Many parameters are generally unknown, thereby it hampers the possibility for obtaining quantitative predictions \cite{balsa2010iterative}. Kinetic parameters characterize the particular reaction catalyzed by a specific enzyme in particular conditions. Therefore, the deviation from the standard value could be correlated with a mutation, an epigenetics or a change of the environment. In other words, determining values of parameters for the considered biological systems could be used for further understanding and analysis of the system.

In traditional systems biology approaches the kinetic parameters in a dynamic model can be identified by fitting the model to experimental data or are measured for individual reactions separately. Such models can be used to confirm hypotheses, to draw predictions and to find those (time varying) stimulation conditions that result in a particular desired behavior \cite{ideker2001new, nielsen2017systems, westerhoff2004evolution}. We propose to go a step forward and we aim at using established models to predict a perturbation in the biological system and to point out the step (reaction) where it occurs via identification of the kinetic parameters of differential equations. Therefore, the goal of our work is to develop a computational-based framework for: (i) identifying the non-measurable parameters so as to reproduce, insofar as it is possible, the experimental data, and (ii) predicting kinetic parameters of any reaction in the biochemical network based on the timecourse of only input and output metabolites. 

\begin{figure}[!htbp]
    \centering
    \includegraphics[width=0.75\columnwidth]{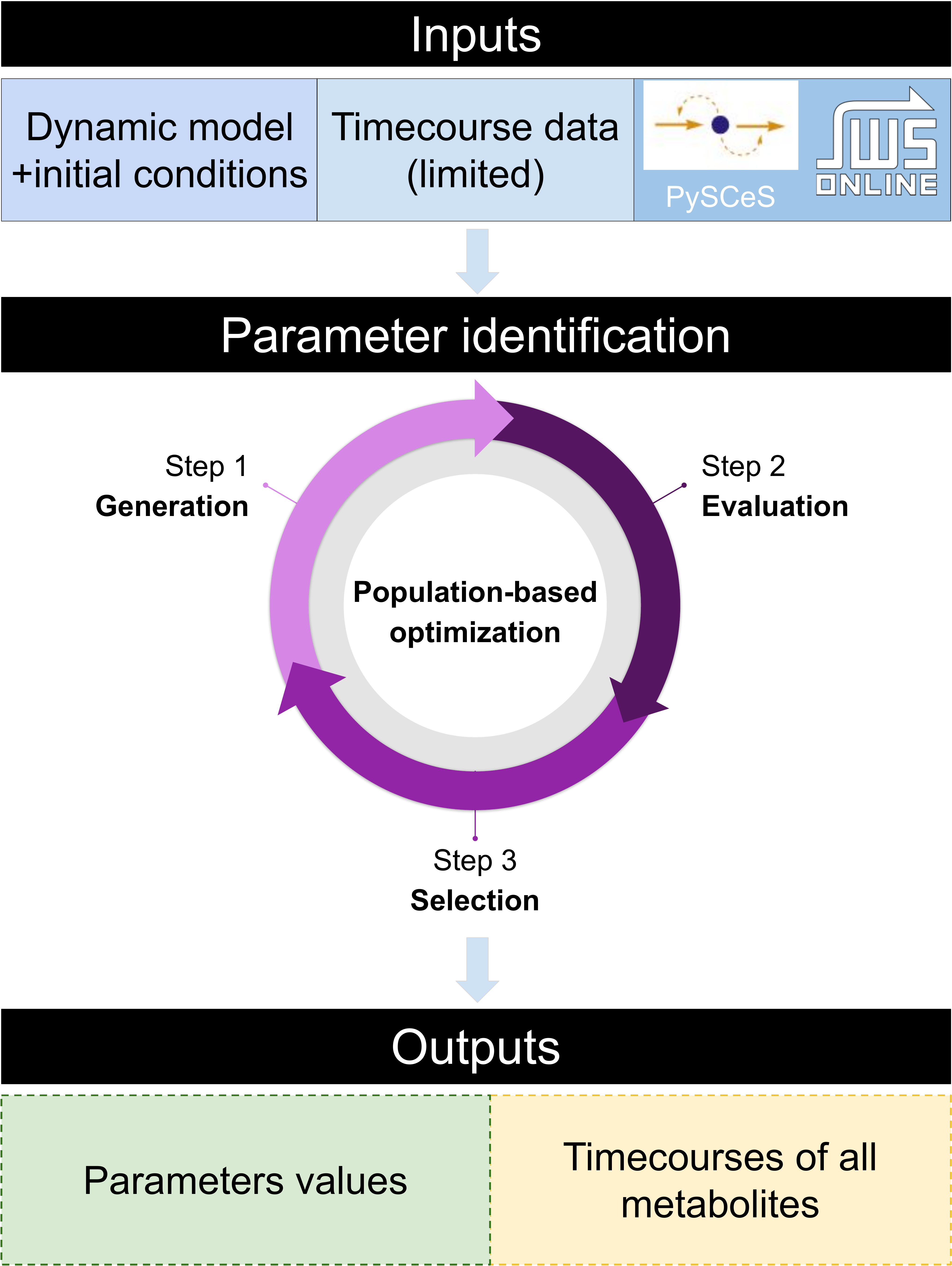}
    \caption{A schematic representation of our approach.}
    \label{fig:our_approach}
\end{figure}

Among many branches of computational methods, the optimization algorithms seemed for us to be most promising for achieving our goal. The optimization strategies, namely, deterministic, stochastic and heuristics have been already applied in systems biology for parameter identification for diabetes dynamics \cite{morbiducci2011identification}, biomarker identification, \textit{in-silico} simulations of biological phenomena, and for a variety of statistical inferences and time course estimations \cite{reali2017optimization}.

In general, optimization is about finding a solution that minimizes (or maximizes) an objective function for given constraints, \textit{i.e.}, possible values that solutions can take. A subset of optimization problems with \textit{non-differentiable} or \textit{black-box} objective functions constitute \textit{derivative-free optimization} (DFO) problems. In general, a black-box is any process that for given input, returns an output, but its analytical description is unavailable or it is non-differentiable \cite{audet2017derivative}. Moreover, the big advantage of the black-box optimization and the derivative-free optimization is that they make almost no assumptions about the problem \cite{eiben2015evolutionary}. Therefore, they are widely used in many domains, \eg, in optimizing computer programs \cite{cranmer2020frontier}, biochemical processes \cite{gerard2013evolutionary, moles2003parameter, tomczak2019estimating, toni2009approximate}, bioengineering \citep{yang2020response}, or in evolutionary robotics \cite{doncieux2015evolutionary}. 

There exists a vast of derivative-free optimization (DFO) methods, ranging from classical algorithms like iterative local search or direct search \cite{audet2017derivative} to modern approaches like Bayesian optimization (BO) \cite{shahriari2015taking} and evolutionary algorithms (EA) \cite{back2013contemporary, eiben2003introduction}. The main drawback of classical approaches is that they become infeasible for high-dimensional problems, and they require additional strategies like multiple starts to obtain good solutions. Bayesian optimization is currently the state-of-the-art approach for black-box optimization. This approach combines a surrogate model with an active querying strategy to find high quality candidate solutions. However, typically the surrogate model is non-parametric that leads to a cubic complexity with respect to the number of stored solutions. Therefore, BO is typically employed to optimize expensive-to-evaluate functions. The last group of DFO methods, evolutionary algorithms or, more generally, population-based methods \cite{gallagher2005population}, utilize a population of solutions that share information and point the search to region of high potential. If evaluating the objective function is relatively low, and, thus, one can afford to have a large population, then these approaches give very good results in a large class of optimization problems. Here, since we deal with biological systems \textit{in silico}, we decide to follow this approach and use the population-based optimization methods for the parameter identification, see Figure \ref{fig:our_approach}.

\begin{figure}[!htbp]
    \centering
    \includegraphics[width=1\columnwidth]{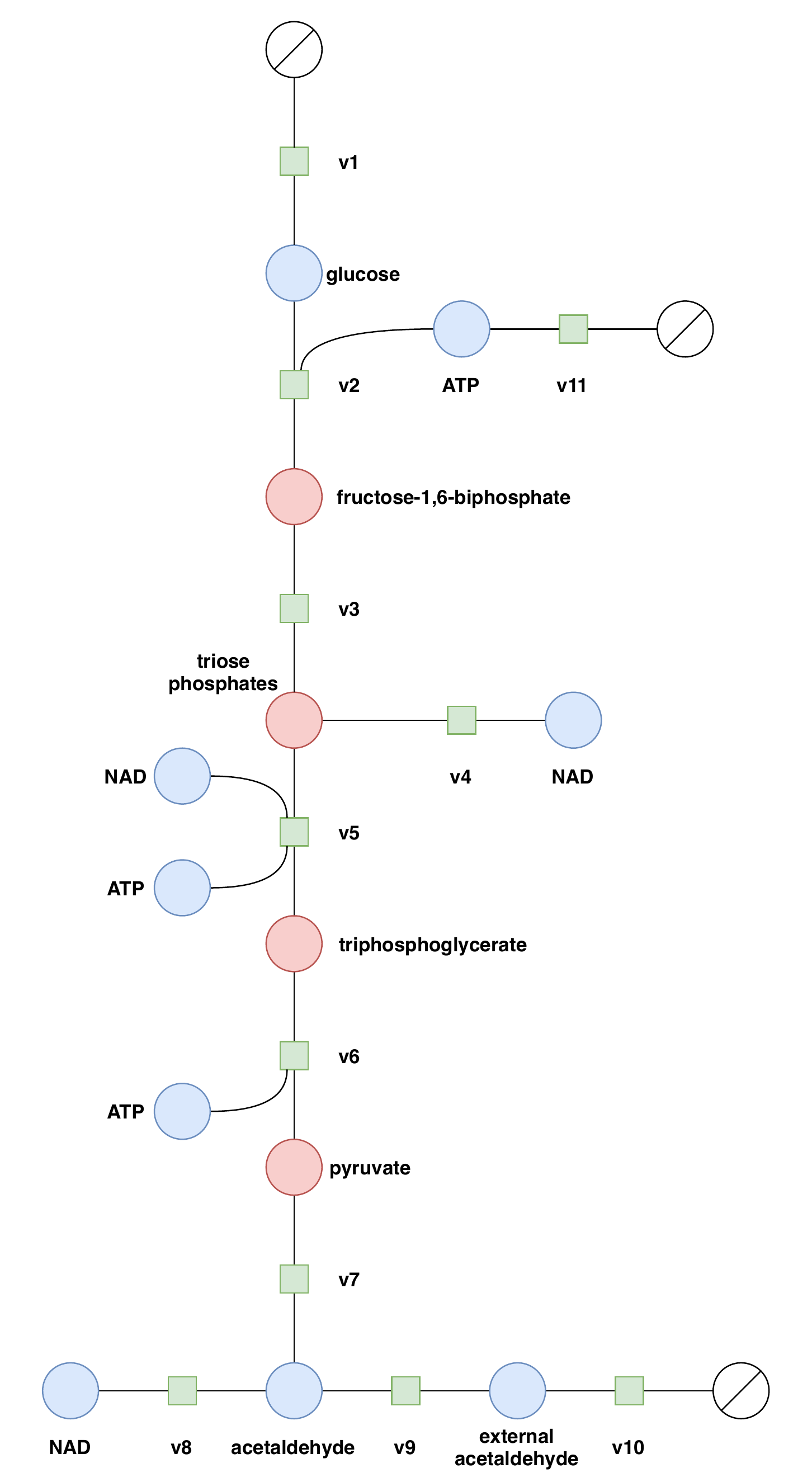}
    \caption{The glycolysis process in the yeast \textit{Saccharomyces cerevisiae} proposed in \cite{wolf2000transduction}. There are $11$ reactions governing the process with $18$ parameters in total, and $9$ metabolites. Blue circles depict observable metabolites, red circles denote unobservable metabolites, and green squares represent reactions. A white circle with a diagonal line corresponds to a sink. The model is taken from the JWS database \cite{olivier2004web}.}
    \label{fig:glycolysis}
\end{figure}

%Therefore, glycolysis is at the heart of classical biochemistry and, as such, in the view of individual steps, it is very well described. However, viewed as a whole, our understanding leaves still much to be desired. 

In this study, we chose \textit{glycolysis} that is a crucial metabolic pathway and its upregulation is correlated with diseases like cancer \cite{gatenby2004cancers, pelicano2006glycolysis}. Nearly all living organisms carry out glycolysis as a part of cellular metabolism. Glycolytic path that consists of a series of reactions breaks down glucose into two three-carbon compounds and extracts energy for cellular metabolism. Therefore, glycolysis is at the heart of classical biochemistry and, as such, it is very well described. 
One of the most intensively studied organisms in context of, among others, glycolysis is \textit{Saccharomyces cerevisiae} species, also known also known as baker’s yeast \cite{duarte2004reconstruction, lee2002transcriptional, mensonides2014kinetic, nielsen2019yeast, orij2012genome}. Whereas, the dynamic model of glycolysis in \textit{Saccharomyces cerevisiae} is of big interest in systems biology dynamic modeling literature \cite{hynne2001full, kourdis2013glycolysis, teusink2000can, van2010measuring, wolf2000transduction}.

%One of the most intensively studied organisms in molecular and cell biology, and systems biology is \textit{Saccharomyces cerevisiae} species, also known as baker’s yeast \cite{duarte2004reconstruction, lee2002transcriptional, mensonides2014kinetic, nielsen2019yeast, orij2012genome}. Therefore, the dynamic model of glycolysis in \textit{Saccharomyces cerevisiae} is of big interest in systems biology dynamic modeling literature \cite{hynne2001full, kourdis2013glycolysis, teusink2000can, van2010measuring, wolf2000transduction}.

We applied our optimization framework to a model of glycolysis in yeast proposed in \cite{wolf2000transduction} that suffices to describe the essence of our research goal, see Figure \ref{fig:glycolysis}. This model contains lumped reactions of the glycolytic pathway and includes production of glycerol, fermentation to ethanol and exchange of acetaldehyde between the cells, and trapping of acetaldehyde by cyanide. 

This paper has a multidisciplinary character. Therefore, we state research goals that are of interest for computational biology, systems biology and derivative-free optimization, namely:
\begin{itemize}
    \item Apply the population-based optimization methods to the parameter identification of the glycolysis process and analyze their performance.
    \item Determine whether it is possible to identify parameters if only a subset of metabolites are observed.
    \item Determine whether it is possible to identify parameters if one parameter in the system is slightly changed, \ie, in the case of a mutation.
    % \item Analyze the performance of the population-based optimization.
\end{itemize}
The first research goal require to implement the population-based optimization methods and combine them with an ODE solver. Moreover, we must be able to express a biological model and process it. For this purpose, we build on top of the Python Simulator for Cellular Systems (PySCeS) library \cite{OliRohHof05}. We also propose two surrogate-assisted population-based optimization methods to reduce computational complexity and enhance exploration. Further, we slightly modify parameter values of the model from \cite{wolf2000transduction} in order to answer our another two research questions. Finally, we conduct extensive experiments \textit{in silico} and provide qualitative and quantitative analysis of the population models.

The contribution of the paper is threefold:
\begin{itemize}
    \item We provide a population-based optimization framework for parameter identification and showcase its performance on the example of the glycolysis of \textit{Saccharomyces cerevisiae}, one of the most studied species in biology.
    \item We analyze the performance of the population-based optimization framework in the considered problem and indicate its high potential for future research.
    \item We extend the Python framework PySCeS \cite{OliRohHof05} by implementing the population-based optimization methods ($4$ methods known in literature, and $2$ new methods) in Python. The code for the methods together with the experiments is available online: \url{https://github.com/jmtomczak/popi}.
\end{itemize}

%MENTION THAT DE COULD BE SEEN AS AN EXTENSION OF NELDER-MEAD

% =====SECTION=====
\section{Methods}

% -----SubSECTION-----
\subsection{Derivative-free optimization}

We consider an optimization problem of a function $f: \mathbb{X} \rightarrow \mathbb{R}$, where $\mathbb{X} \subseteq \mathbb{R}^{D}$ is the search space. In this paper we focus on the minimization problem, namely:
\begin{equation}
    \mathbf{x}^{*} = \arg\min_{\mathbf{x} \in \mathbb{X}} f(\mathbf{x}; \mathcal{D}) ,
\end{equation}
where $\mathcal{D}$ denotes observed data.

Further, we assume that the analytical form of the function $f$ is unknown or cannot be used to calculate derivatives, however, we can query it through a simulation or experimental measurements. Problems of this sort are known as \textit{derivative-free} or \textit{black-box}\footnote{In general, a \textit{black-box} problem means that a formal description of a problem is unknown, however, very often non-differentiable problems with known mathematical representation (\eg, differential equations) are treated as black-box.} optimization problems \cite{audet2017derivative, jones1998efficient}. Additionally, we consider a bounded search space, \textit{i.e.}, we include inequality constraints for all dimensions in the following form: $l_d \leq x_d \leq u_d$, where $l_d, u_d \in \mathbb{R}$ and $l_d < u_d$, for $d=1, 2, \ldots, D$ .

% -----SubSECTION-----
\subsection{Parameter identification in glycolysis}

We consider the glycolysis process in yeast as a biochemical system with inputs and outputs (see Figure \ref{fig:glycolysis}).
The input to the system is glucose (\glu), and the outputs are ATP (\atp), NAD (\nad), acetaldehyde (\ac) and external acetaldehyde (\ace).
The other metabolites, \ie, triose phosphates (\triop), pyruvate (\pyr), fructose-1,6-biphosphate (\fru) and triphoshoglycerate (\tp) are considered to be unobserved quantities.
The system is governed by $11$ reactions with $18$ parameters in total (see Appendix for details).
Each reaction is represented by an ordinary differential equation that is known.
We assume that we have measurements of the inputs and outputs, \ie, \glu, \atp, \nad, \ac, and \ace, and each quantity is represented as a timecourse of length $T$.
We denote these measurements by 
$$\mathcal{D} = \{\glu, \atp, \nad, \ac, \ace\} .$$

Further, following the nomenclature presented in \cite{cranmer2020frontier}, we consider the system of differential equations representing the glycolysis process as the \textbf{simulator} that for given values of parameters and initial conditions provides timecourses of all metabolites.
Then, we can denote parameters by $\mathbf{x}$ and the simulator by $\mathrm{sim}: \mathcal{X} \rightarrow \mathbb{R}^{9 \times T}$, \ie, $\mathrm{sim}$ takes parameters $\mathbf{x}$ and simulates timcourses of length $T$ for all $9$ metabolites, including $\glu, \atp, \nad, \ac, \ace$.
In order to calculate the objective (or the fitness) of the parameter values, we use the following function:
\begin{equation}\label{eq:fitness}
    f(\mathbf{x}; \mathcal{D}) = \sum_{i=1}^{5} \frac{1}{\gamma\cdot T} \sum_{t=1}^{T} \|\mathbf{y}_{i,t} - \mathrm{sim}_{i,t}(\mathbf{x}) \|_{2}^{2},
\end{equation}
where $\mathbf{y}_{i,t}$ corresponds to one of the $5$ observed metabolites at the $t$-th time step, and $\mathrm{sim}_{i,t}(\mathbf{x})$ is the corresponding synthetically generated signal given by the simulator with parameters $\mathbf{x}$, $\gamma > 0$ specifies the strength of penalizing a mistake. 
Notice that this is the (unnormalized) logarithm of the product of Gaussian distributions with means given by $\mathrm{sim}(\mathbf{x})$ and the diagonal covariance matrix with shared variance $\gamma$.

\begin{figure*}[tbp]
    \centering
    \includegraphics[width=0.24\textwidth]{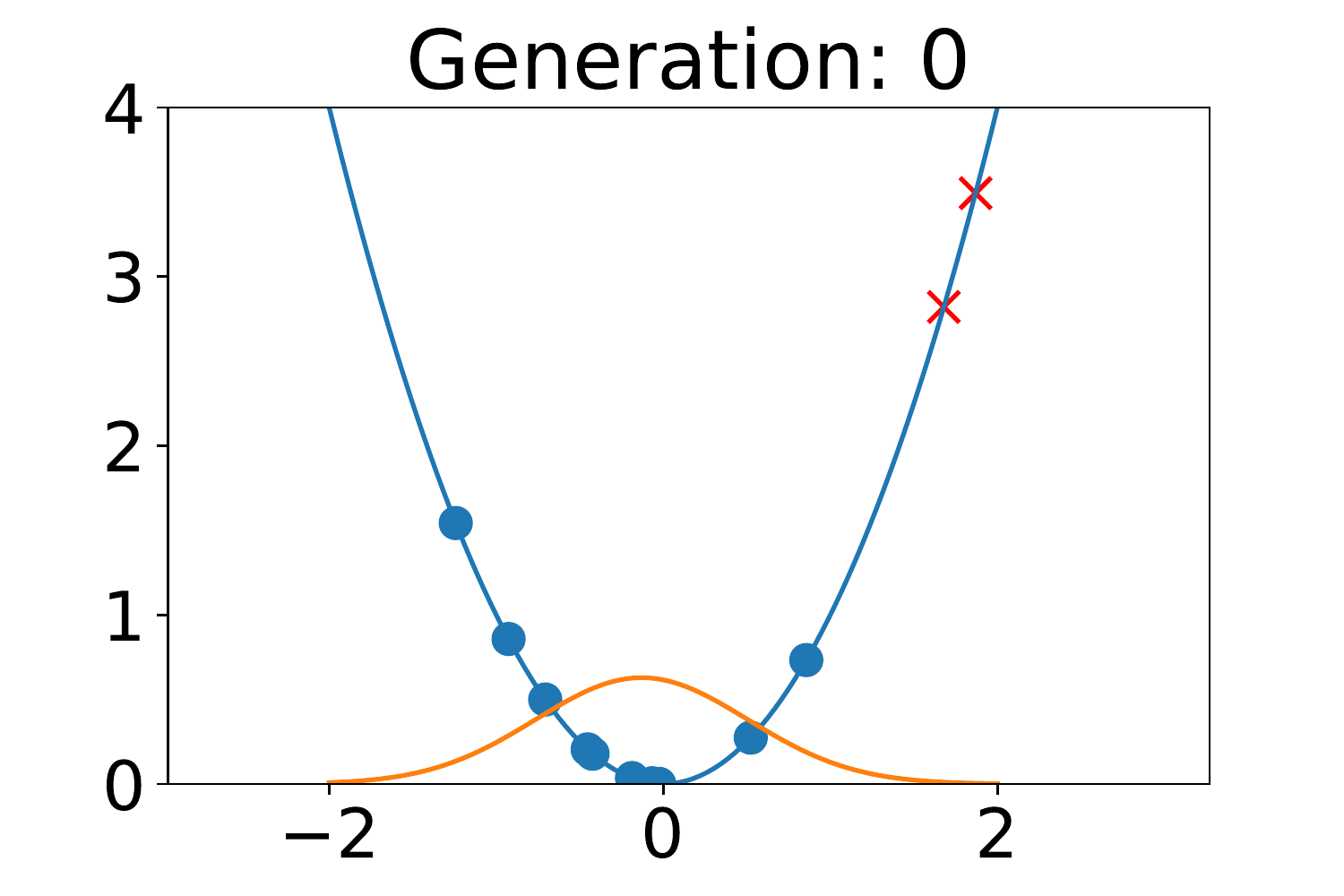}\ 
    \includegraphics[width=0.24\textwidth]{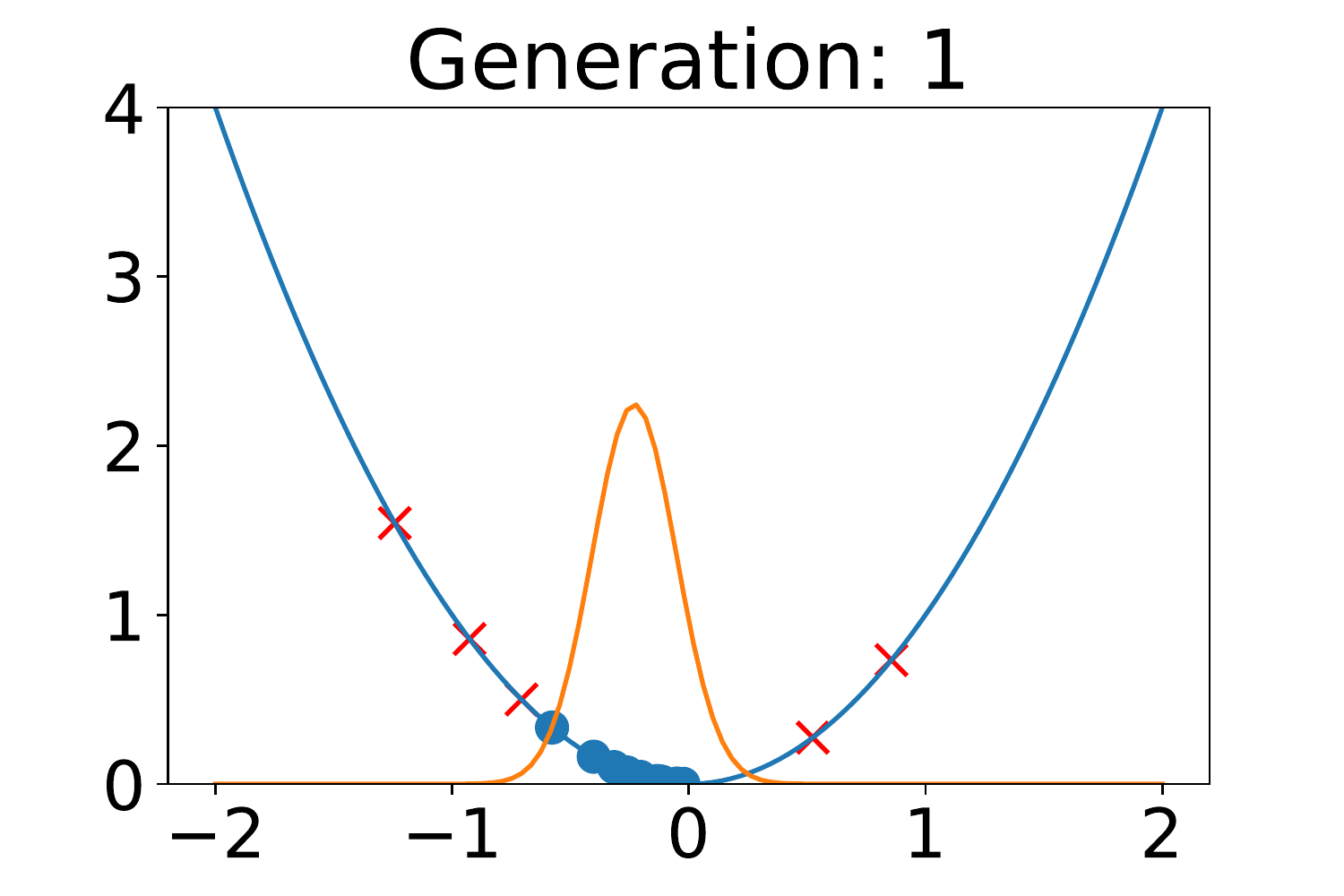}\ 
    \includegraphics[width=0.24\textwidth]{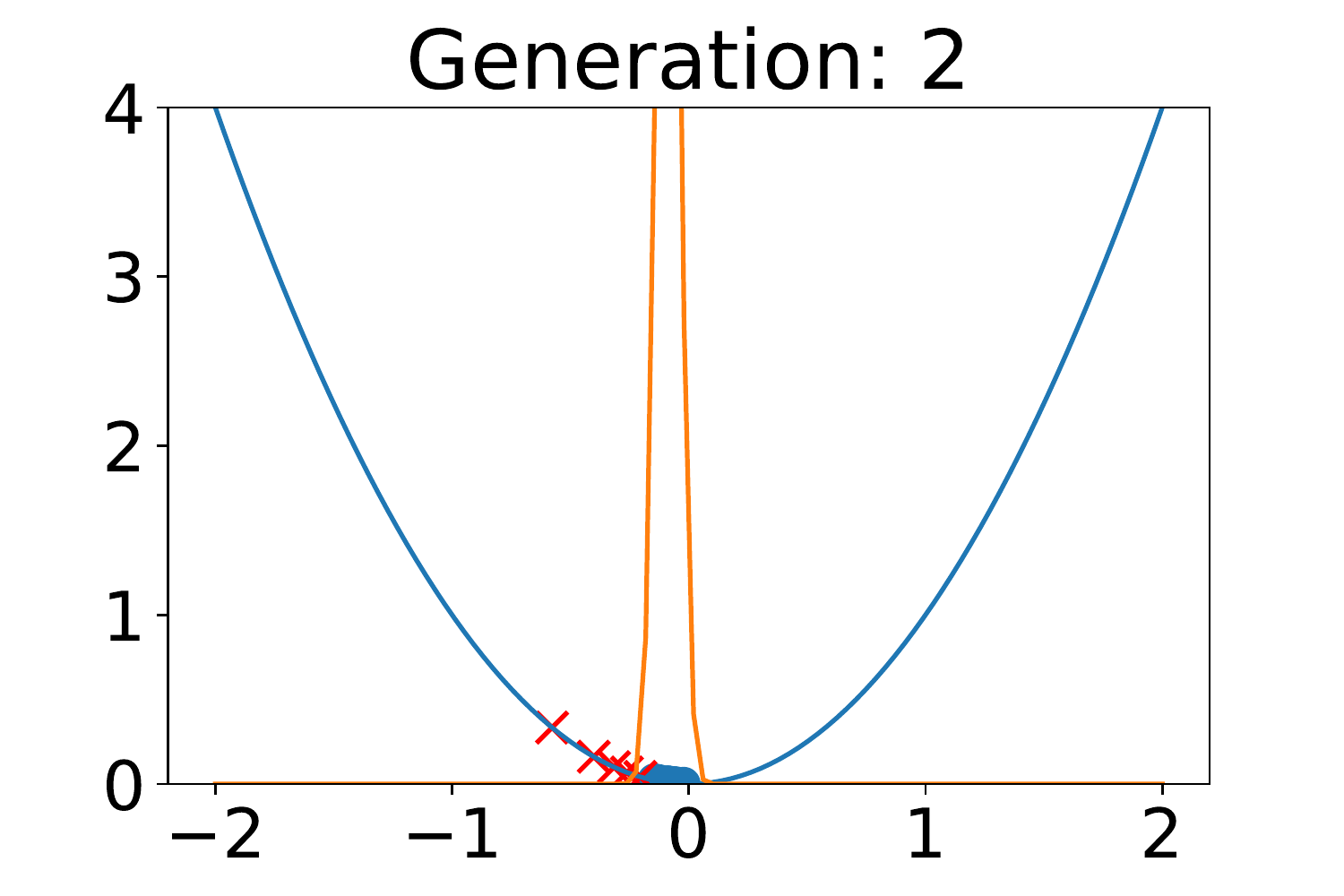}\ 
    \includegraphics[width=0.24\textwidth]{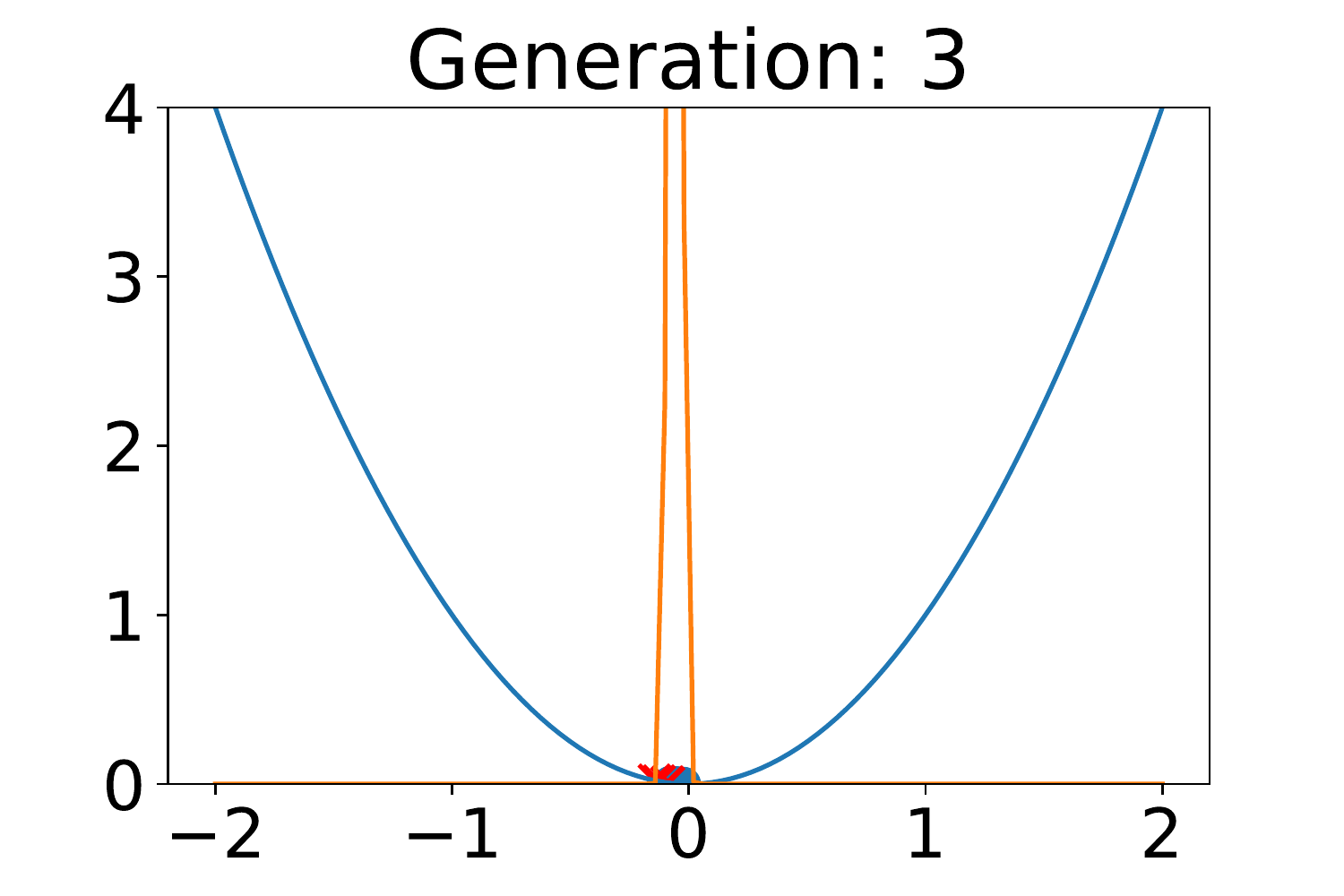}
    \caption{An illustration of a population-based optimization of a quadratic function (blue solid line). At each generation a population is selected (blue nodes) and weakest individuals are discarded (red crosses). New candidate solutions are generated by sampling from the normal distribution fit to the previous population (orange solid line).}
    \label{fig:population_based_optimization}
\end{figure*}

% -----SubSECTION-----
\subsection{Population-based optimization methods}

One group of widely-used methods for derivative-free optimization problems is population-based optimization algorithms. 
The idea behind these methods is to use a \textit{population} of \textit{individuals}, \ie, a collection of candidate solutions $\mathcal{X} = \{\mathbf{x}_{1}, \ldots, \mathbf{x}_{N}\}$, instead of a single individual in the iterative manner. 
The premise of utilizing the population over a single candidate solution is to obtain better exploration of the search space and exploiting potential local optima \cite{gallagher2005population, eiben2003introduction}. 

In the essence, every population-based algorithm consists of three following steps that utilize a procedure for generating new individuals $G$, and a selection procedure $S$, that is:
\begin{itemize}[ ]
    \item \textbf{(Init)} Initialize $\mathcal{X} = \{\mathbf{x}_{1}, \ldots, \mathbf{x}_{N}\}$ and evaluate all individuals $\mathcal{F}_x = \{f_n: f_n = f(x_n),\ x_n \in \mathcal{X}\}$.
    \item \textbf{(Generation)} Generate new candidate solutions using the current population, $\mathcal{C} = G(\mathcal{X}, \mathcal{F}_x)$.
    \item \textbf{(Evaluation)} Evaluate all candidates solutions: 
    $$\mathcal{F}_{c} = \{f_n: f_n = f(x_n),\ x_n \in \mathcal{C}\} .$$
    \item \textbf{(Selection)} Select a new population using the candidate solutions and the old population $$\mathcal{X} := S(\mathcal{X}, \mathcal{F}_x, \mathcal{C}, \mathcal{F}_{c}) .$$ Go to \textbf{Generate} or terminate.
\end{itemize}
An exemplary population-based optimization approach is depicted in Figure \ref{fig:population_based_optimization}.

In general, the population-based optimization methods are favorable over standard DFO algorithms in problems when querying the objective function is relatively cheap. 
If time required to obtain a value of the objective function (or the fitness function in the context of EA) is low, then their computational complexity is linear with respect to the size of the population $N$.
Bayesian Optimization, for instance, is known to give good performance, but its complexity typically scales cubicly with respect to the number of queries \cite{shahriari2015taking}.
Here, we take advantage of very low execution time of running a simulator (the glycolysis model) and propose to use the population-based methods for the parameter identification task.

There is a plethora of population-based DFO algorithms \cite{eiben2003introduction}, however, our goal is to verify whether this approach in general could be successfully used in the considered task. Therefore, we decide to choose four instances of group of methods that are easy-to-use and are proven to work well in practice: evolutionary strategies (ES), differential evolution (DE), estimation of distribution algorithms (EDA), and recently proposed reversible differential evolution (RevDE). Moreover, we propose to enhance EDA and RevDE with a surrogate model to allow better exploration and speed up calculations.

% -----SubSubSECTION-----
\subsubsection{Evolutionary Strategies (ES)}

Evolutionary strategies can be seen as a specialization of evolutionary algorithms with very specific choices of $G$ and $S$.
The core of ES is to formulate $G$ using the multivariate Gaussian distribution.
Here, we follow the widely-used (1+1)-ES that generates a new candidate using the Gaussian mutation parameterized by $\sigma > 0$, namely:
\begin{equation}\label{eq:gaussian_mutation}
    \mathbf{x}' = \mathbf{x} + \sigma \cdot \varepsilon,
\end{equation}
where $\varepsilon \sim \mathcal{N}(0,\mathrm{I})$, and $\mathcal{N}(0, \mathrm{I})$ denotes the Gaussian distribution with zero mean and the identity covariance matrix $\mathrm{I}$.
Next, if the fitness value of $\mathbf{x}'$ is smaller than the value of fitness function of $\mathbf{x}$, the new candidate is accepted and the old one is discarded.

The crucial element of this approach is determining the value of $\sigma$.
In order to overcome possibly time-consuming hyperparameter search, the following adaptive procedure is proposed \cite{back2013contemporary}:
\begin{equation}
    \sigma := \left\{
    \begin{matrix}
    \sigma \cdot c & \text{if } p_{s} < 1/5,\\
    \sigma / c  & \text{if } p_{s} > 1/5,\\
    \sigma & \text{if } p_{s} = 1/5 .
    \end{matrix}
    \right.
\end{equation}
where $p_{s}$ is the number of accepted individuals of the offspring divided by the population size $N$, and $c$ is equal $0.817$ following the recommendation in \cite{schwefel1977numerische}.

% -----SubSubSECTION-----
\subsubsection{Differential Evolution (DE)}

\textit{Differential evolution} is another population-based method that is loosely based on the Nelder-Mead method \cite{storn1997differential, price2006differential}.
A new candidate is generated by randomly picking a triple from the population, $(\mathbf{x}_i, \mathbf{x}_j, \mathbf{x}_k) \in \mathcal{X}$, and then $\mathbf{x}_i$ is perturbed by adding a scaled difference between $\mathbf{x}_j$ and $\mathbf{x}_k$, that is:
\begin{equation}\label{eq:mutation}
    \mathbf{y} = \mathbf{x}_i + F ( \mathbf{x}_j - \mathbf{x}_k ) ,
\end{equation}
where $F \in \mathbb{R}_{+}$ is the scaling factor. This operation could be seen as an adaptive \textit{mutation operator} that is widely known as \textit{differential mutation} \cite{price2006differential}.

Further, the authors of \cite{storn1997differential} proposed to sample a binary mask $\mathbf{m} \in \{0, 1\}^{D}$ according to the Bernoulli distribution with probability $p = P(m_d = 1)$ shared across all $D$ dimensions, and calculate the final candidate according to the following formula:
\begin{equation}\label{eq:crossover}
    \mathbf{v} = \mathbf{m} \odot \mathbf{y} + (1 - \mathbf{m}) \odot \mathbf{x}_{i},
\end{equation}
where $\odot$ denotes the element-wise multiplication. In the evolutionary computation literature this operation is known as \textit{uniform crossover operator} \cite{eiben2003introduction}. In this paper, we fix $p=0.9$ following general recommendations in literature \cite{pedersen2010good} and use the uniform crossover in all methods.

The last component of a population-based method is a selection mechanism. There are multiple variants of selection \cite{eiben2003introduction},  however, here we use the ``survival of the fittest'' approach, \ie, we combine the old population with the new one and select $N$ candidates with highest fitness values, \ie, the deterministic $(\mu + \lambda)$ selection.

This variant of DE is referred to as “DE/\textit{rand}/\textit{1}/\textit{bin}”, where \textit{rand} stands for randomly selecting a base vector, \textit{1} is for adding a single perturbation and \textit{bin} denotes the uniform crossover. Sometimes it is called \textit{classic DE} \cite{price2006differential}.

% -----SubSubSECTION-----
\subsubsection{Reversible Differential Evolution (RevDE)}

The mutation operator in DE perturbs candidates using other individuals in the population to generate a single new candidate. 
As a result, having too small population could limit exploration of the search space. 
In order to overcome this issue, a modification of DE was proposed that utilized all three individuals to generate three new points in the following manner \cite{tomczak2020differential}:
\begin{align}\label{eq:revde_raw}
    \mathbf{y}_1 &= x_i + F(\mathbf{x}_j - \mathbf{x}_k) \nonumber \\
    \mathbf{y}_2 &= x_j + F(\mathbf{x}_k - \mathbf{y}_1) \\
    \mathbf{y}_3 &= x_k + F(\mathbf{y}_1 - \mathbf{y}_2) .\nonumber
\end{align}
New candidates $\mathbf{y}_1$ and $\mathbf{y}_2$ could be further used to calculate perturbations using points outside the population. 
This approach does not follow a typical construction of an EA where only evaluated candidates are mutated. 
Further, we can express (\ref{eq:revde_raw}) as a linear transformation  using matrix notation by introducing matrices as follows:
\begin{equation}\label{eq:r}
    \begin{bmatrix}
        \mathbf{y}_1 \\
        \mathbf{y}_2 \\
        \mathbf{y}_3
    \end{bmatrix}
     = 
    \underbrace{\begin{bmatrix}
        1       & F              & -F \\
        -F      & 1 - F^2        & F + F^2 \\
        F + F^2 & -F + F^2 + F^3 & 1 - 2 F^2 - F^3
    \end{bmatrix}}_{=\mathbf{R}} 
    \begin{bmatrix}
        \mathbf{x}_1 \\
        \mathbf{x}_2 \\
        \mathbf{x}_3
    \end{bmatrix} .
\end{equation}
In order to obtain the matrix $\mathbf{R}$, we need to plug $\mathbf{y}_1$ to the second and third equation in (\ref{eq:revde_raw}), and then $\mathbf{y}_2$ to the last equation in (\ref{eq:revde_raw}). 
As a result, we obtain $M = 3N$ new candidate solutions.
This version of DE is called \textit{Reversible Differential Evolution}, because the linear transformation $\mathbf{R}$ is reversible \cite{tomczak2020differential}.

% -----SubSubSECTION-----
\subsubsection{Estimation of Distribution Algorithms (EDA)}

Most of the population-based optimization methods aim at finding a solution and the information about the distribution of the search space and the fitness function is represented implicitly by the population.
However, this distribution could be modeled explicitly using a probabilistic model \cite{gallagher2005population}.
These methods have become known as estimation of distribution algorithms \cite{larranaga2001estimation, muhlenbein1996recombination, pelikan2015estimation}.

The key difference between EDA and EA is the generation step.
While an EA uses evolutionary operators like mutation and cross-over to generate new candidate solutions, EDA fits a probabilistic model to the population, and then new individuals are sampled from this model.

Therefore, fitting a distribution to the population is the crucial part of an EDA.
There are various probabilistic models that could be used for this purpose.
Here, we propose to fit the multivariate Gaussian distribution $\mathcal{N}(\mathbf{\mu}, \Sigma )$ to the population $\mathcal{X}$.
For this purpose, we can use the empirical mean and the empirical covariance matrix:
\begin{equation}
    \hat{\mathbf{\mu}} = \frac{1}{N} \sum_{n=1}^{N} \mathbf{x}_{n},
\end{equation}
and
\begin{equation}
    \hat{\Sigma} = \frac{1}{N} \sum_{n=1}^{N} (\mathbf{x}_{n} - \hat{\mathbf{\mu}}) (\mathbf{x}_{n} - \hat{\mathbf{\mu}})^{\top} .
\end{equation}
An efficient manner of sampling new candidates is to first calculate the Cholesky decomposition of the covariance matrix, $\hat{\Sigma} = \mathbf{L}\mathbf{L}^{\top}$, where $\mathbf{L}$ is the lower-triangular matrix, and then computing:
\begin{equation}\label{eq:eda}
    \mathbf{x}' = \mathbf{\mu} + \mathbf{L} \varepsilon,
\end{equation}
where $\varepsilon \sim \mathcal{N}(0, \mathrm{I})$. The Eq. \ref{eq:eda} is repeated $M$ times to generate a new set of candidate solutions.
Here, we set $M$ to the size of the population, \ie, $M=N$.

Once new candidate solutions are generated, the selection mechanism is applied.
In this paper, we use the same selection procedure as the one used for DE.

% -----SubSubSECTION-----
\subsubsection{Population-based methods with surrogate models (RevDE+ \& EDA+)}

A possible drawback of population-based methods is the necessity of evaluating large populations that, even though we assume a low time cost per a single evaluation, could significantly slow down the whole optimization process.
In order to overcome this issue, a surrogate model could be used to partially replace querying the fitness function \cite{jin2011surrogate}.
The surrogate model is either a probabilistic model or a machine learning model that gathers previously evaluated populations, and allows to mimic the behavior of the fitness function.
It is assumed that the computational costs is lower or even significantly lower than the computational cost of running the simulator.

There are multiple possible surrogate models, however, non-parametric models, \eg, Gaussian processes \cite{shahriari2015taking}, are preferable, because they do not suffer from catastrophic forgetting (\ie, overfitting to last population and forgetting first populations).
Here, we consider \textit{K}-Nearest-Neighbor (\textit{K}-NN) regression model that stores all previously seen individuals with evaluations, and the prediction of a new candidate solution is an average over $K$ (\eg, $K=3$) closest previously seen individuals.
Current implementations of the \textit{K}-NN regressor provide efficient search procedures that result in the computational complexity better than $N\cdot D$, \eg, using KD-trees results in $O(D\log N)$.

\paragraph{RevDE+}
In the RevDE approach we generate $3N$ new candidate solutions and all of them are further evaluated.
However, this introduces and extra computational cost of running the simulator.
This issue could be alleviated by using the \textit{K}-NN regressor to approximate the fitness values of the new candidates.
Further, we can select $N$ most promising points.
We refer to this approach as RevDE+.

\paragraph{EDA+}
The outlined procedure of EDA produces $M$ new candidate solutions and in order to keep a similar computational cost as ES and DE, we set $M$ to $N$.
However, this could significantly limit the potential of modeling a search space, because sampling in high-dimensional search spaces requires a significantly large number of point.
A potential solution to this problem could be the application of the \textit{K}-NN regressor to quickly verify of the $L$ new points.
As long as the time cost of providing the approximated value of the fitness function is lower than the running time of the simulator, we can afford to take $L > N$ (\eg, $L = 5N$).
We refer to this approach as EDA+.

\begin{figure*}[!htbp]
    \begin{tabular}{lcc}
        \textbf{A} &  \\
        \includegraphics[width=145px,height=113px]{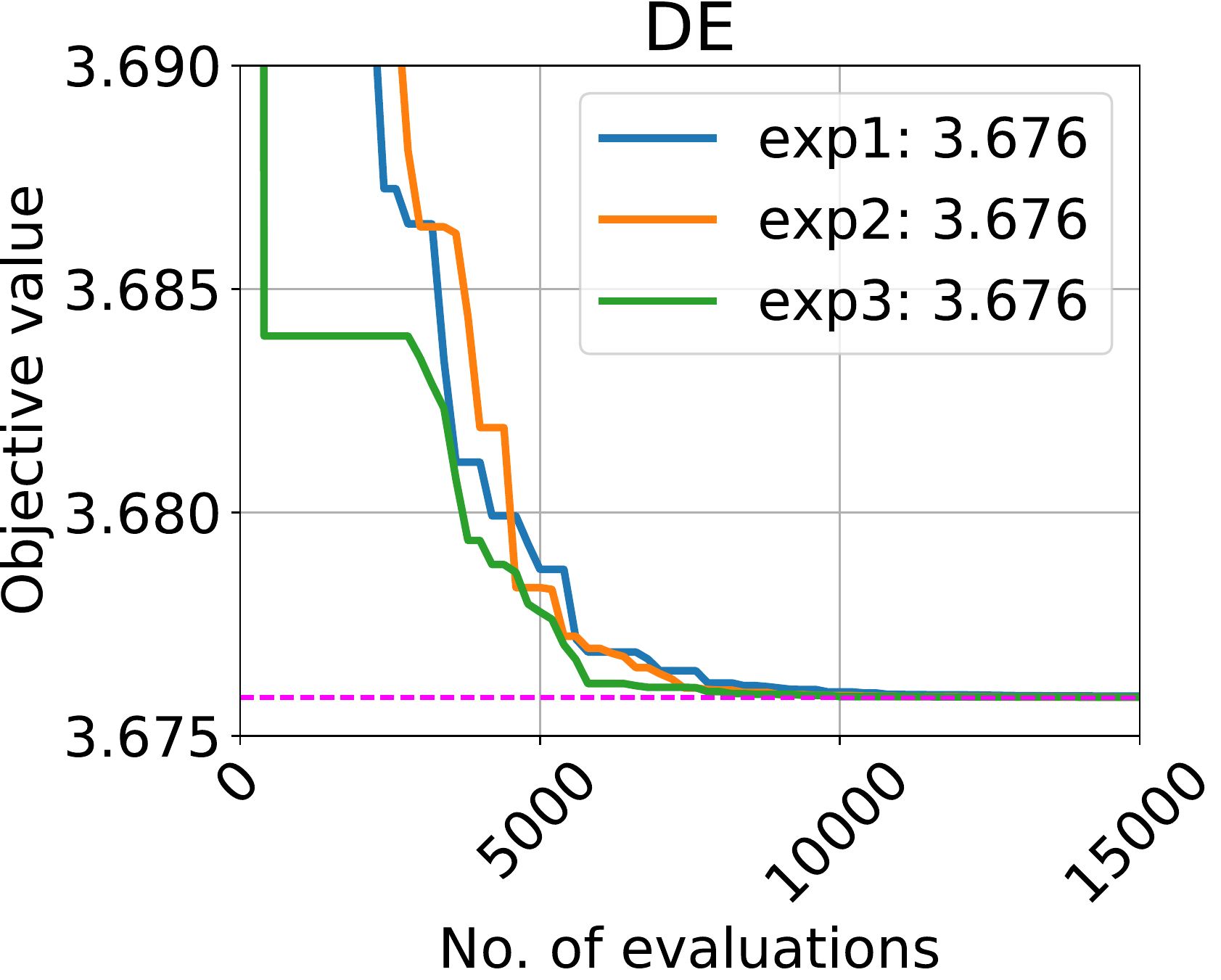} &
        \includegraphics[width=145px,height=113px]{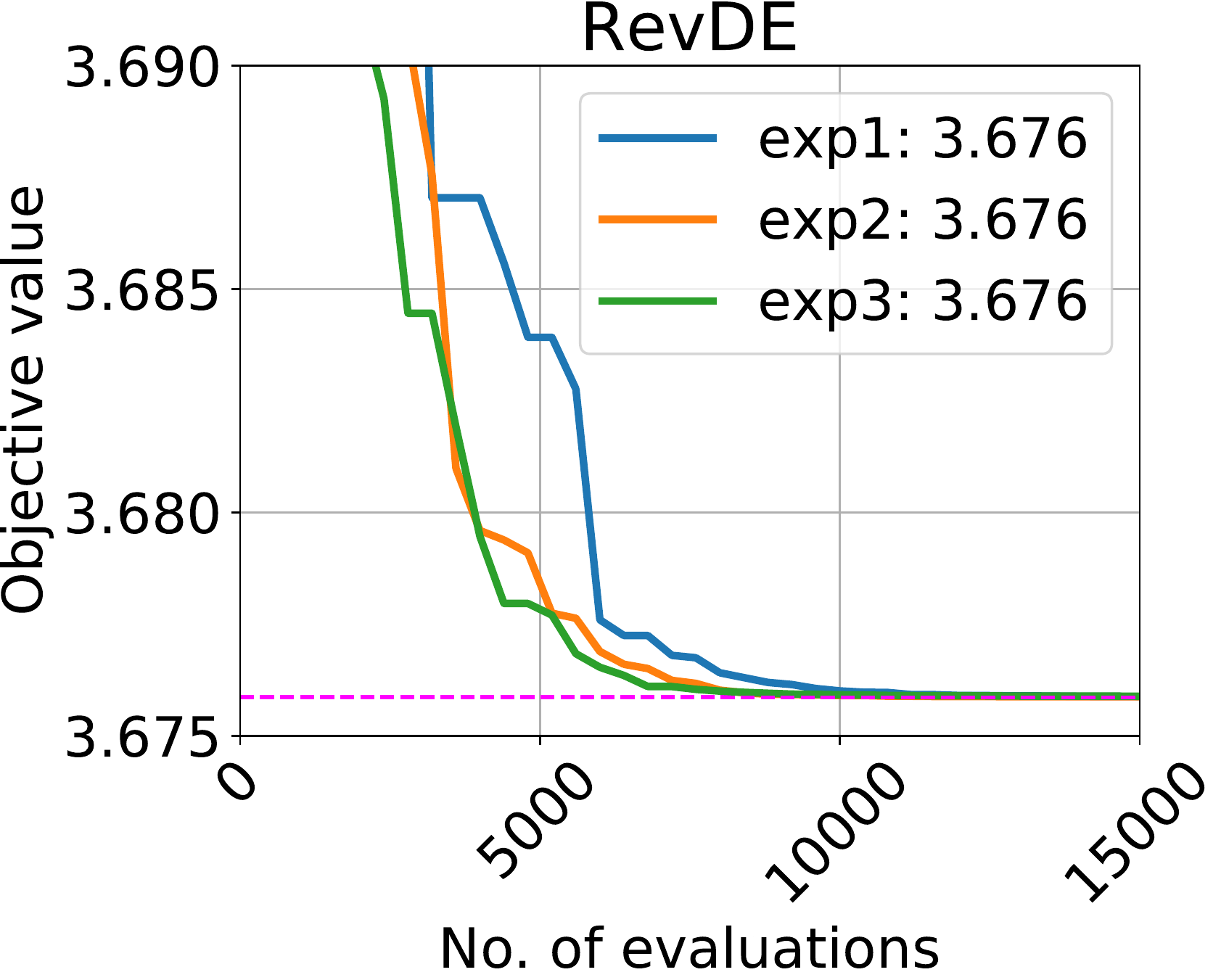} &
        \includegraphics[width=145px,height=113px]{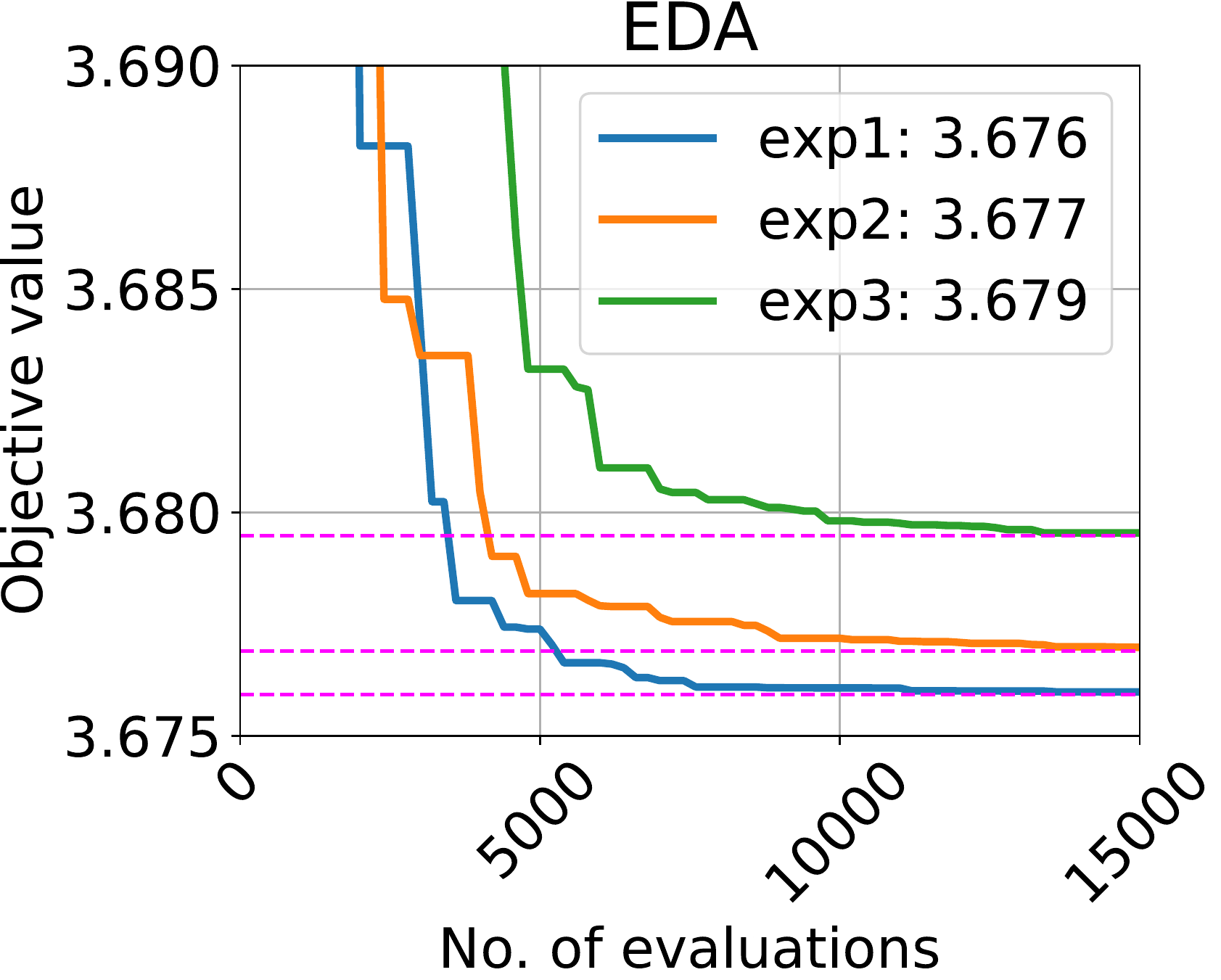} \\
        \includegraphics[width=145px,height=113px]{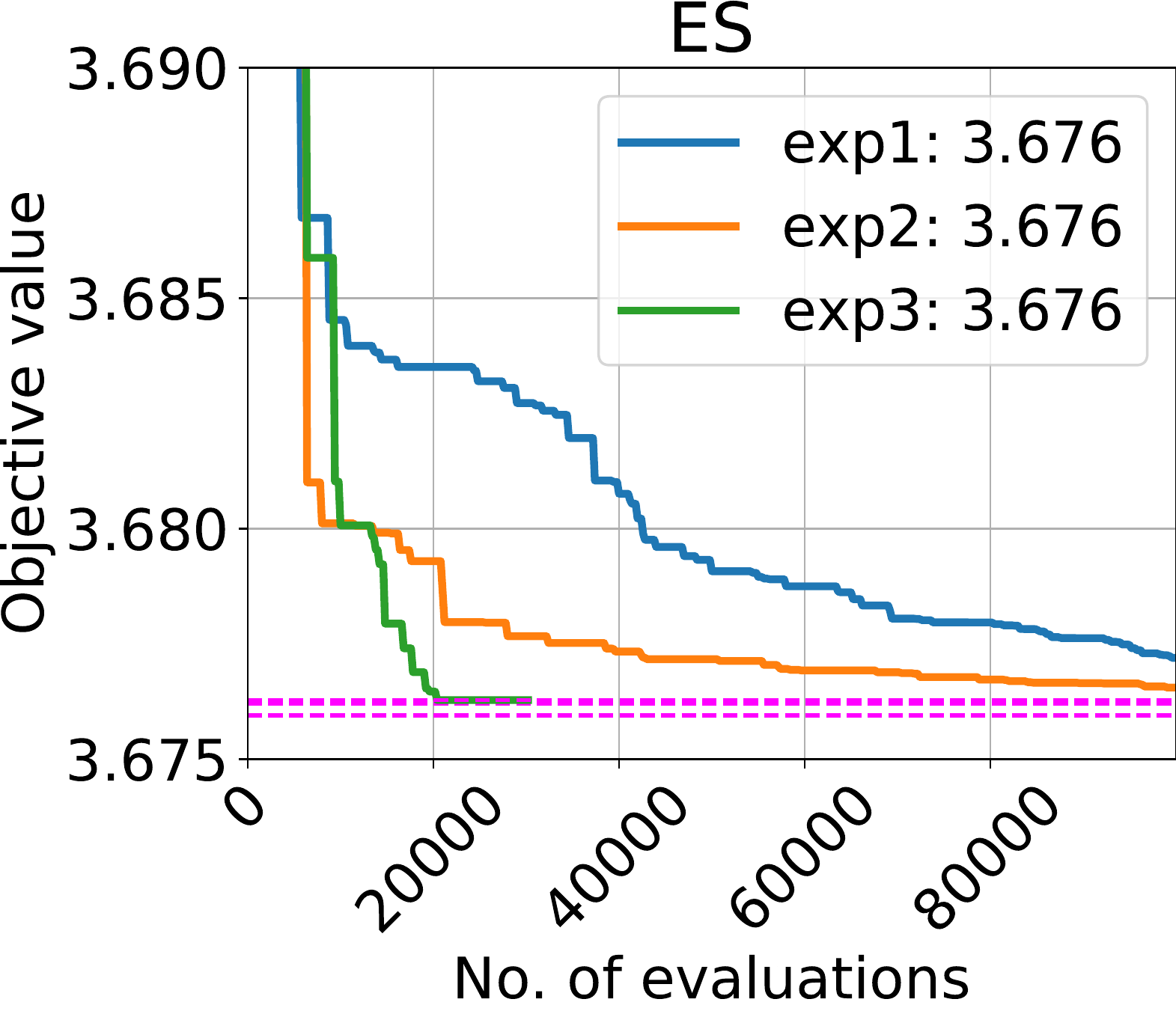} &
        \includegraphics[width=145px,height=113px]{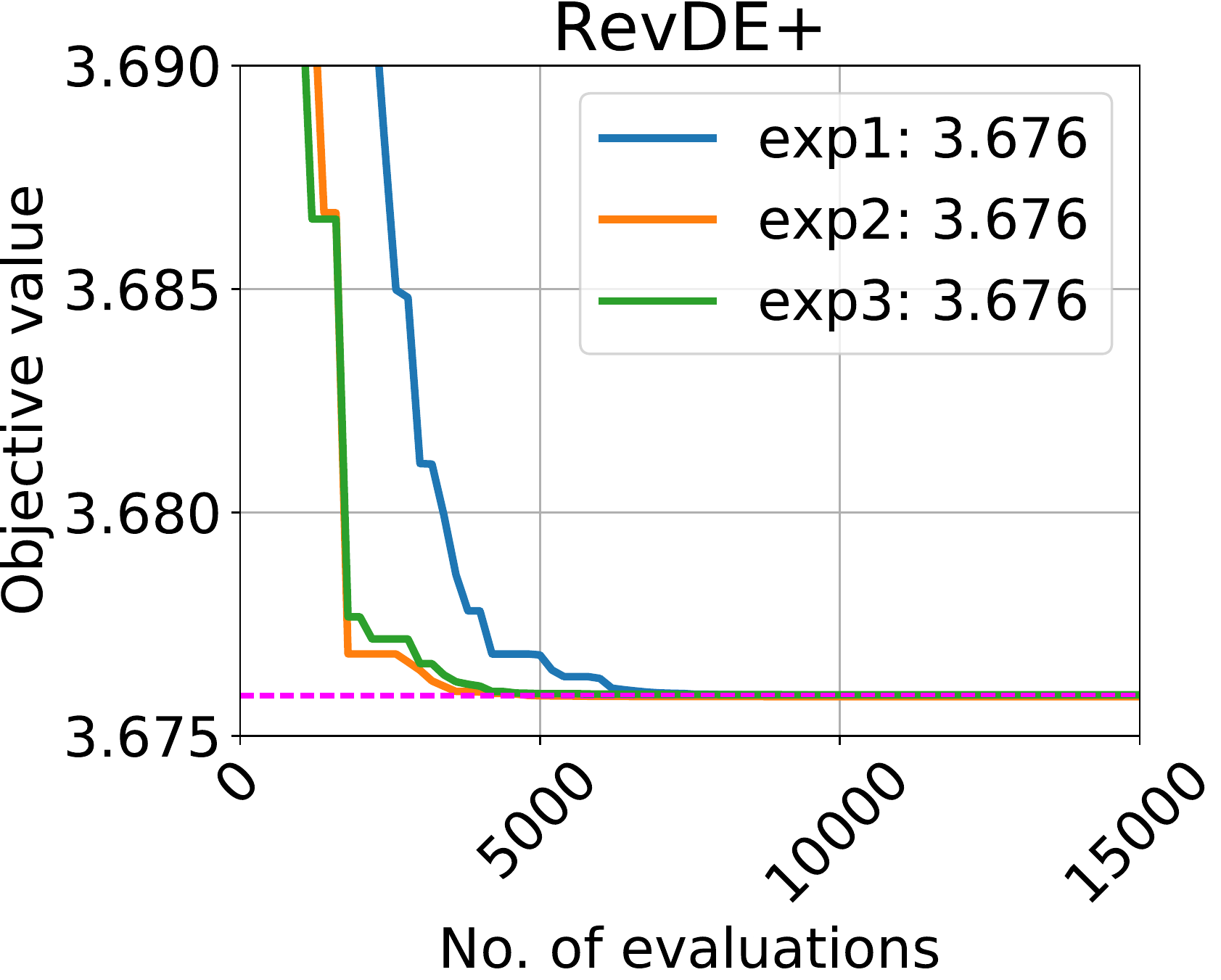} &
        \includegraphics[width=145px,height=113px]{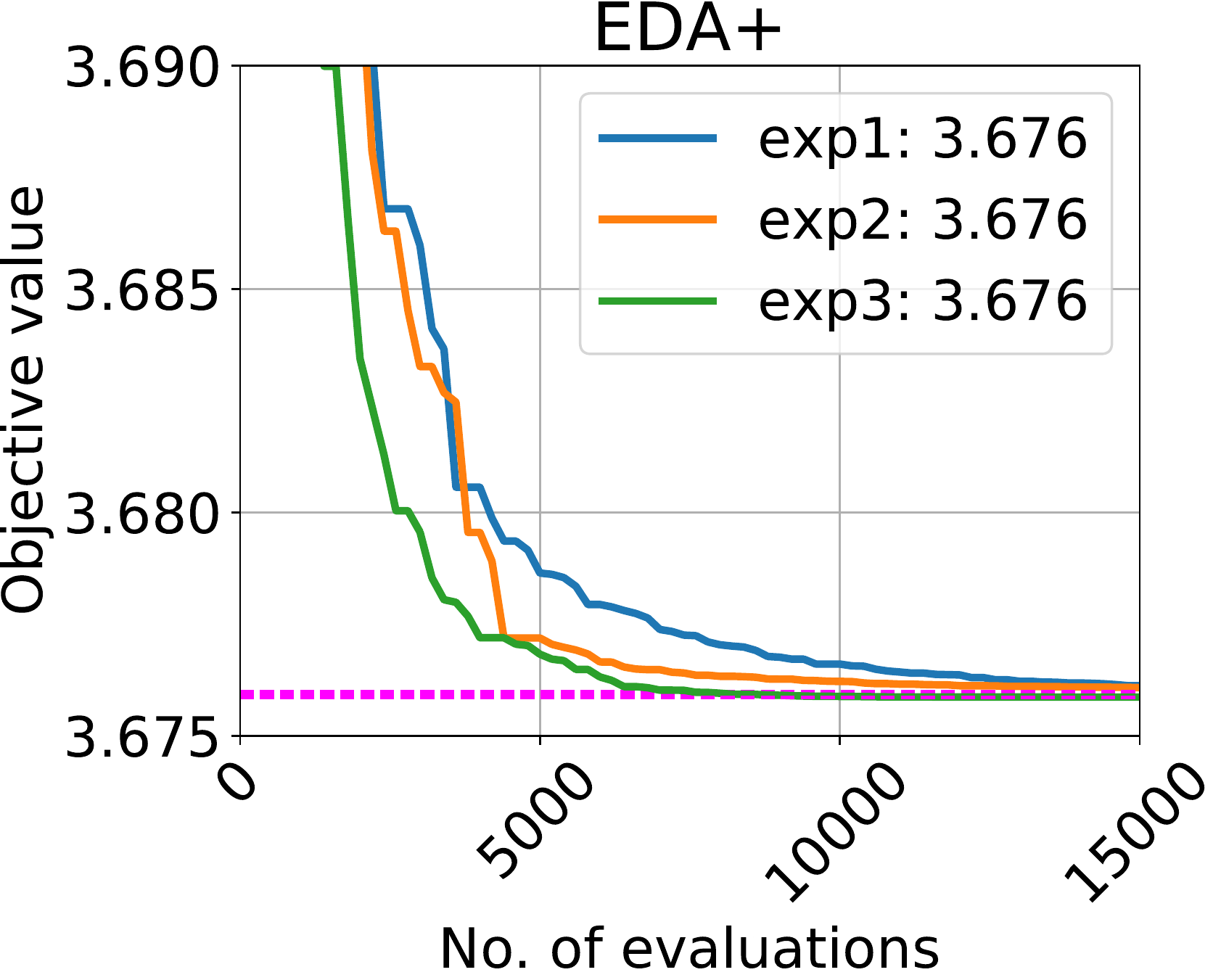} \\
        \textbf{B} & \\
        \includegraphics[width=145px,height=113px]{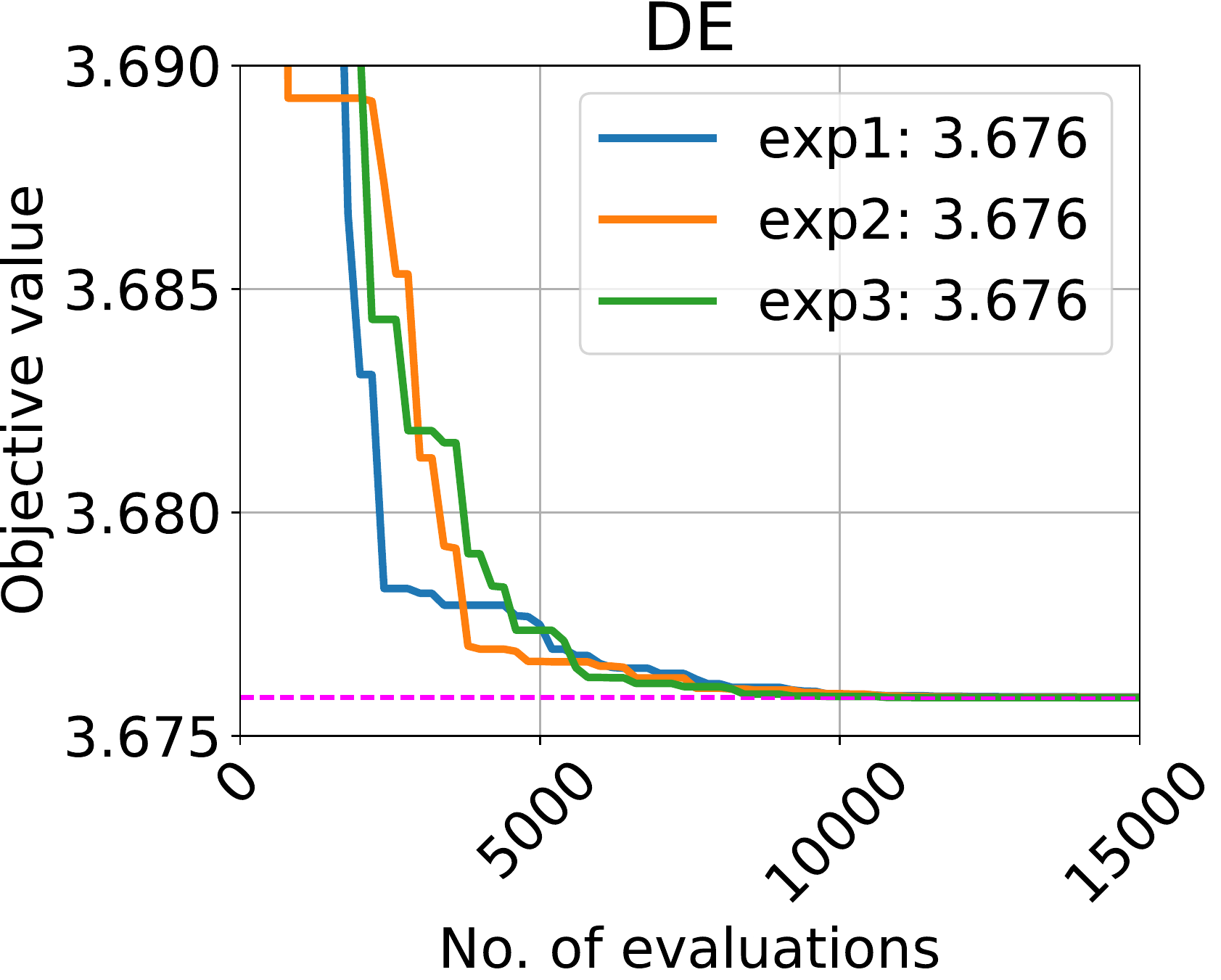} &
        \includegraphics[width=145px,height=113px]{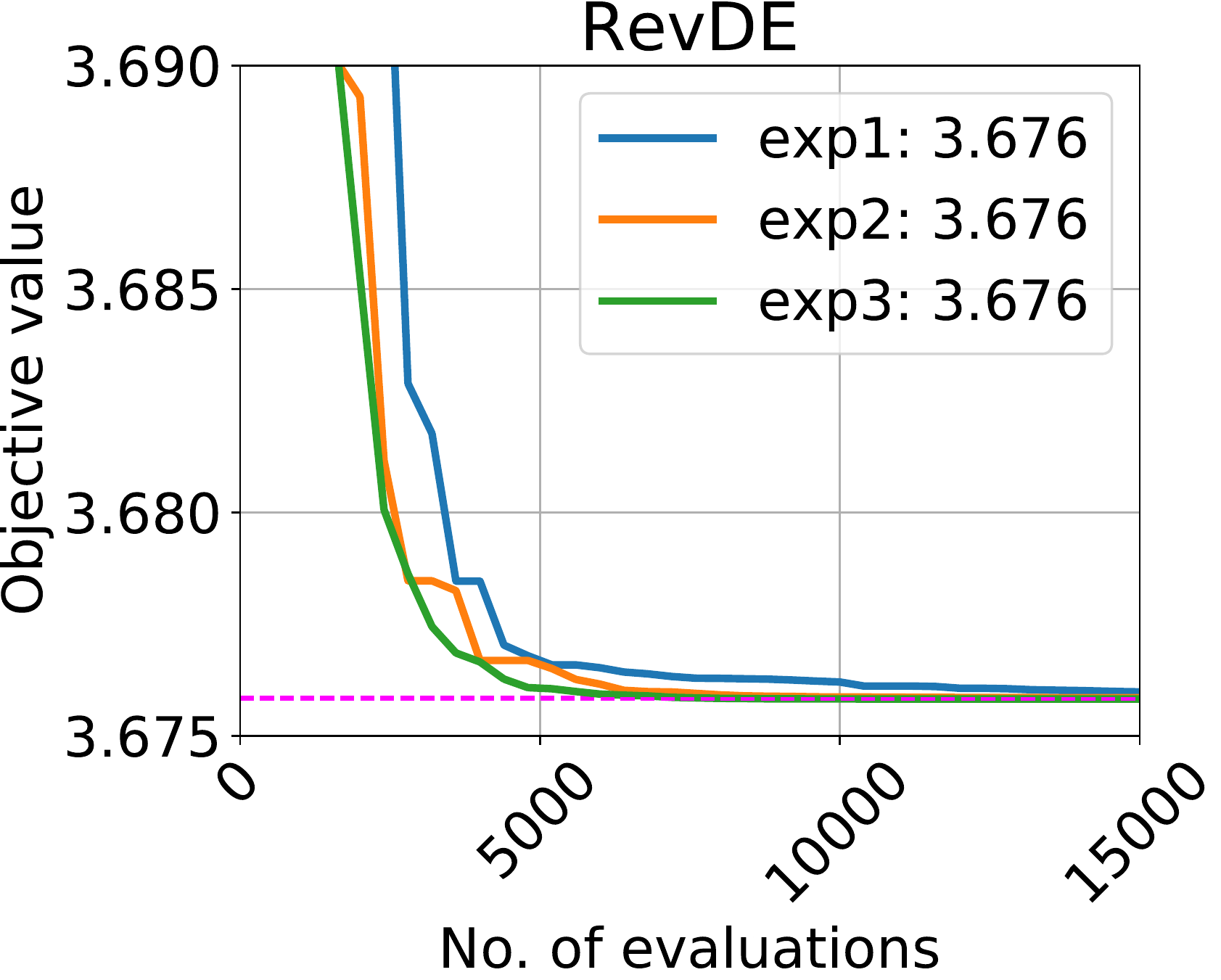} &
        \includegraphics[width=145px,height=113px]{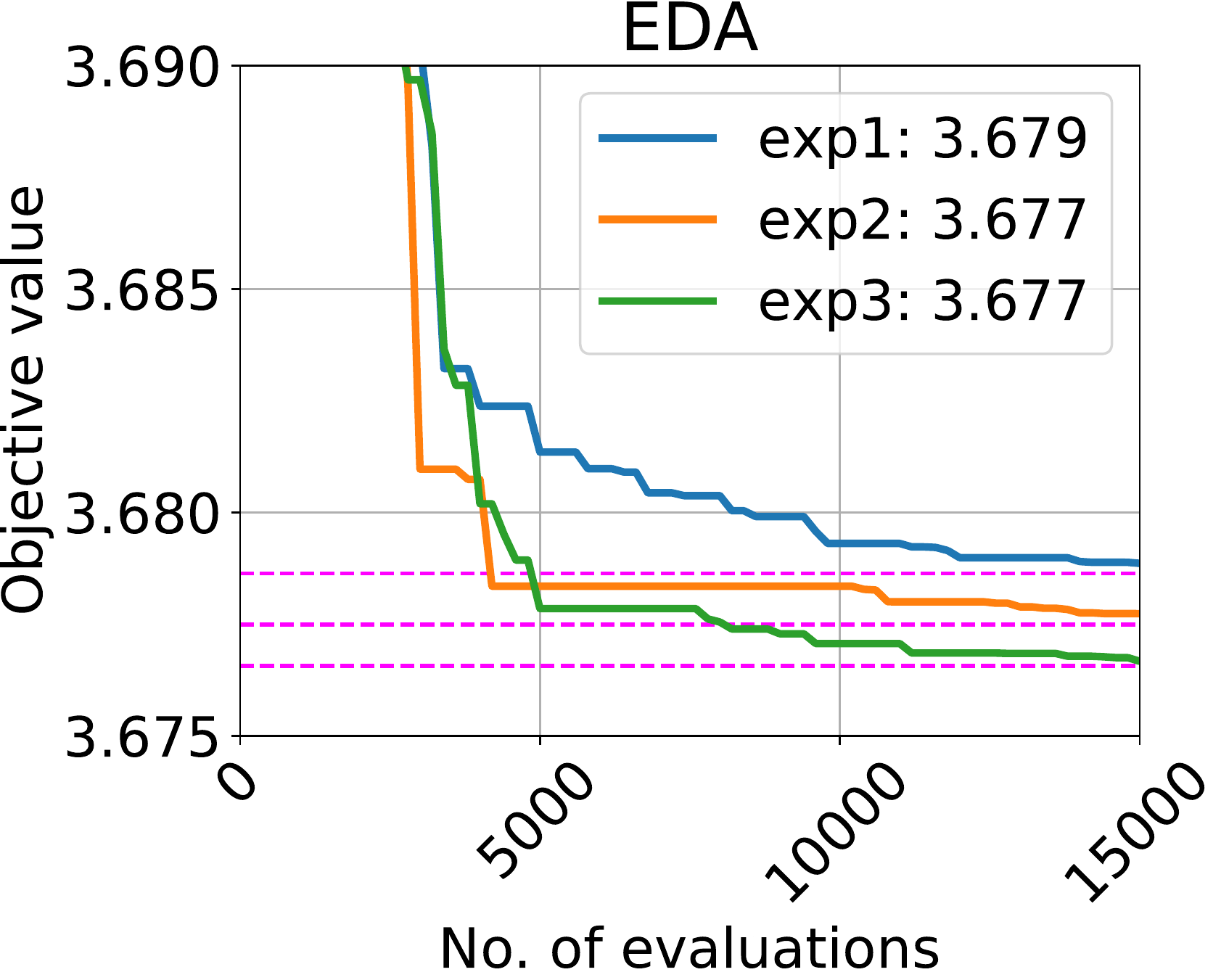} \\
        \includegraphics[width=145px,height=113px]{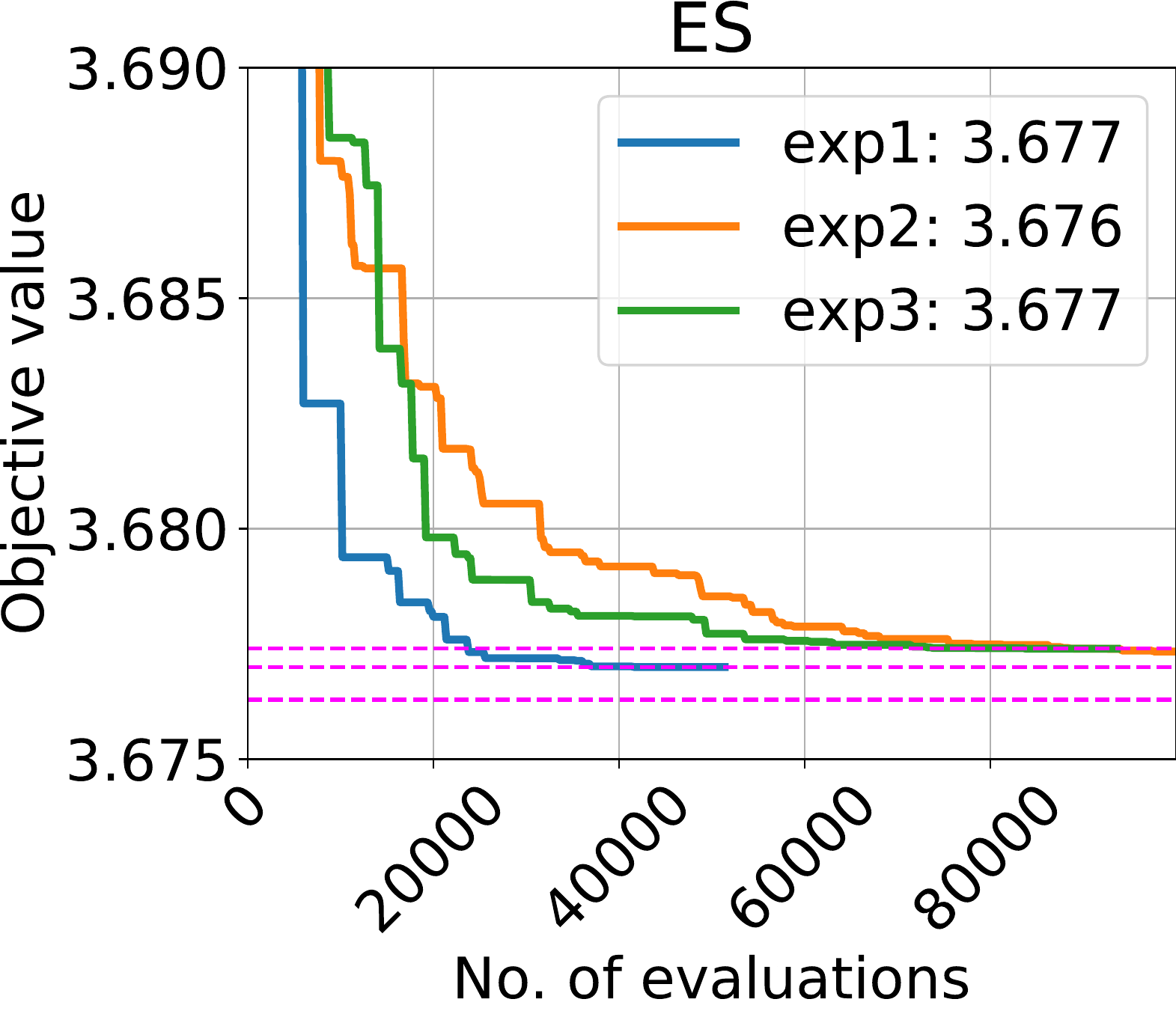} &
        \includegraphics[width=145px,height=113px]{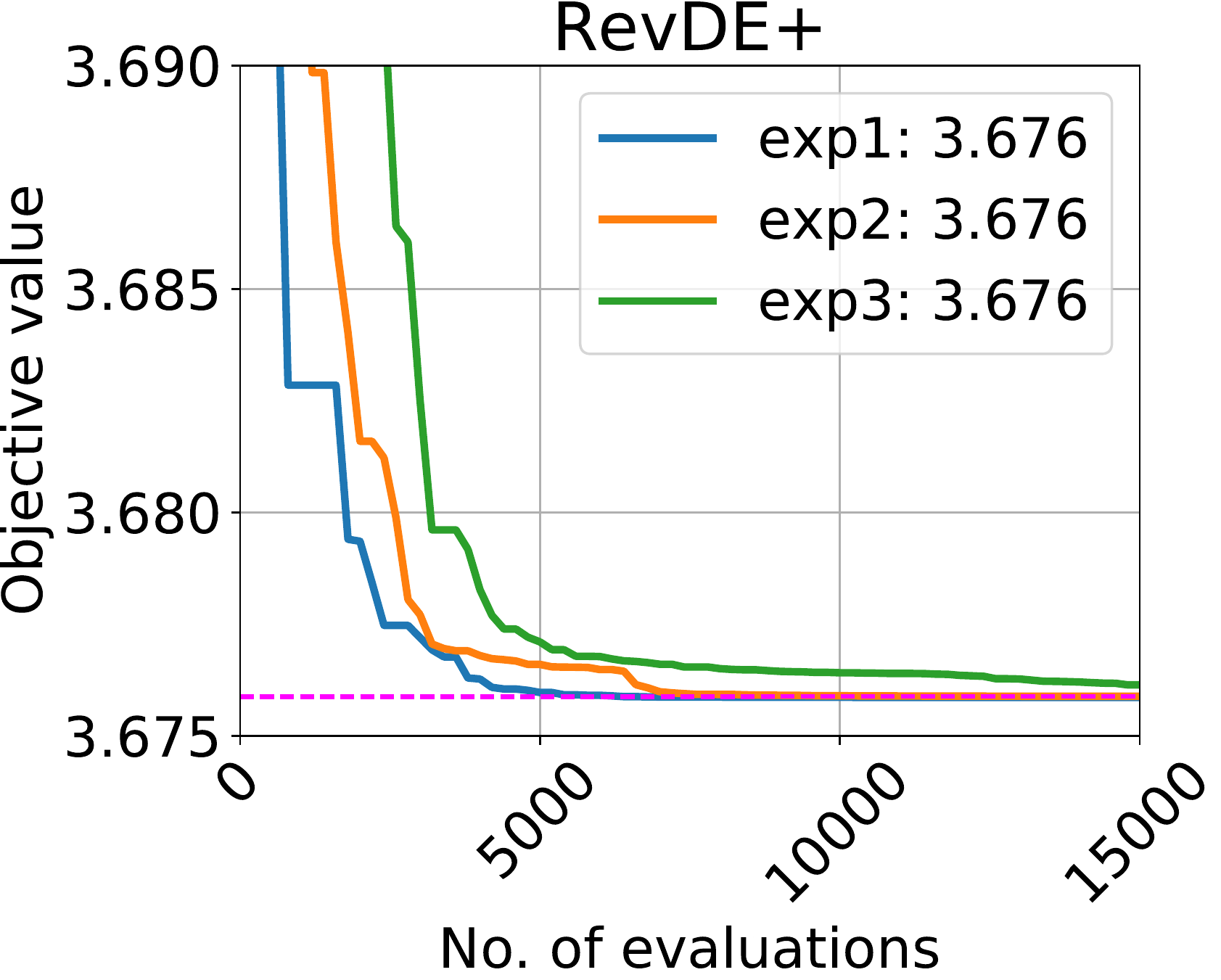} &
        \includegraphics[width=145px,height=113px]{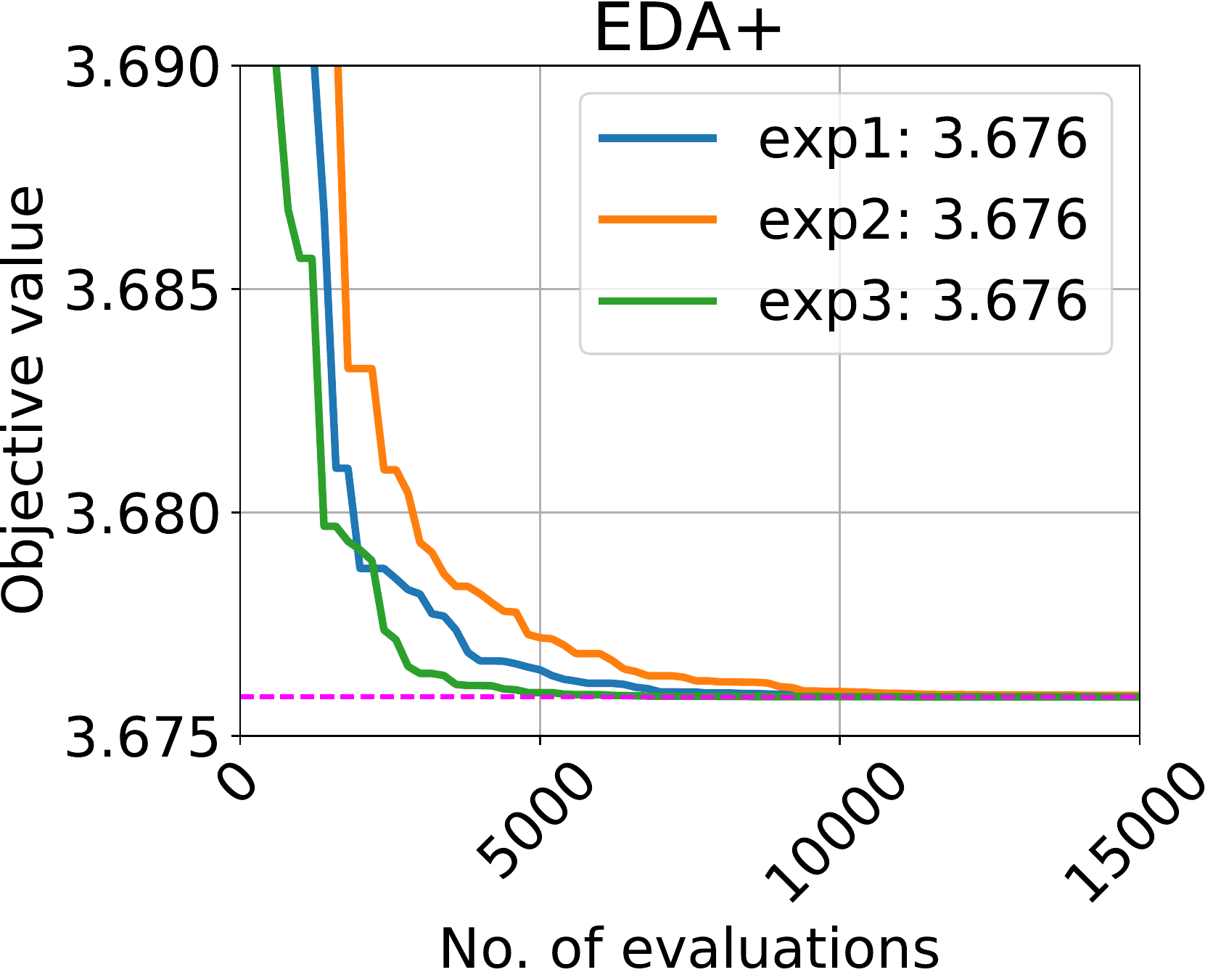}
    \end{tabular}
    \caption{The convergence of the population-based optimization methods over $3$ runs: \textbf{A} Case 1, \textbf{B} Case 2 (\textit{mutation}). In the legends, we indicate the value of the fitness function after the methods converged.}
    \label{fig:energies}
\end{figure*}

% =====SECTION=====
\section{Results}

% -----SubSECTION-----
\subsection{Experimental setup}

\paragraph{Model} 
In order to verify whether it is possible to identify parameters in the glycolysis process in \textit{Saccharomyces cerevisiae} by observing only a subset of metabolites, we consider a model presented in \cite{wolf2000transduction}.
The model consists of $9$ ordinary differential equations and $11$ reaction with $18$ parameters.
In the Appendix \ref{app:model}, we present details about the model, as well as initial conditions, and parameter values measured in \cite{wolf2000transduction} (\textit{real parameter values}).
We treat the system of ordinary differential equations as the simulator.

In the original model in \cite{wolf2000transduction}, the authors were focused on oscillatory character of the system, therefore, they assume a constant injection of \glu (see reaction $v_1$ in Figure \ref{fig:glycolysis}, \ie, the glucose transporter).
However, we consider other scenario where there is only an initial input of glucose. 
For this purpose, we set $k_0$ in $v_1$ equal $0$.

\paragraph{Observations} 
In the experiments we assume only \glu, \atp, \nad, \ac, and \ace\ are observed.
We generate the observed metabolites by running the simulator with the real parameter values.
In order to mimic real measurements that are typically noisy, we add a Gaussian noise with zero mean and the standard deviation equal $3\%$ of a generated value of a metabolite at a given time step.
We notice that adding noise prohibits finding a solution (\ie, values of parameters) that achieves error defined in Eq. \ref{eq:fitness} equal zero.

We repeat experiments three times.
For each repetition, we set the length of a timecourse to $T=30$.

\paragraph{Two cases}
Our main research goal is the parameter identification of a partially observable system of multiple biochemical reactions.
However, as highlighted in the introduction, a proper identification of parameters could be used for fingerprinting normal and abnormal biochemical processes. 
Therefore, we aim at developing optimization methods that allow to identify parameters of reactions that are not directly observed.
For this purpose, we distinguish two cases.
In the \textit{Case 1} we use the model as described before.
In the \textit{Case 2} we assume a \textit{mutation} of the reaction $v_3$ that combines two unobserved metabolites (\fru\ and \triop, see Figure \ref{fig:glycolysis}). 
We simulate the perturbation by changing the value of the parameter $k_2$ from $9.8$ to $4.9$. 
It this difficult to predict how such relatively small modification influences the whole system, and, thus, it serves as an import case study for optimization methods.

\paragraph{Quantitative and qualitative evaluation}
In order to answer our research questions, we use the following evaluation measures.
First, we monitor a convergence of the optimization methods by plotting the error in Eq. \ref{eq:fitness}.
We are interested whether an optimizer converges to a minimum, and how fast it is achieved.
The speed of convergence is defined by the number of evaluated individuals by a population-based methods.

Second, we qualitatively inspect the difference between the simulator output of unobserved metabolites and the real metabolites.
The qualitative evaluation is given by the Eq. \ref{eq:fitness}, however, it is also important to obtain an insight into how a potentially misidentified parameter result in a metabolite timecourse.

Third, since we know the real parameter values, we can also evaluate a difference between them and the best values found by the optimization methods.
We use the absolute value of the difference of two values.
We calculate the mean and the standard deviations of the difference from three runs, and use the cumulative distribution function of the folded normal distribution\footnote{The difference between two real-valued random variable is normally distributed. However, taking the absolute value of a normally distributed random variable results in the folded normal distribution.} to visualize the distribution of differences (the ideal case is $0$).

\paragraph{Hyperparameters of optimizers}
For all optimization methods, we set the population size to $N=100$. All optimizers run maximally $1000$ generations.
In the case of ES, we use the initial value of $\sigma$ equal $0.1$.
For DE, RevDE, and RevDE+, we use $F=0.5$, and $p=0.9$.
For EDA we take $M=100$.
In the case of EDA+ and RevDE+, we use the $K$-NN as the surrogate model with $K=3$, and we do not store more than $10,000$ evaluated individuals.

\paragraph{Implementation details} All computational methods are implemented in Python using standard packages (\eg, \texttt{numpy}, \texttt{scipy}). 
We also use the Python Simulator for Cellular Systems (PySCeS) library \cite{OliRohHof05} that is an extendable toolkit for the analysis and investigation of cellular systems.
It allows to import a model represented in a human-readable manner, and solve a system of differential equation using a built-in solver.
In the experiments, we downloaded the model proposed in \cite{wolf2000transduction} from the JWS Online Database \cite{olivier2004web}, available under the following link \cite{wolf}.
All implemented population-based optimization methods as well as the experiments are available online: \url{http://XXX}.

\begin{figure}[!htbp]
    \begin{tabular}{lc}
        \textbf{A} &  \\
        \includegraphics[width=110px,height=90px]{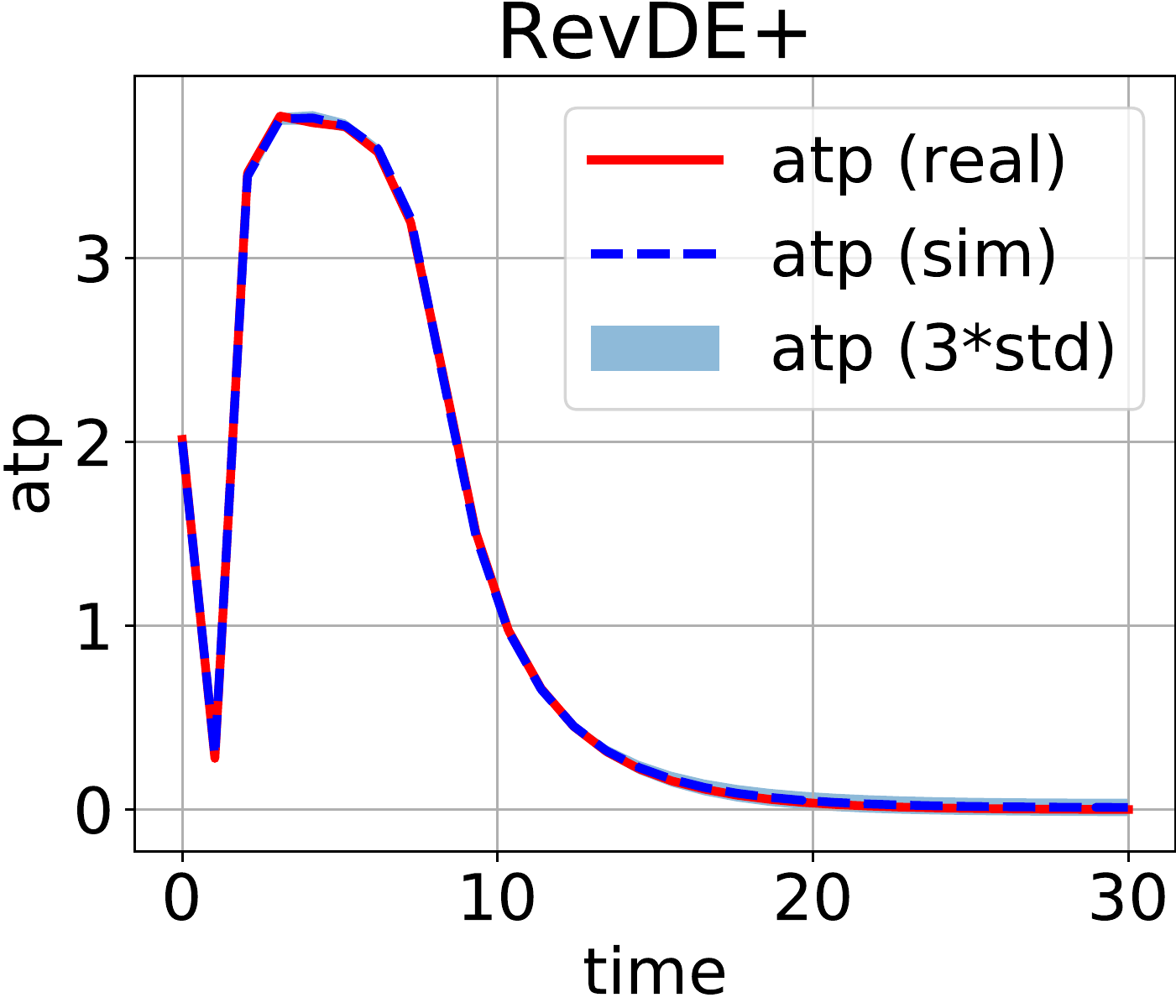} & 
        \includegraphics[width=110px,height=90px]{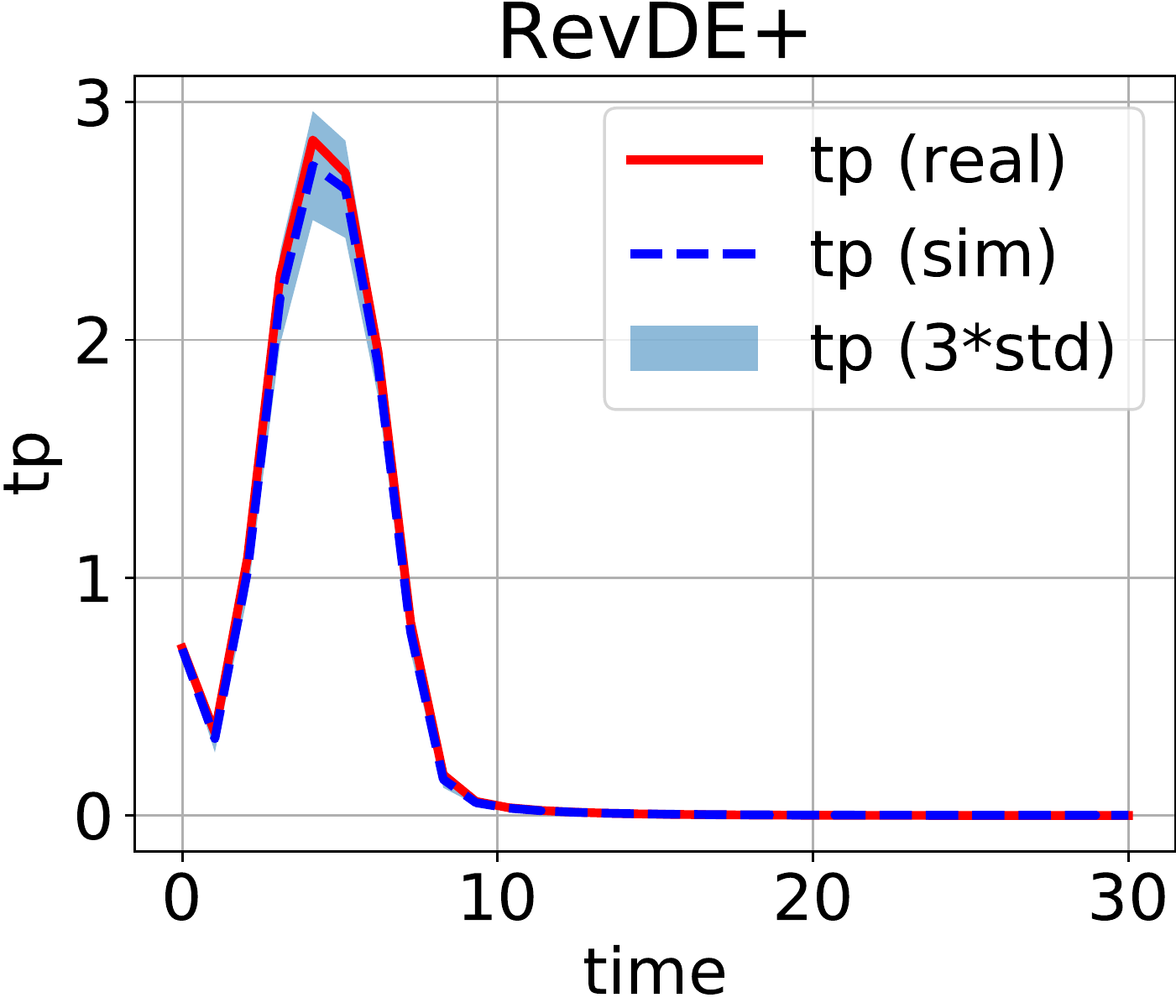}  \\
        \textbf{B} & \\
        \includegraphics[width=110px,height=90px]{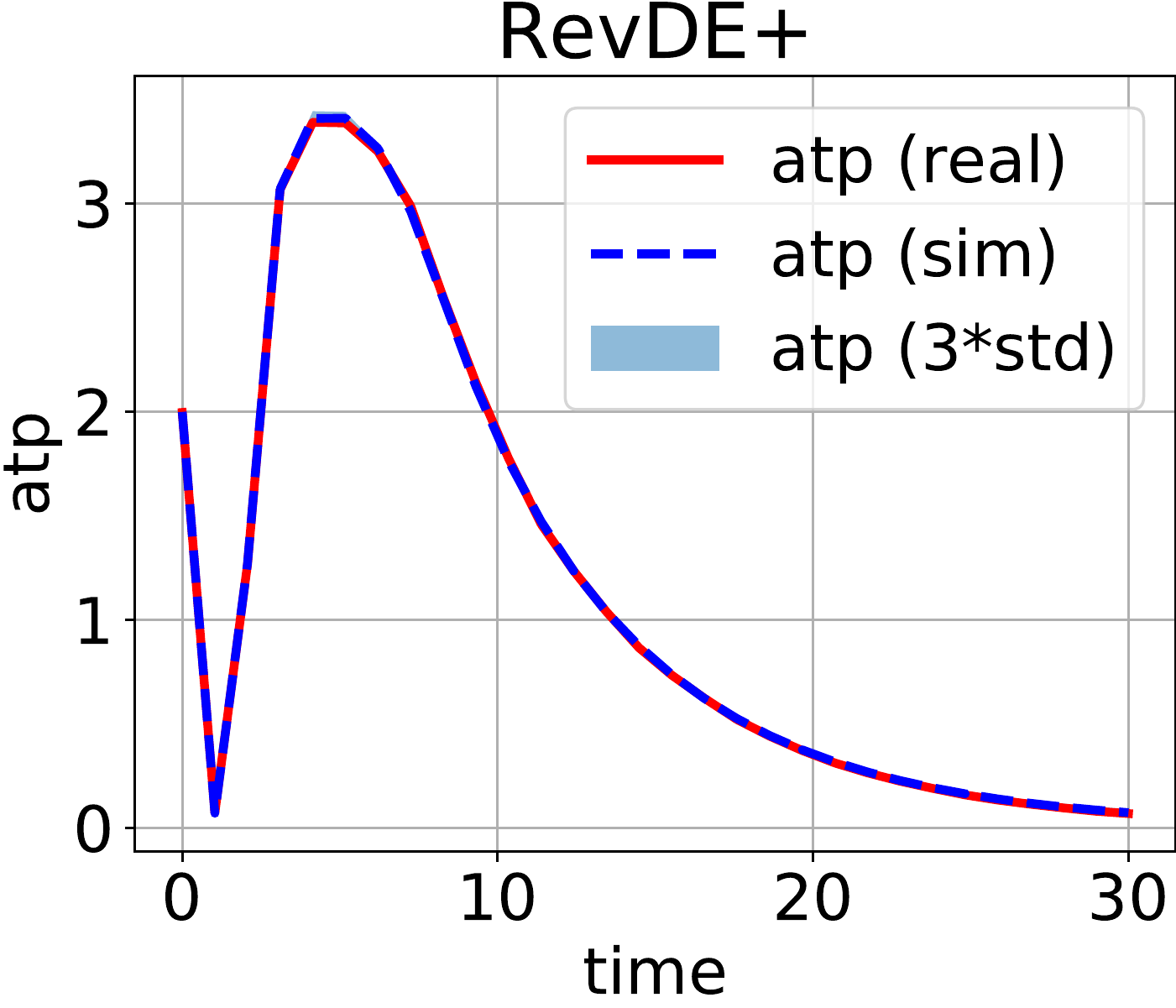} & 
        \includegraphics[width=110px,height=90px]{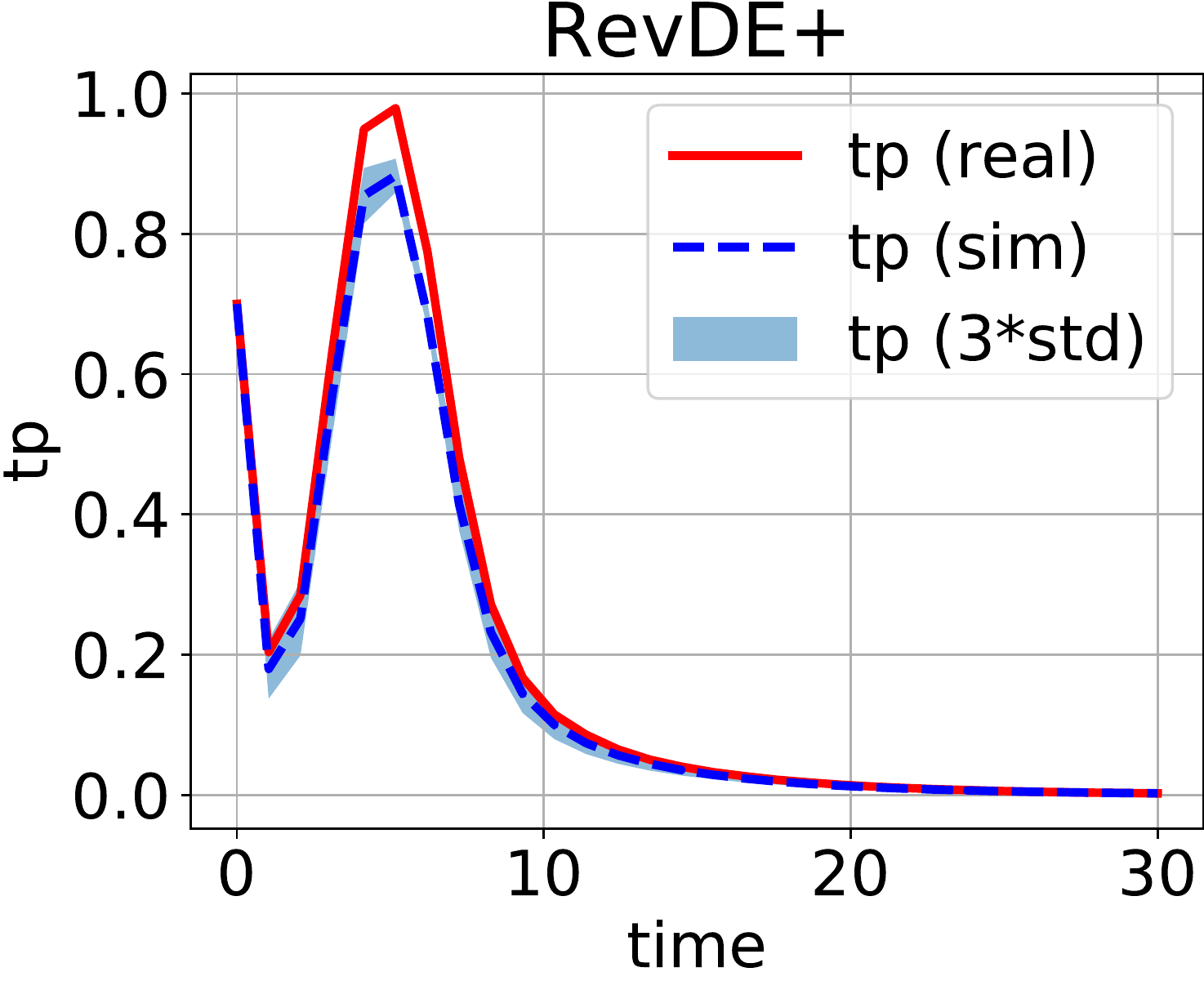} 
    \end{tabular}
    \caption{A comparison of the timecourses of the selected two metabolite for RevDE+: \textbf{A}: Case 1, \textbf{B}: Case 2 (\textit{mutation}). Real timecourses are depicted in red, and the average value and a confidence interval ($3\times$ standard deviation) over $3$ runs of the simulator are in blue.}
    \label{fig:obs}
\end{figure}

% -----SubSECTION-----
\subsection{Discussion}

\paragraph{Fitness value} In Figure \ref{fig:energies} we present convergence of the methods for Case 1 (Figure \ref{fig:energies}.A) and Case 2 (Figure \ref{fig:energies}.B).
We notice that all methods were able to converge and achieve very similar fitness values.
However, the (1+1)-ES method was slowest due to the slow exploration capabilities.
EDA also required more evaluations to obtain better results.
Interestingly, DE, RevDE, RevDE+ and EDA+ achieved almost identical values of the fitness function (the differences were beyond the three digit precision).
An important observation is that application of the surrogate model (the $K$-NN regressor) allowed to speed up the convergence of RevDE+ compared to RevDE significantly. Moreover, in the case of EDA, the surrogate model allowed a better exploration ($M = 5N$) and, thus, EDA+ obtained better results in a significantly less number of evaluations.
We conclude that all population-based methods were able to converge and achieved almost identical scores, and our proposition of applying the surrogate model led to improving RevDE and EDA

\paragraph{Timecourses} The final value of the fitness function tells us how well the simulator models the observed timecourses for given parameters provided by an optimizer.
Additionally, we can also qualitatively inspect the timecourses both the observed and unobserved metabolites.
In Figure \ref{fig:obs} we present exemplary timecourses for \atp (the observed metabolite), and \tp (the unobserved metabolite), for parameter values found by RevDE+.
Since all methods obtained very low errors close to noise in data (see Figure \ref{fig:energies}), we show the timecourses of the unobserved metabolites in Figures \ref{fig:obs_1} and \ref{fig:obs_2} for Case 1 and Case 2, respectively, in the Appendix \ref{app:results}.

In all cases except ES for \tp, the best parameter values found by the optimizers resulted in timecourses that are almost identical to the real observations.
For all unobserved metabolites the average over $3$ repetitions of the experiments overlapped with the real value, or lied within the confidence interval ($3\times$ standard deviation).
This is a result that we hoped for since being able to generate unobserved metabolite is extremely important for analysing biological systems.

\begin{figure*}[tbp]
    \begin{tabular}{lccc}
        \textbf{A} & & & \\
        \includegraphics[width=110px,height=90px]{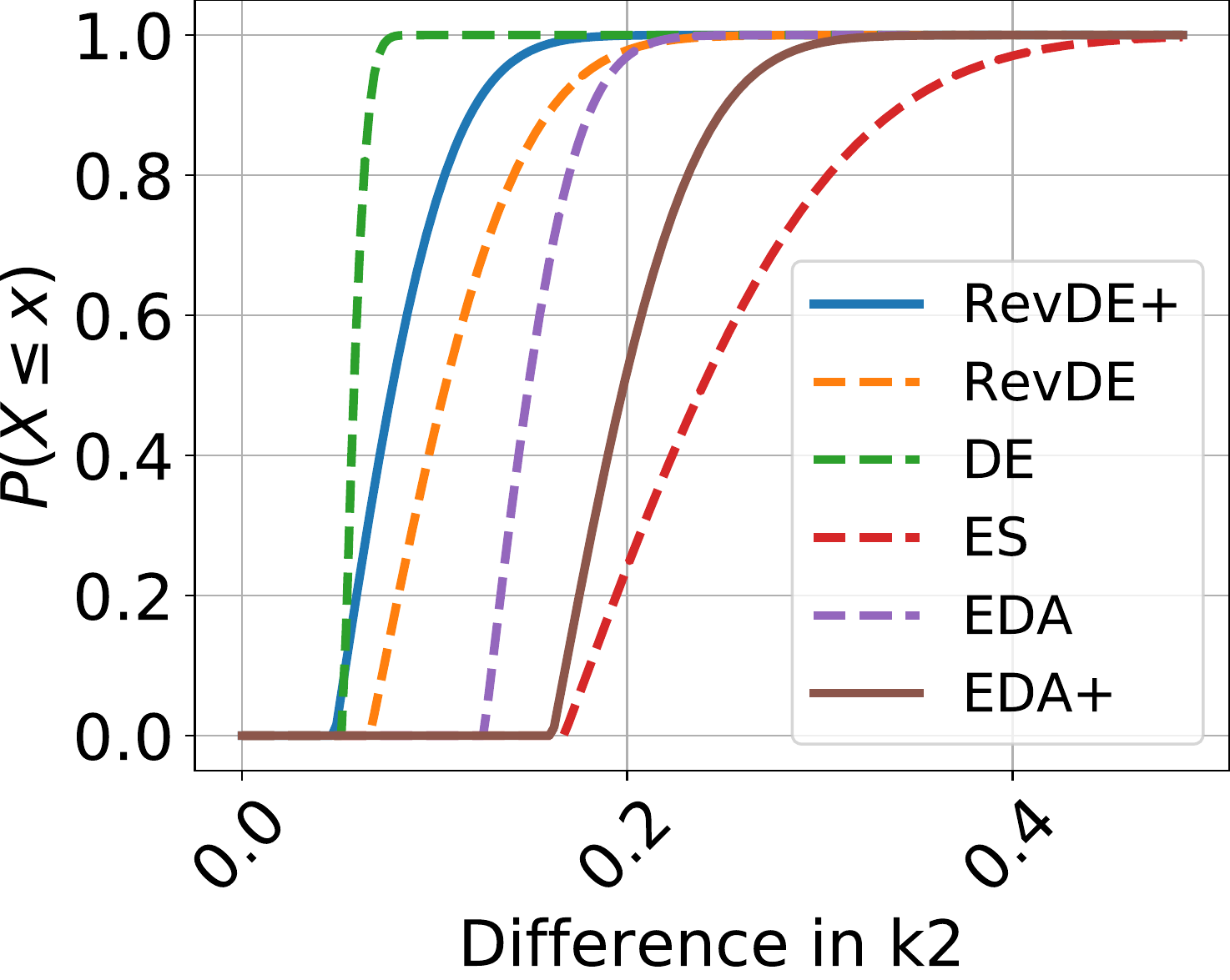} & 
        \includegraphics[width=110px,height=90px]{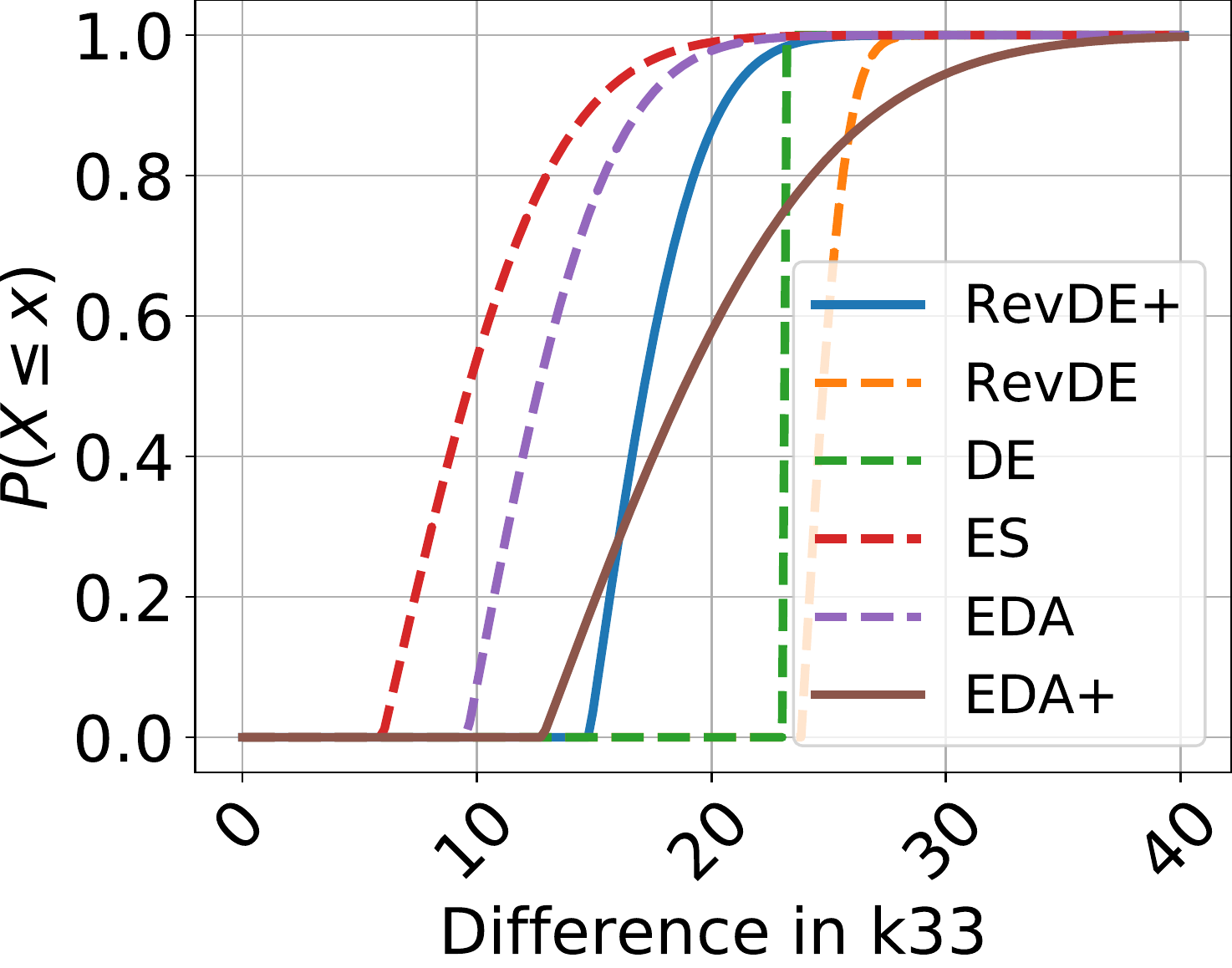} &
        \includegraphics[width=110px,height=90px]{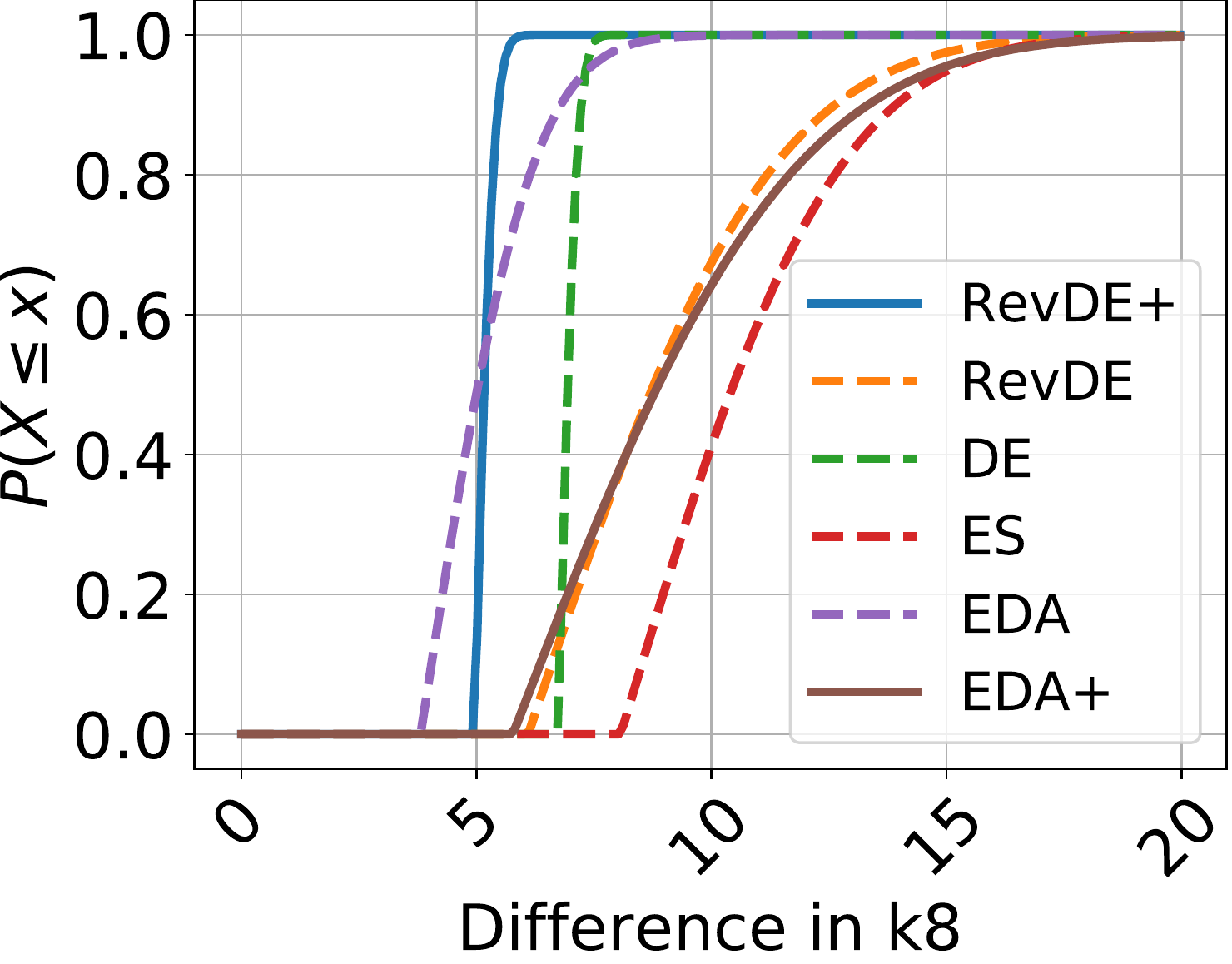} &
        \includegraphics[width=110px,height=90px]{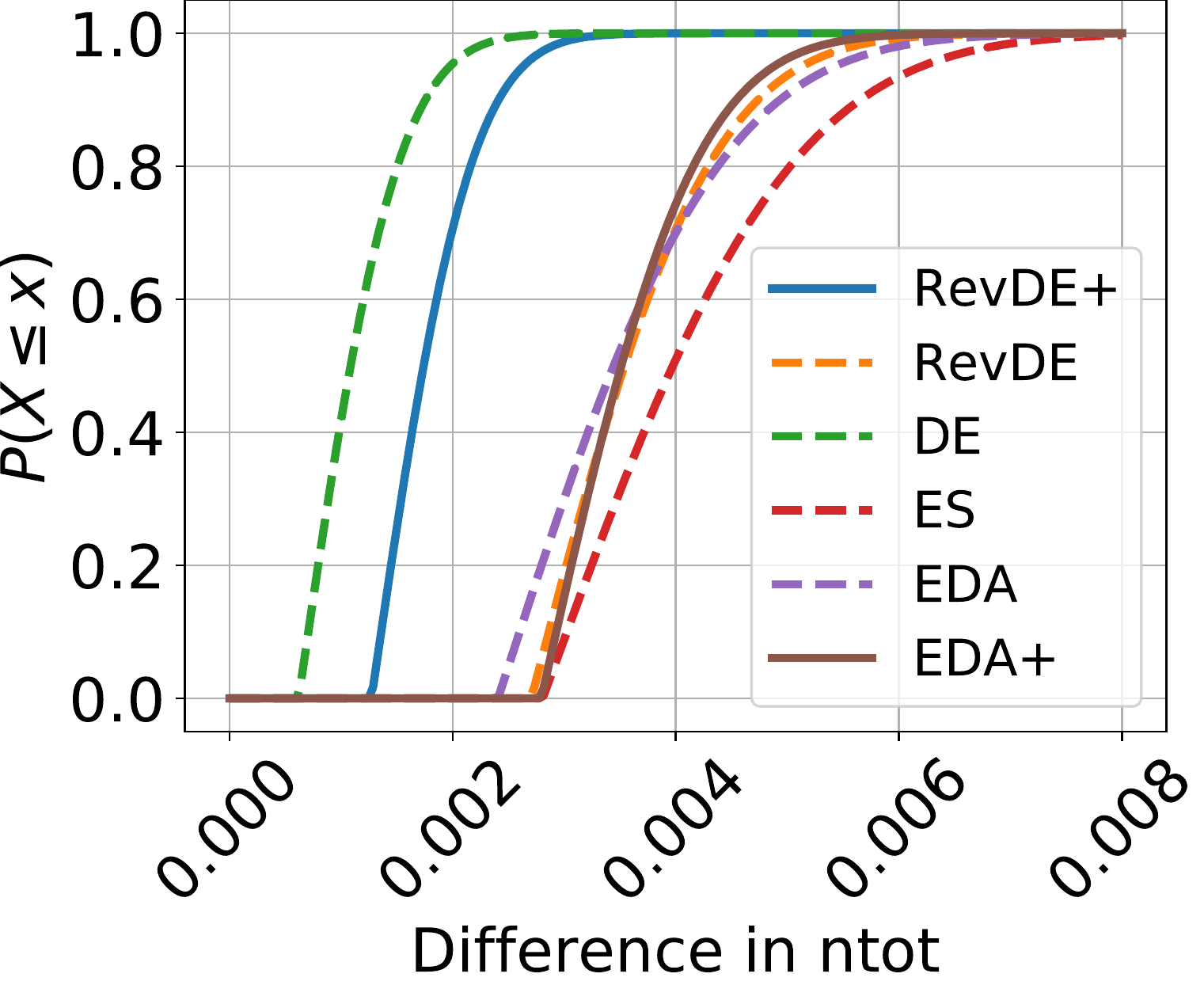} \\
        \textbf{B} & & & \\
        \includegraphics[width=110px,height=90px]{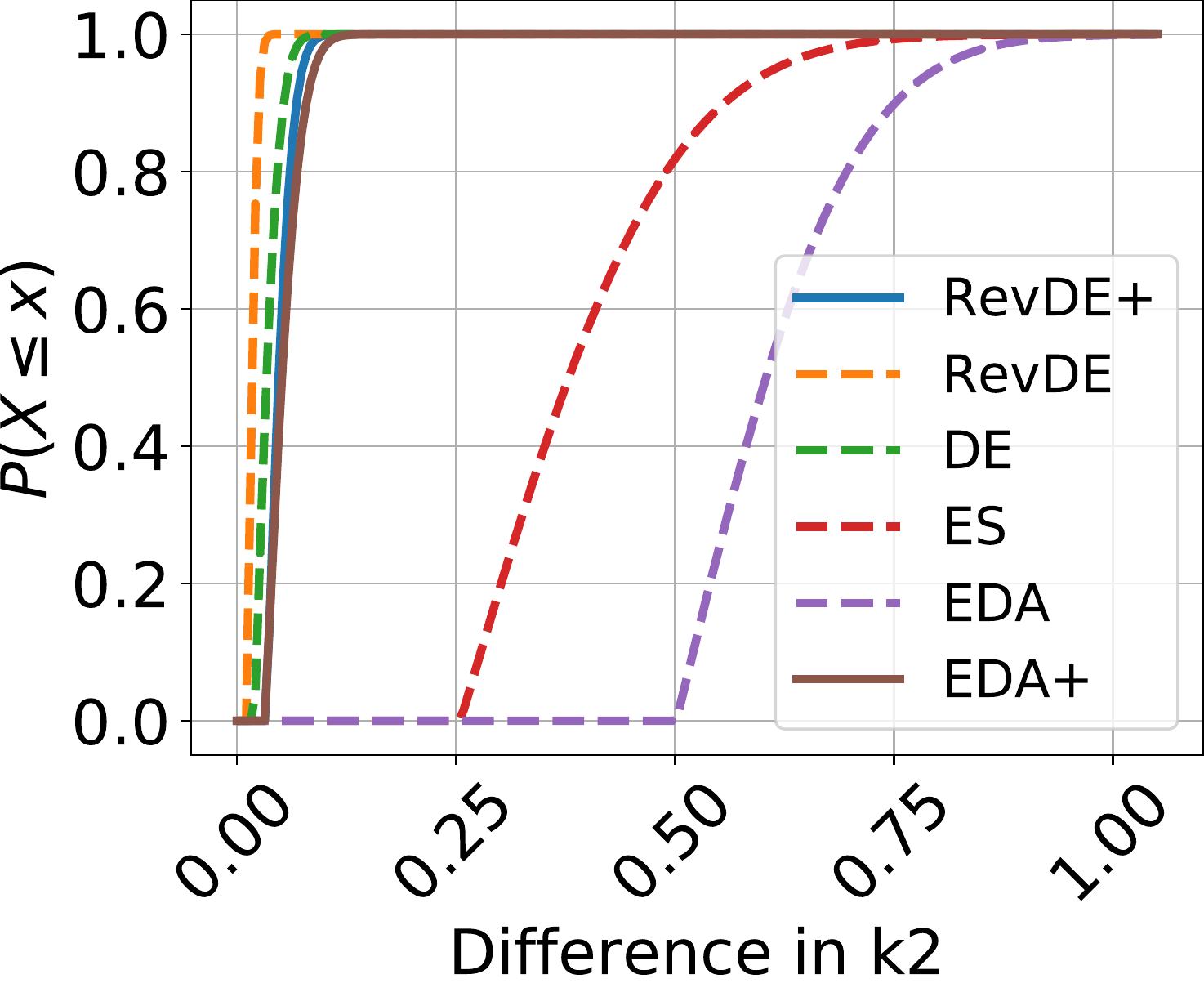} & 
        \includegraphics[width=110px,height=90px]{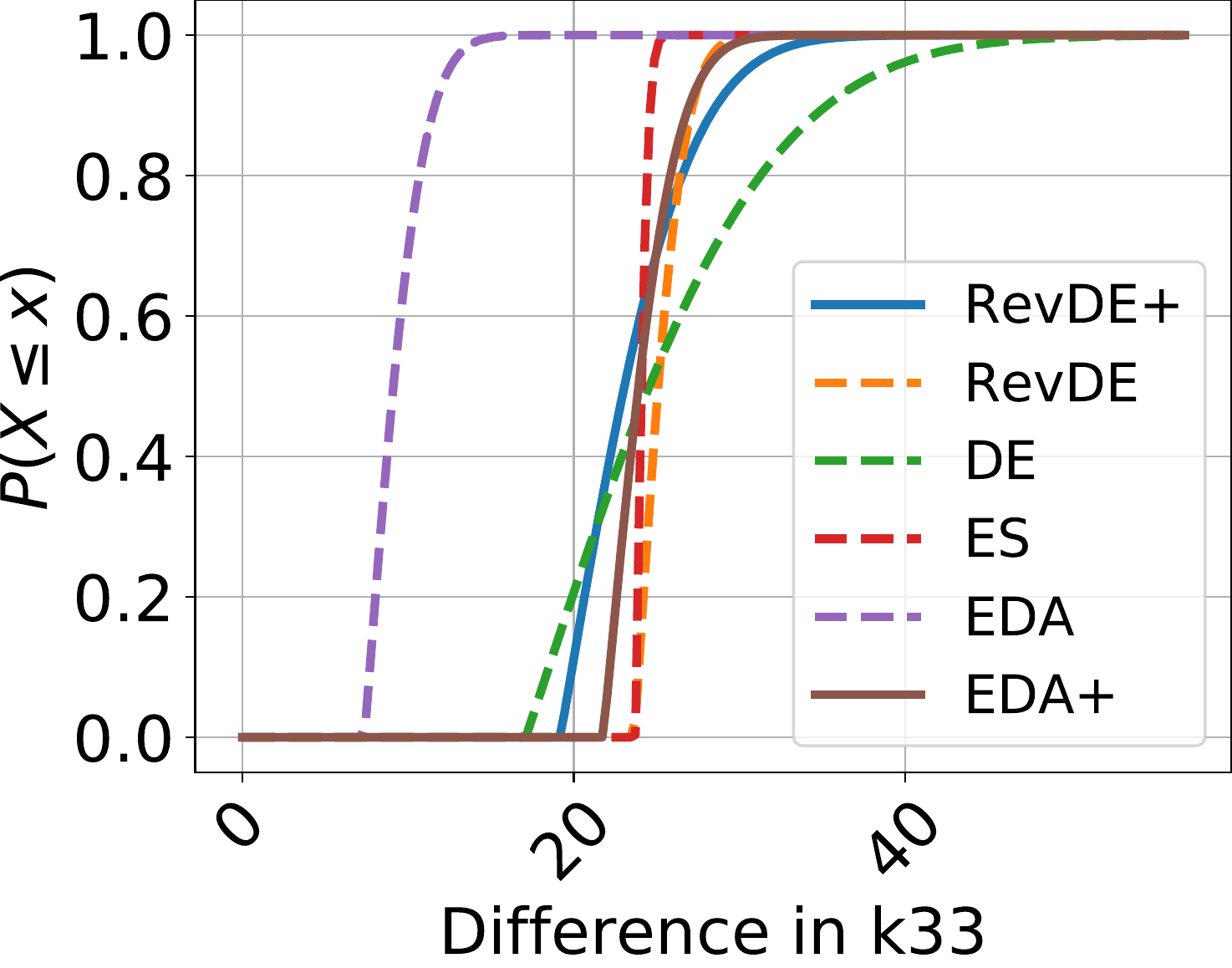} &
        \includegraphics[width=110px,height=90px]{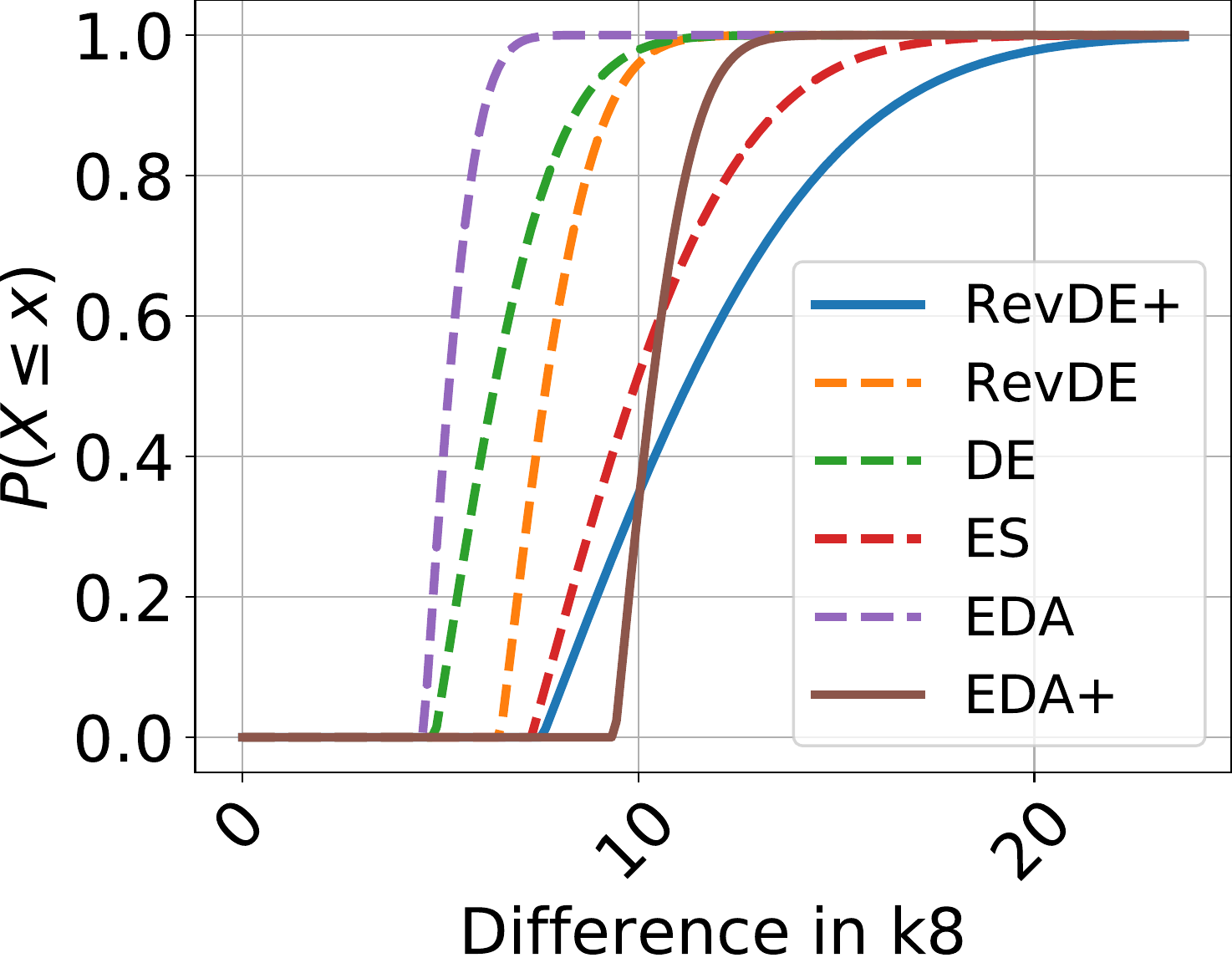} &
        \includegraphics[width=110px,height=90px]{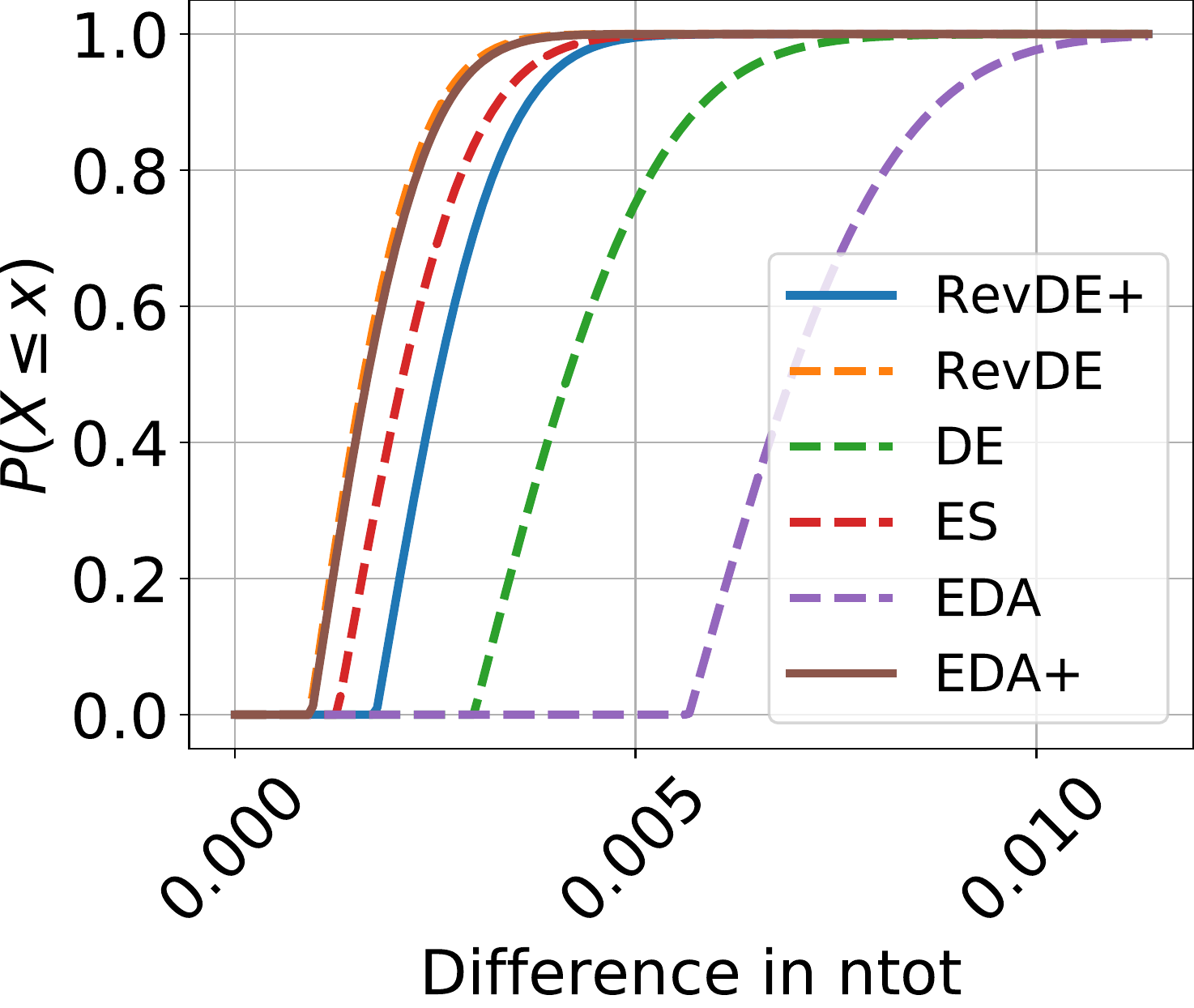}
    \end{tabular}
    \caption{The cumulative distribution functions (cdfs) of the differences for selected parameters: \textbf{A}: Case 1, \textbf{B}: Case 2 (\textit{mutation}). Ideally, a cdf of an optimization method should resemble a step-function centered at $0$. The averages and the scales are calculated over $3$ repetitions of the experiment.}
    \label{fig:differences}
\end{figure*}

\paragraph{Differences in parameters}
In this paper, we know precisely the values of the parameters since they were measured in \cite{wolf2000transduction} and, in Case 2, we modify one parameter by hand.
Hence, we can compare the parameter values found by the optimization methods with the real parameter values.
In Figure \ref{fig:differences} we present four parameters for which we see a clear difference between Case 1 and Case 2.
Difference of all parameters are included in the Appendix \ref{app:results}, in Figures \ref{fig:differences_1} and \ref{fig:differences_2}.
In general, the differences are marginal and we can conclude that all parameter values were properly identified.
The biggest problems though appear for parameters that have very large values, \eg, $k_8$ or $k_{33}$. 

Interestingly, we noticed some significant differences for parameter values found by the methods between Case 1 and Case 2.
For instance, for $k_{33}$, most methods achieved a difference around $10-15$ in Case 1, while it was almost doubled in Case 2.
This result shows that parameter identification in complex biological systems is a challenging task and its behavior cannot be easily predicted upfront.

% =====SECTION=====
\section{Conclusion}
Here, we propose a new optimization framework for parameter identification of biological systems. We assume a dynamical model of the system with initial conditions, measurements of selected metabolites, and an ODE solver together with other tools to represent the model in machine-readable format as inputs in our framework (see Figure \ref{fig:our_approach}). Since the cost of solving the model is relatively low, we propose to utilize the population-based optimization methods to identify the parameters of the system. As a result, once we have found the parameters values, we can run simulations and generate timecourses of all metabolites, including the ones that are unobserved, and further analyze the biological system. In this paper, we provide a proof-of-concept of one of the most important metabolic processes, namely, the glycolysis pathway, of the well-studied \textit{Saccharomyces cerevisiae}, known also as baker's yeast.

We outlined the general scheme of population-based optimization methods, followed by a description of three classic population-based DFO algorithms, namely, differential equation (DE), an evolutionary strategy, and the univariate Gaussian estimation of distribution algorithm (EDA). Next, we described recently published extension of the differential evolution called Reversible Differential Evolution (RevDE) \cite{tomczak2020differential}. Further, in order to decrease the computational complexity of the RevDE, we proposed to utilize the $K$-NN regressor as a surrogate model. Similarly, we used the $K$-NN based surrogate model to increase exploration capabilities of the EDA. In the experiments, we showed that all population-based methods could be successfully used to identify parameters of a complex biological networks. However, it seems that too simplistic approaches (\eg, (1+1)-ES) could be slow and not accurate enough (see Figures \ref{fig:obs_1} and \ref{fig:obs_2}). The surrogate-based methods indeed achieved better scores than their vanilla counterparts with almost negligible computational burden.

In the introduction, we stated three research goals and in the experiments we achieved them all. First, we applied various population-based optimization algorithms to the parameter identification of the glycolysis pathway. In the experiments, we analyzed the performance of the methods and we noticed that: (i) all algorithms converged to (local) minima, however, ES and EDA needed more evaluations, (ii) in both cases (\ie, with the real parameter values and with the mutation) the methods converged, (iii) enhancing RevDE and EDA with surrogate models led to speeding up convergence and increasing exploration capabilities.

Second, we show in the experiments that indeed it is possible to identify parameter values while only a subset of metabolites are observed. This result is important and encouraging for further studies with larger networks. 

Last, we consider a case study with a mutation of one reaction, \ie, a value of a parameter is modified. We chose the reaction $v_3$ since it combines two unobserved metabolites (\fru and \triop) that makes the problem more challenging. In this case, the population-based optimization methods were able to find good solutions anyway that again is an essential indication of usefulness of the presented optimizers.

From the biological perspective, our work is among first that showed that the parameter identification problem of complex biological systems with limited access of observed metabolites is possible. Our paper is a positive proof-of-concept and should be further investigated. A possible research direction is to consider larger dynamical models (\ie, more reactions and more parameters), and a more thorough analysis of the parameter identification with various number of observed metabolites. The possible application of our approach is the identification of single mutation as well as phenotypic heterogeneity within a population. Therefore, our method opens up a new perspective in the field of medicine, and biology, and may result in an innovative computational-aided diagnostic and / or analytical tool.

From the computational perspective, our work indicates a great potential of population-based optimization methods in the field of biology and biochemistry. In the case of relatively low computational costs of obtaining an evaluation of parameters, the population-based methods seem to be sufficient to solve the parameter identification problem. Moreover, our results for applying surrogate models to the optimizers can be highly effective. It is a well-known fact (\eg, see \cite{jin2011surrogate}), nevertheless, we believe that the optimization with surrogate models has a great future and should be investigated in more detail.

\printcredits

\section*{Acknowledgments}
\noindent EW-T is financed by a grant within Mobilno\'s\'c Plus V from the Polish Ministry of Science and Higher Education (Grant No. 1639/MOB/V/2017/0).

% ============REFERENCES=============================
\bibliographystyle{cas-model2-names}

%\bibliography{cas-refs}

\newpage

\appendix

\begin{figure*}[!tbp]
    \centering
    \includegraphics[width=110px,height=90px]{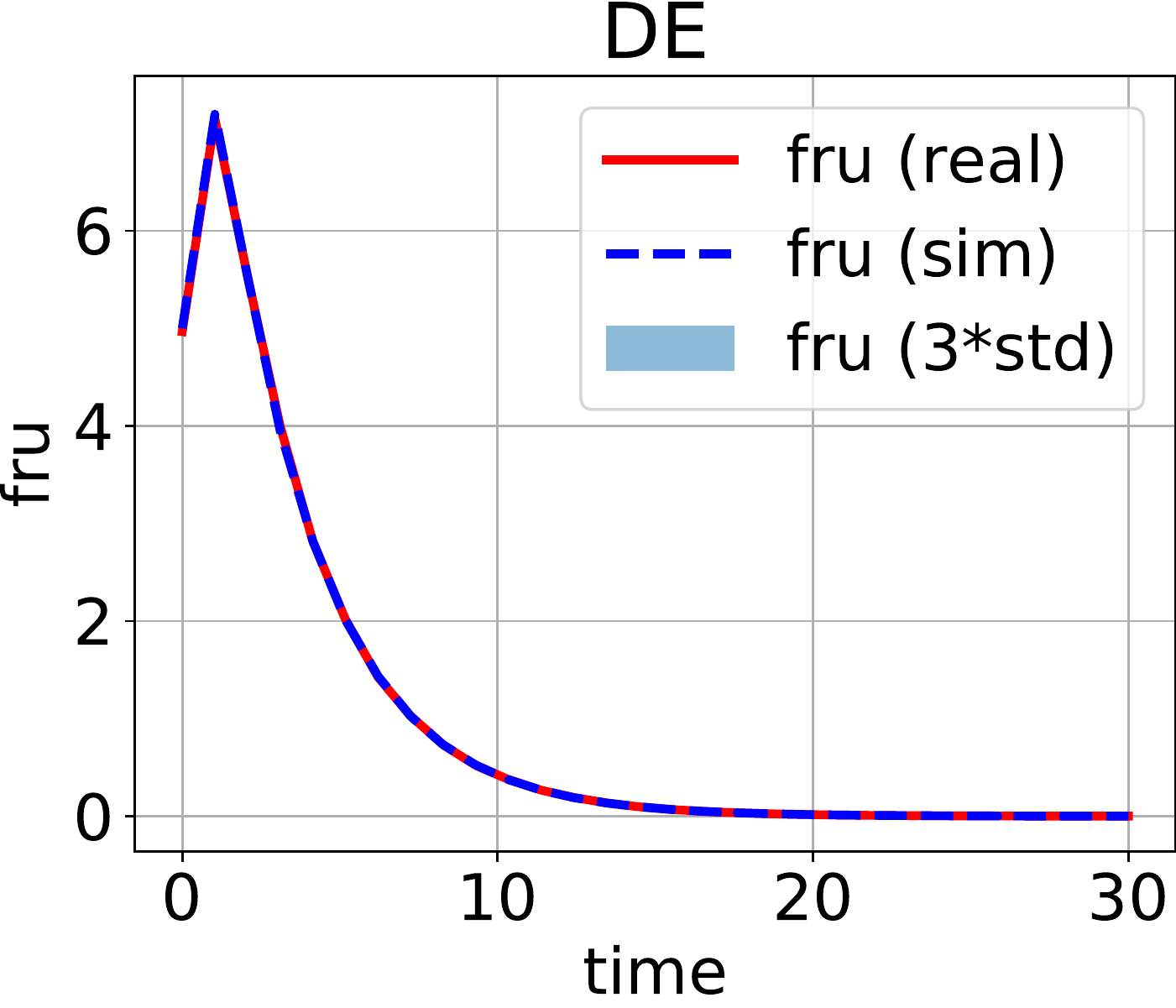}\quad
    \includegraphics[width=110px,height=90px]{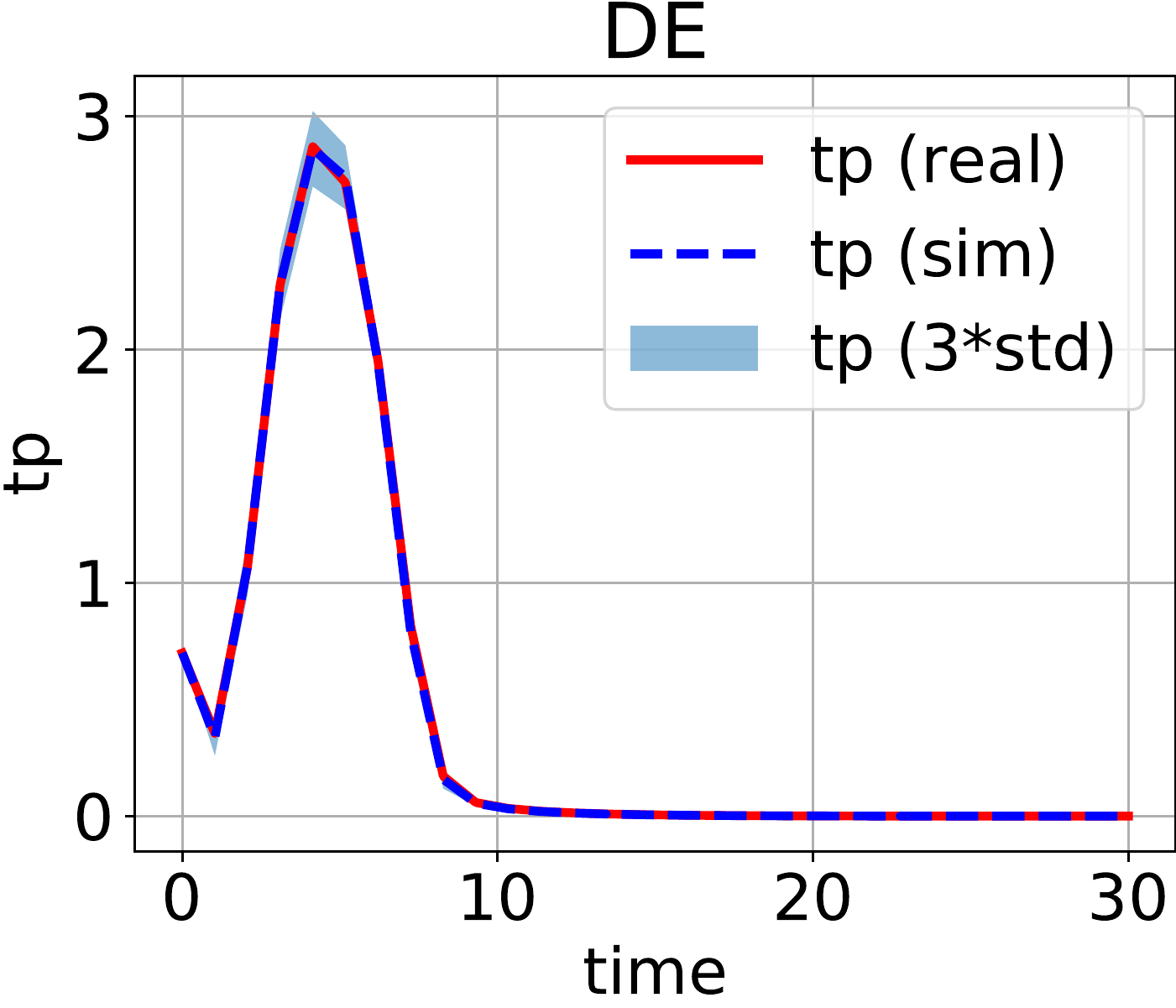}\quad
    \includegraphics[width=110px,height=90px]{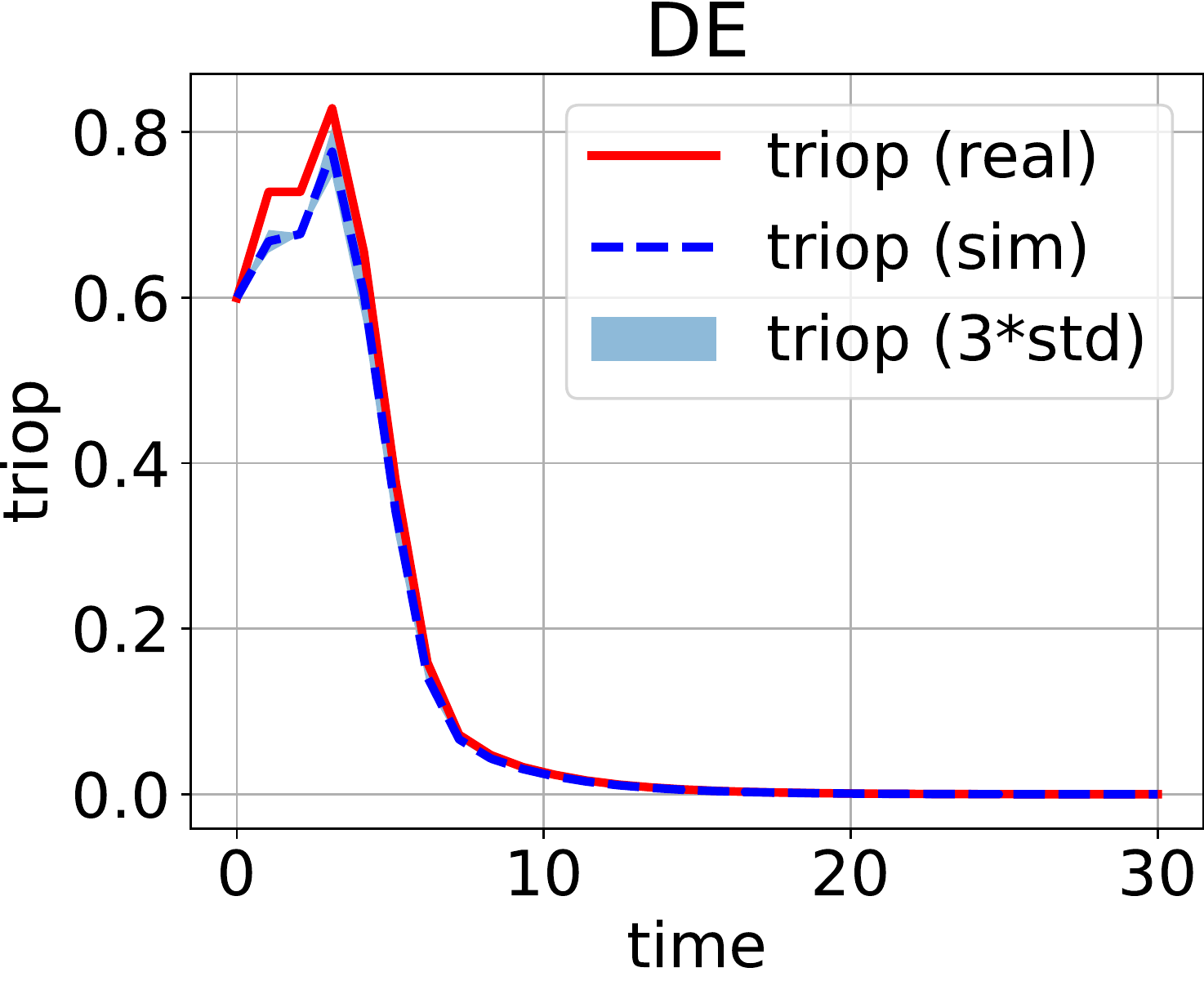}\quad
    \includegraphics[width=110px,height=90px]{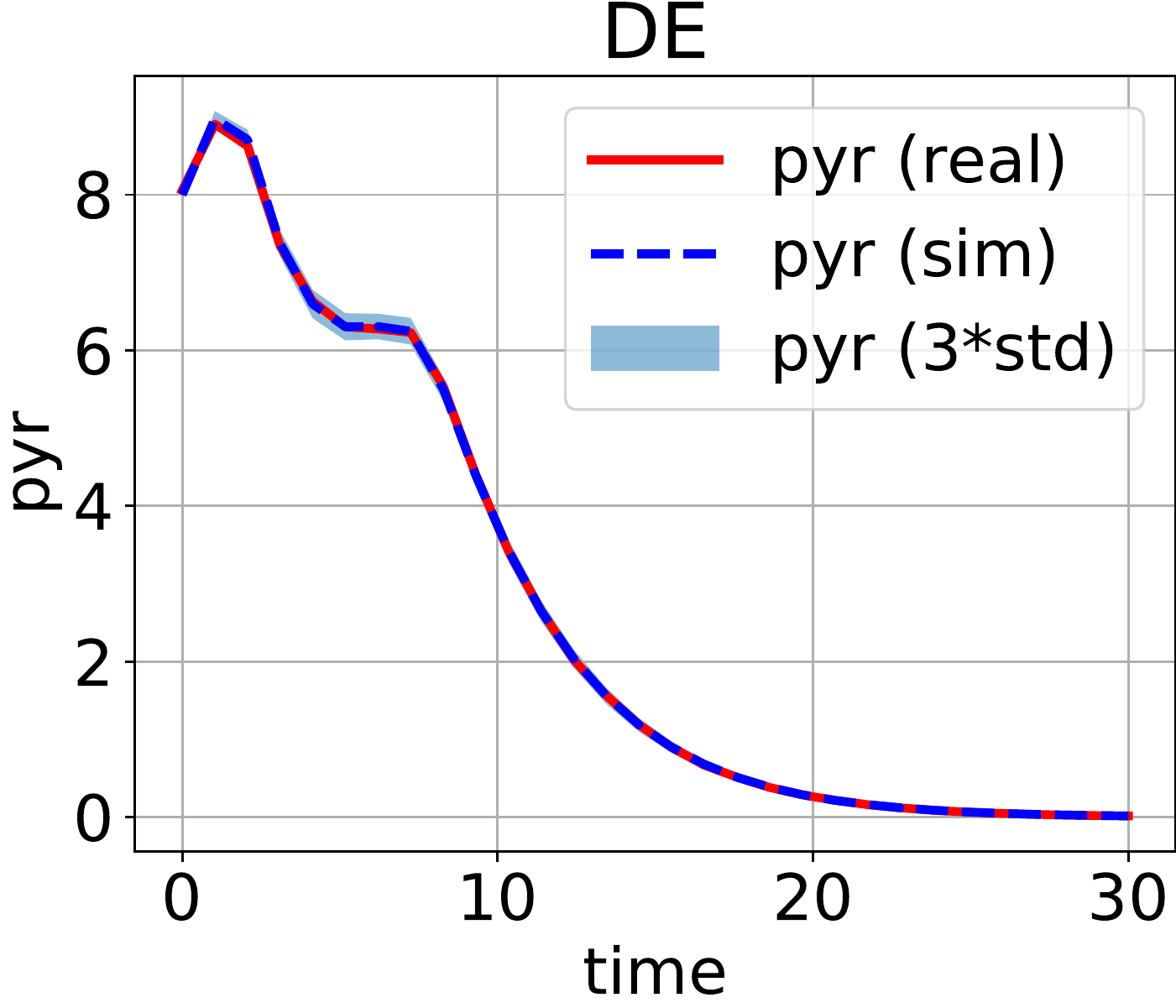} \\
        \vskip 3mm
    \includegraphics[width=110px,height=90px]{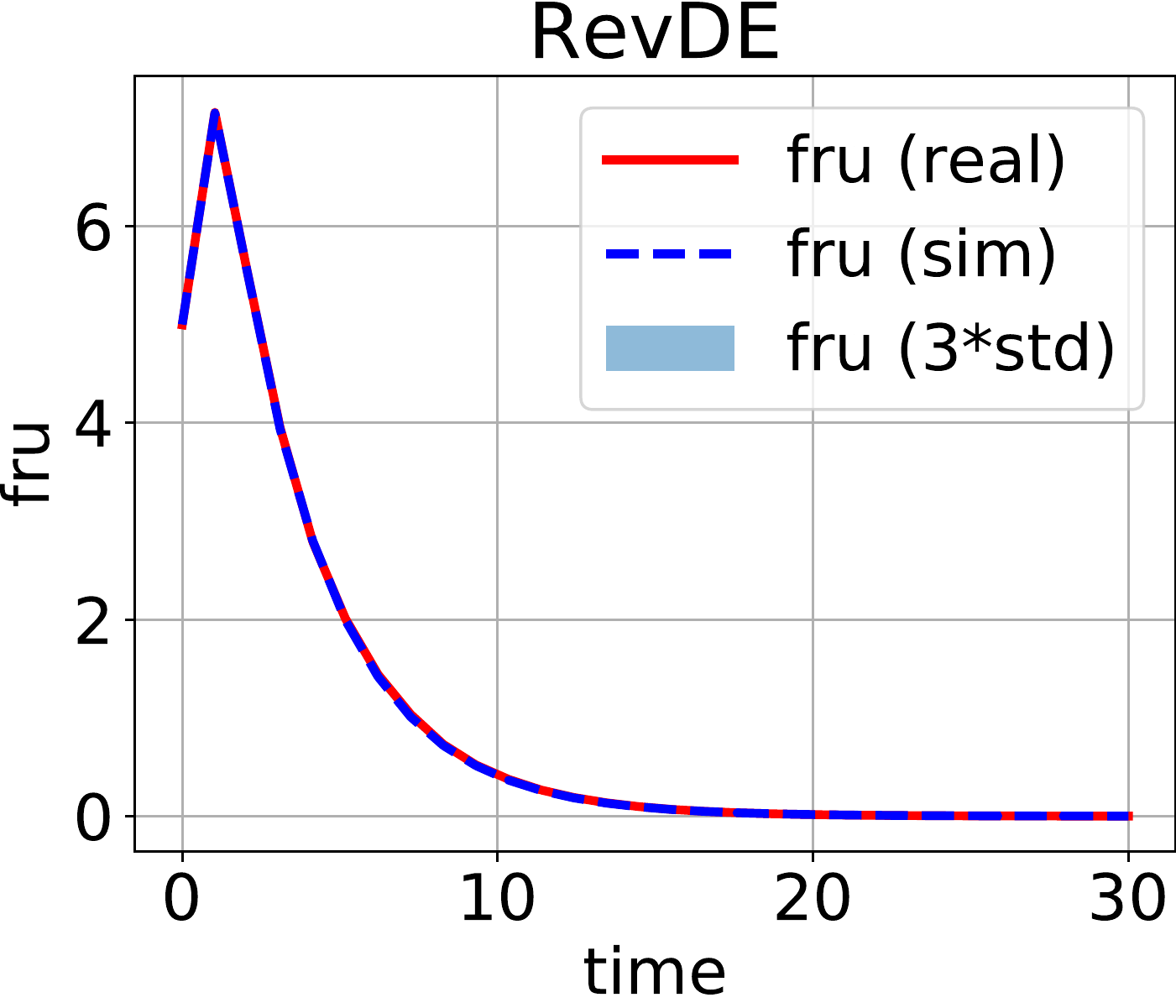}\quad
    \includegraphics[width=110px,height=90px]{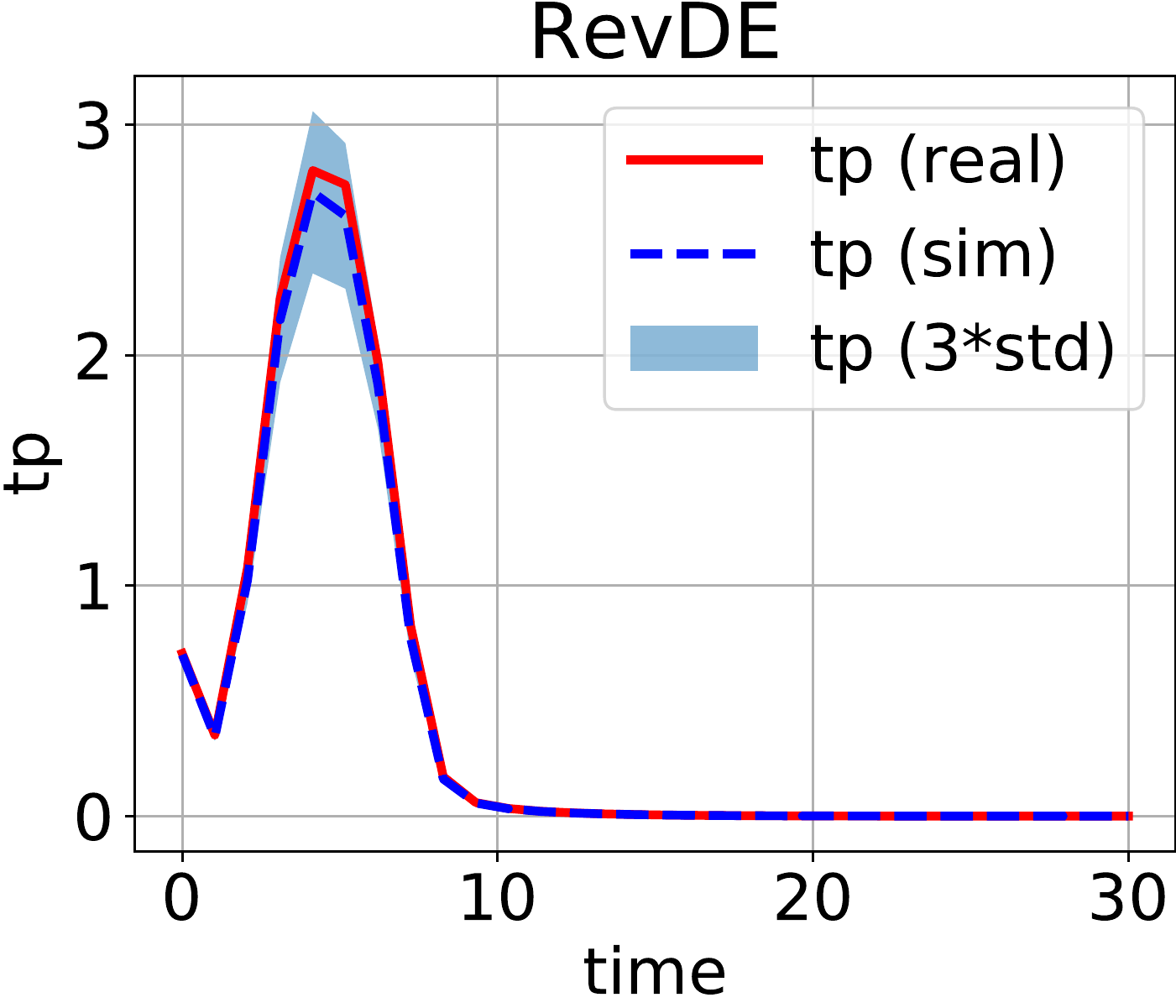}\quad
    \includegraphics[width=110px,height=90px]{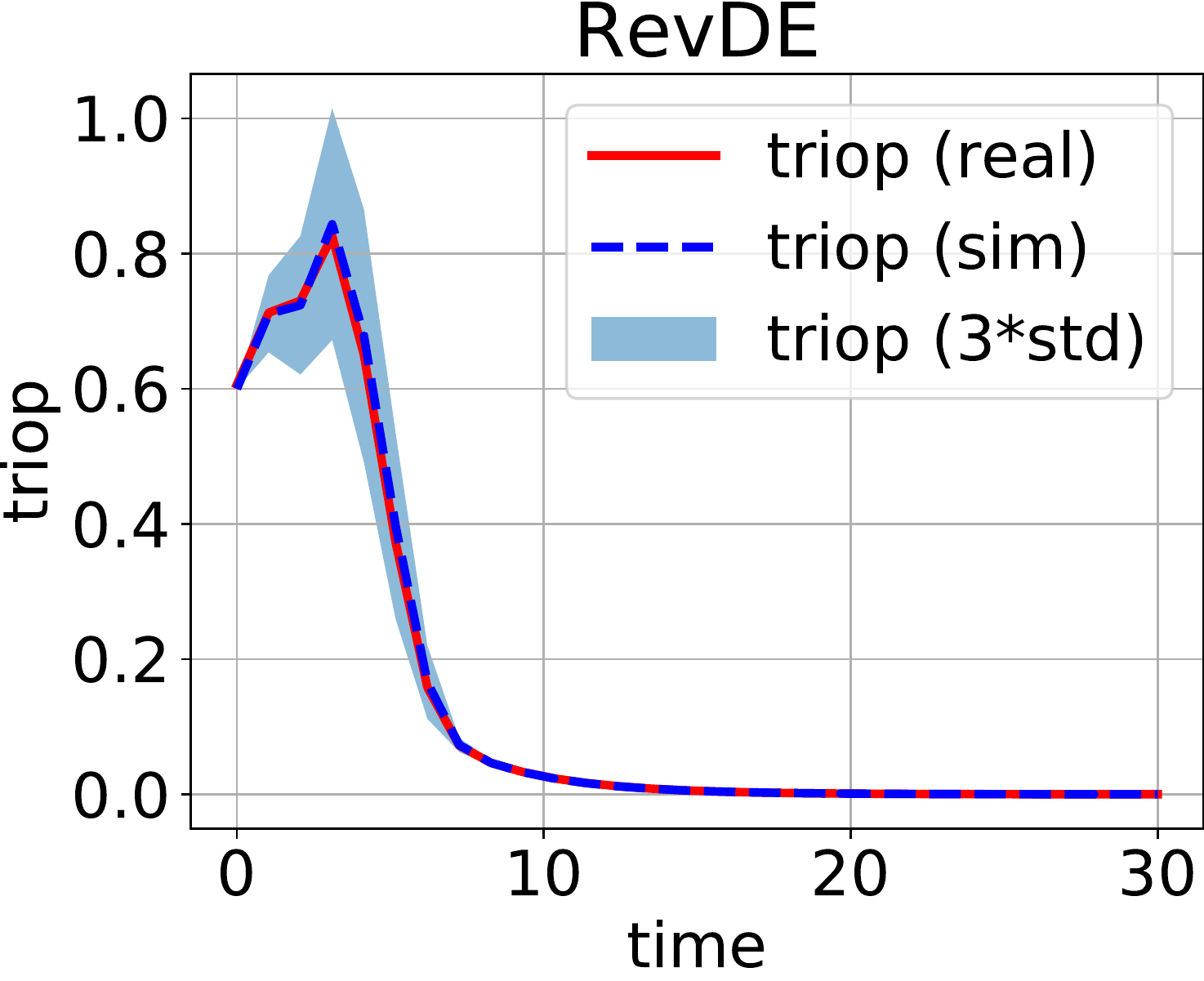}\quad
    \includegraphics[width=110px,height=90px]{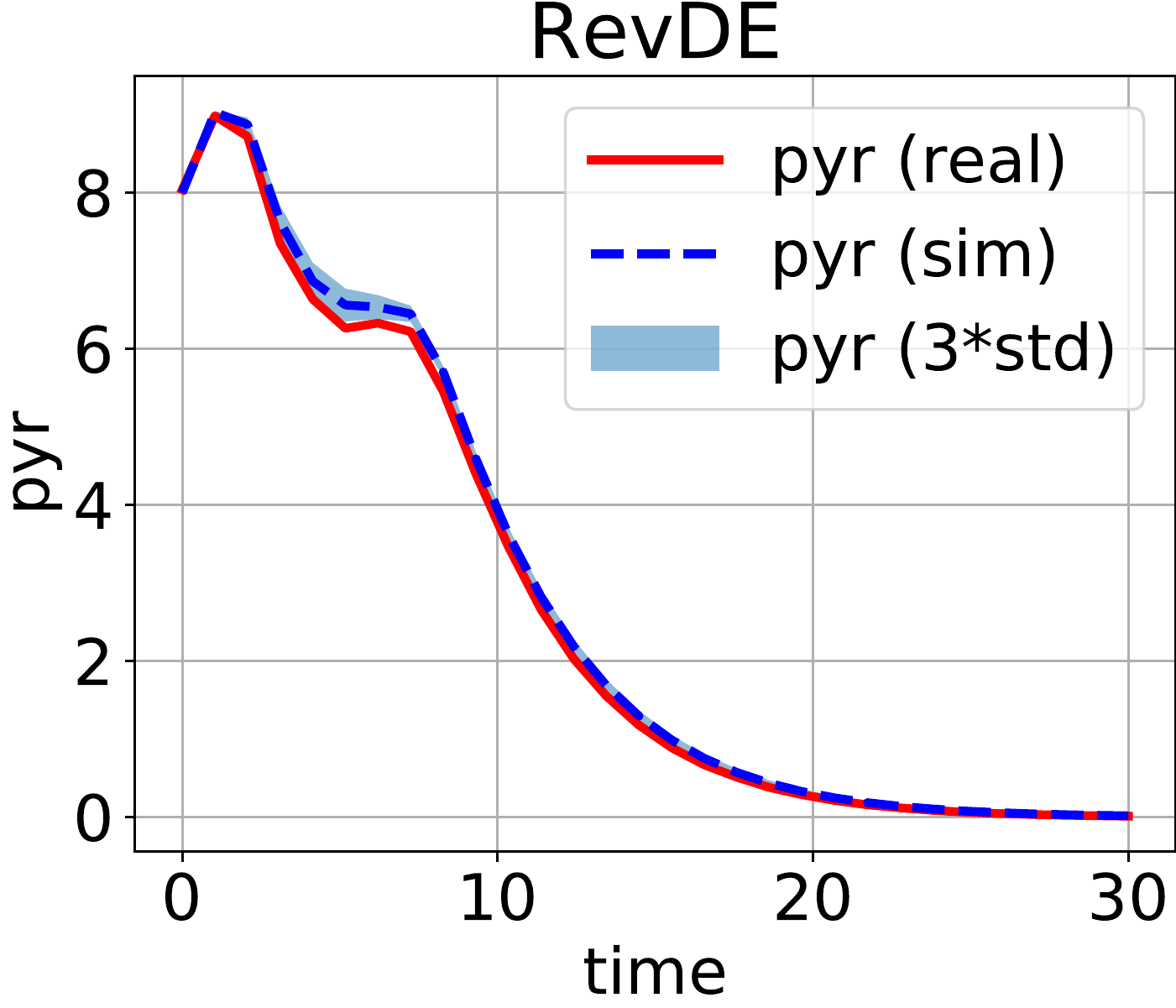} \\
        \vskip 3mm
    \includegraphics[width=110px,height=90px]{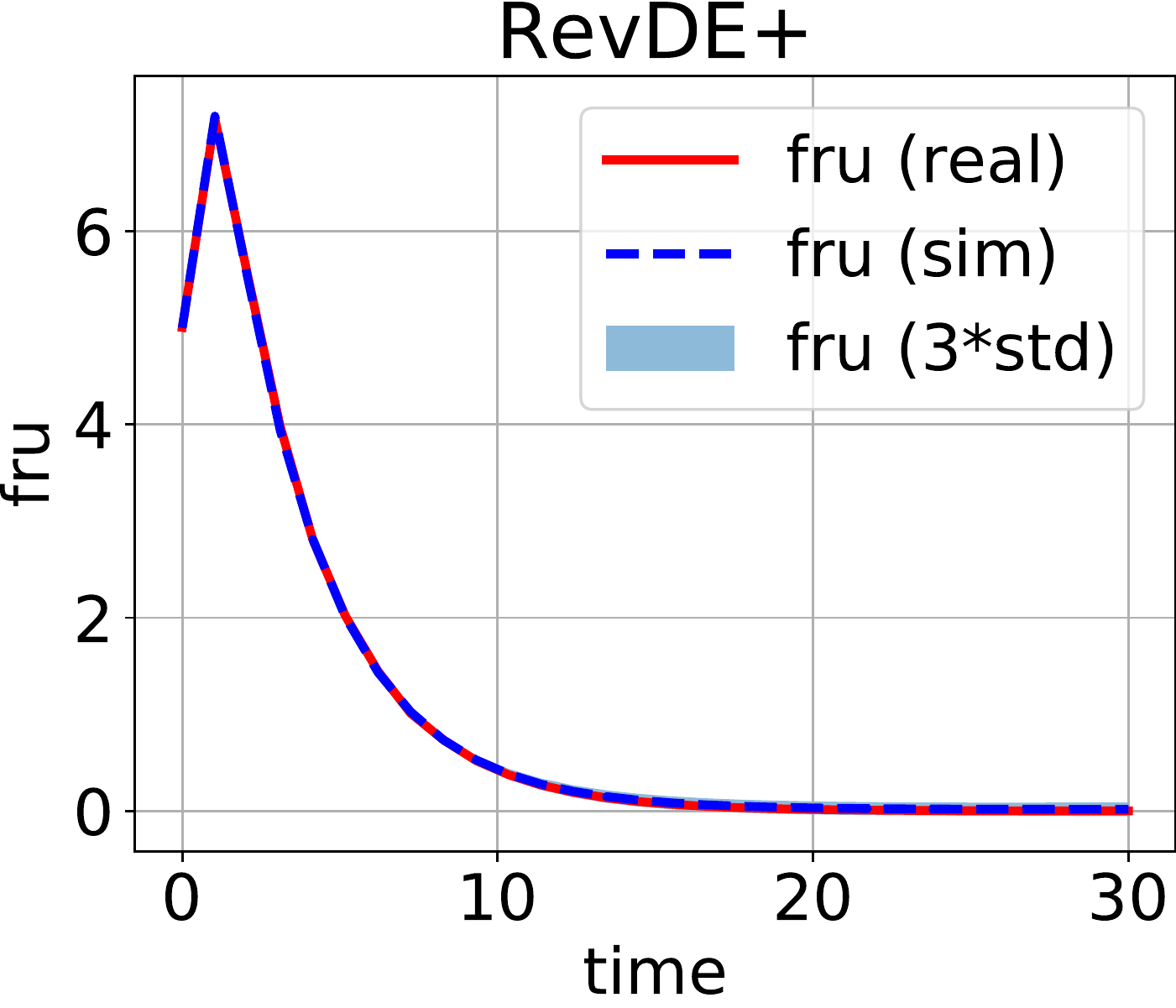}\quad
    \includegraphics[width=110px,height=90px]{figs/results_wolf/wolf__RevDEknn__obs_tp.pdf}\quad
    \includegraphics[width=110px,height=90px]{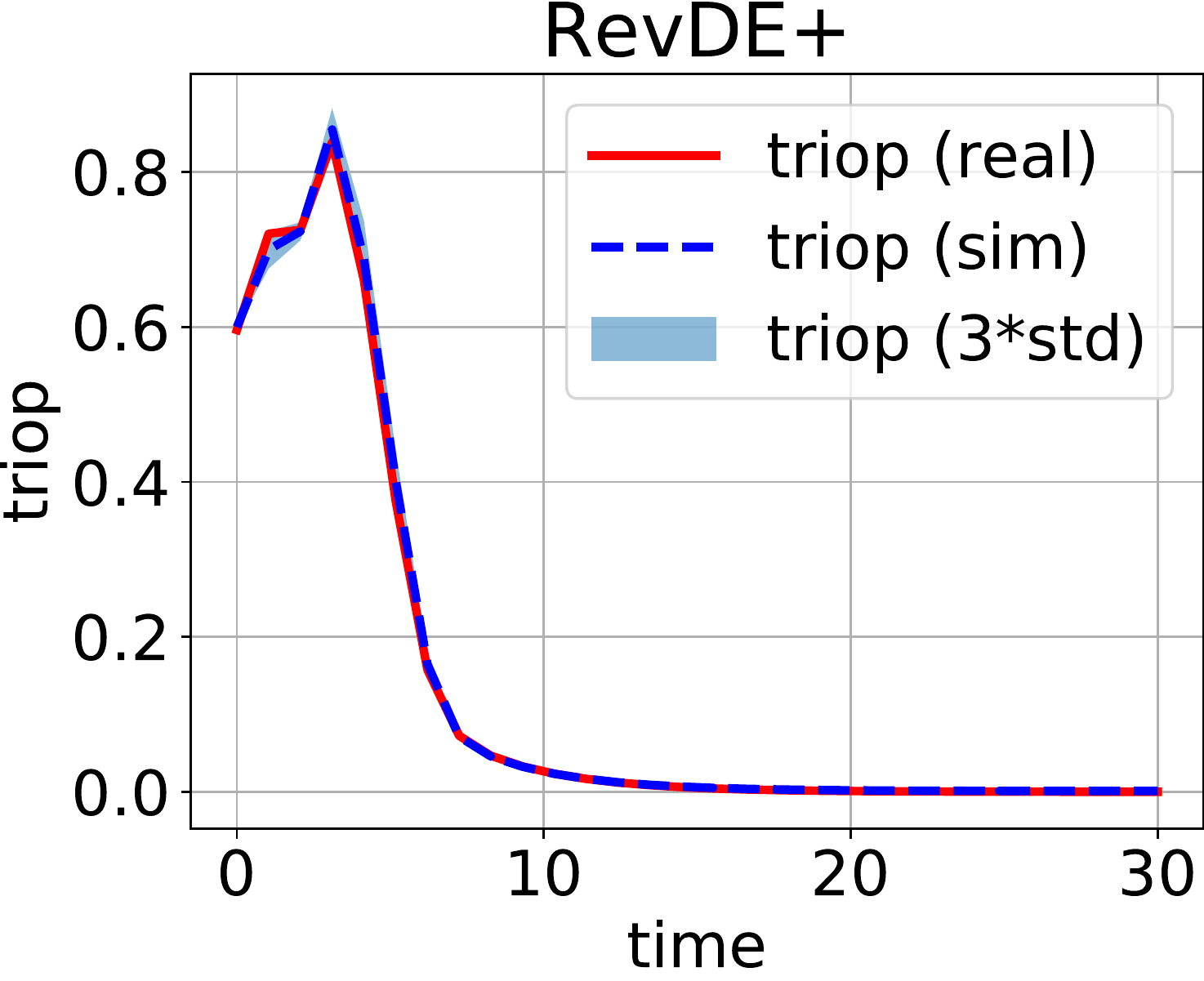}\quad
    \includegraphics[width=110px,height=90px]{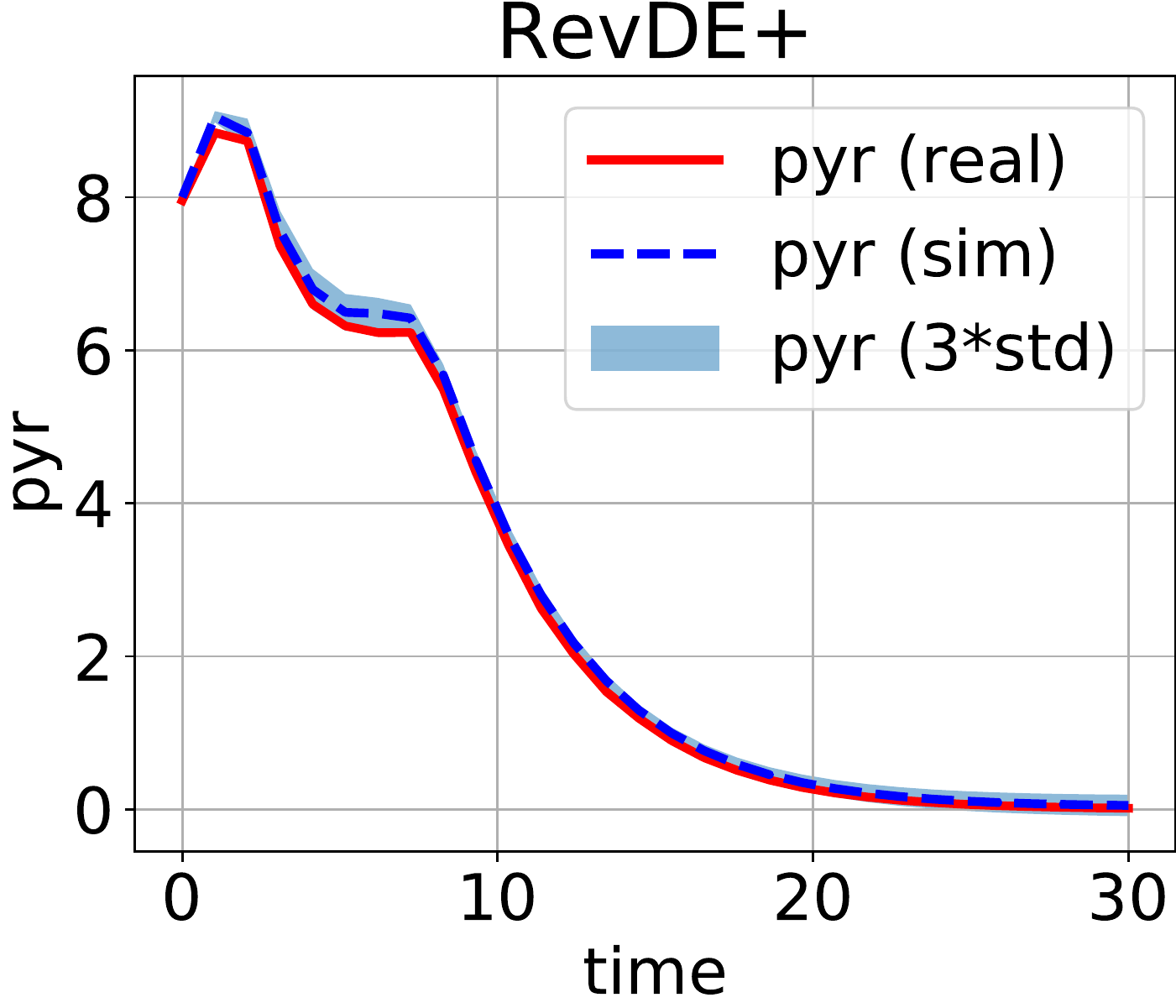} \\
        \vskip 3mm
    \includegraphics[width=110px,height=90px]{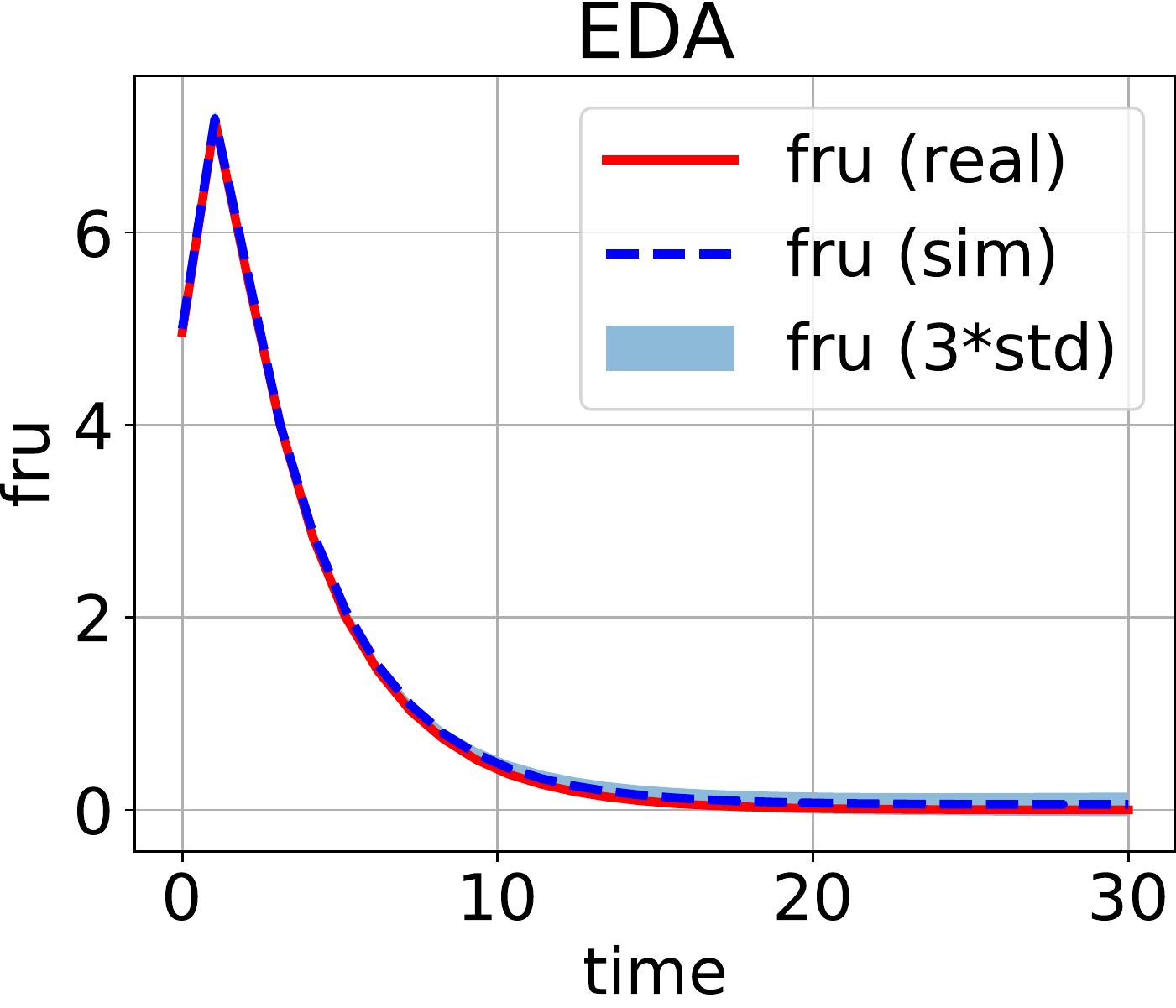}\quad
    \includegraphics[width=110px,height=90px]{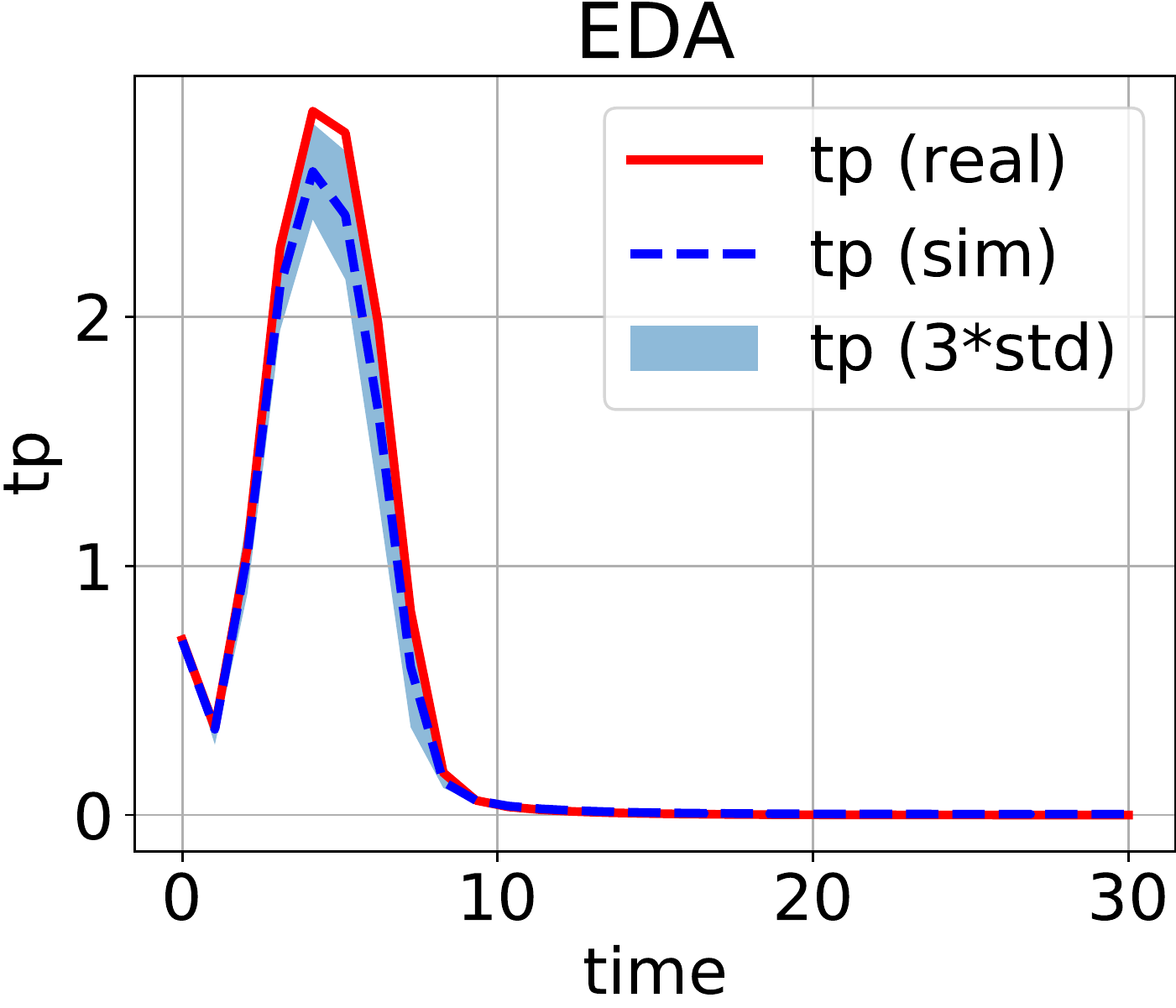}\quad
    \includegraphics[width=110px,height=90px]{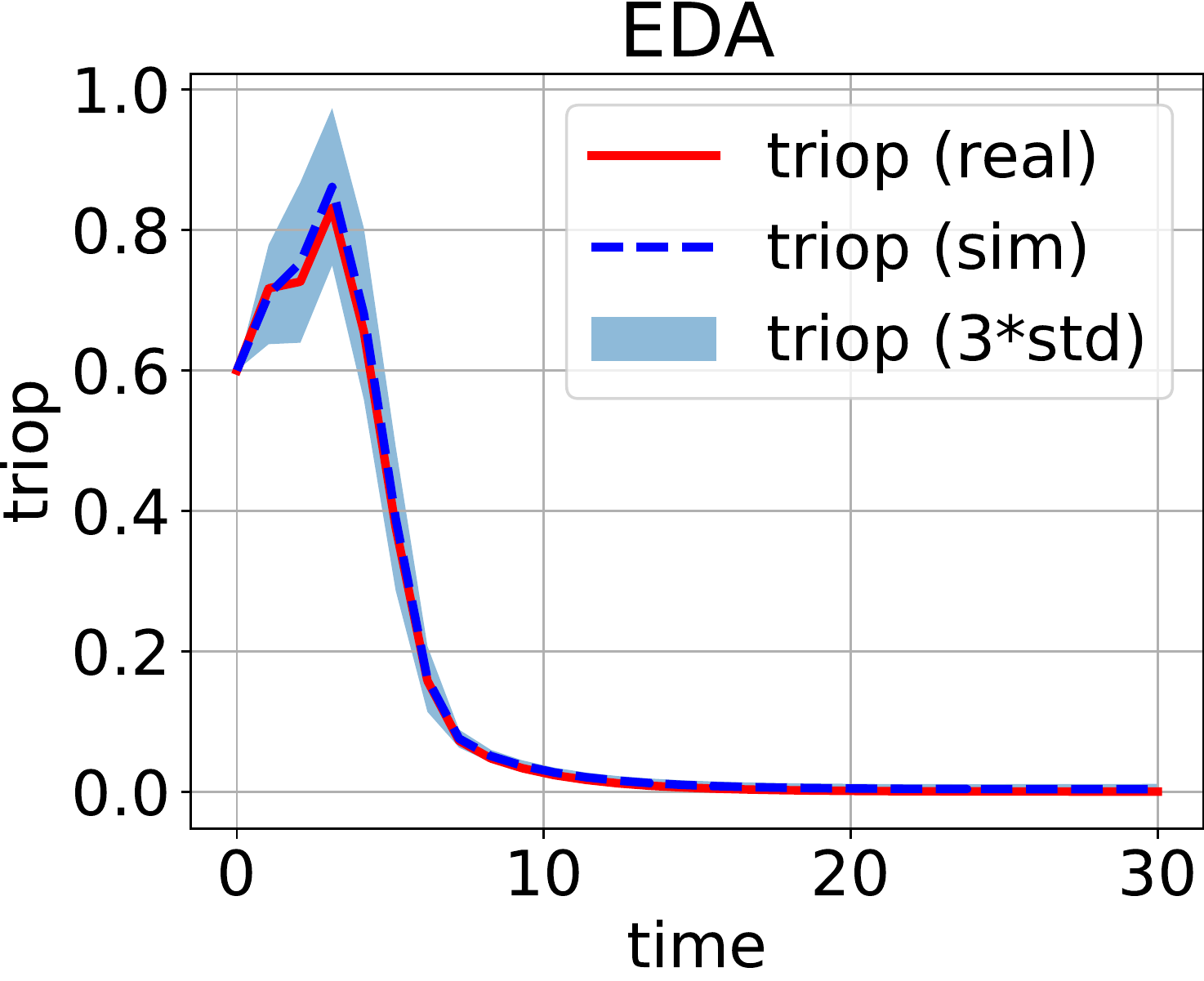}\quad
    \includegraphics[width=110px,height=90px]{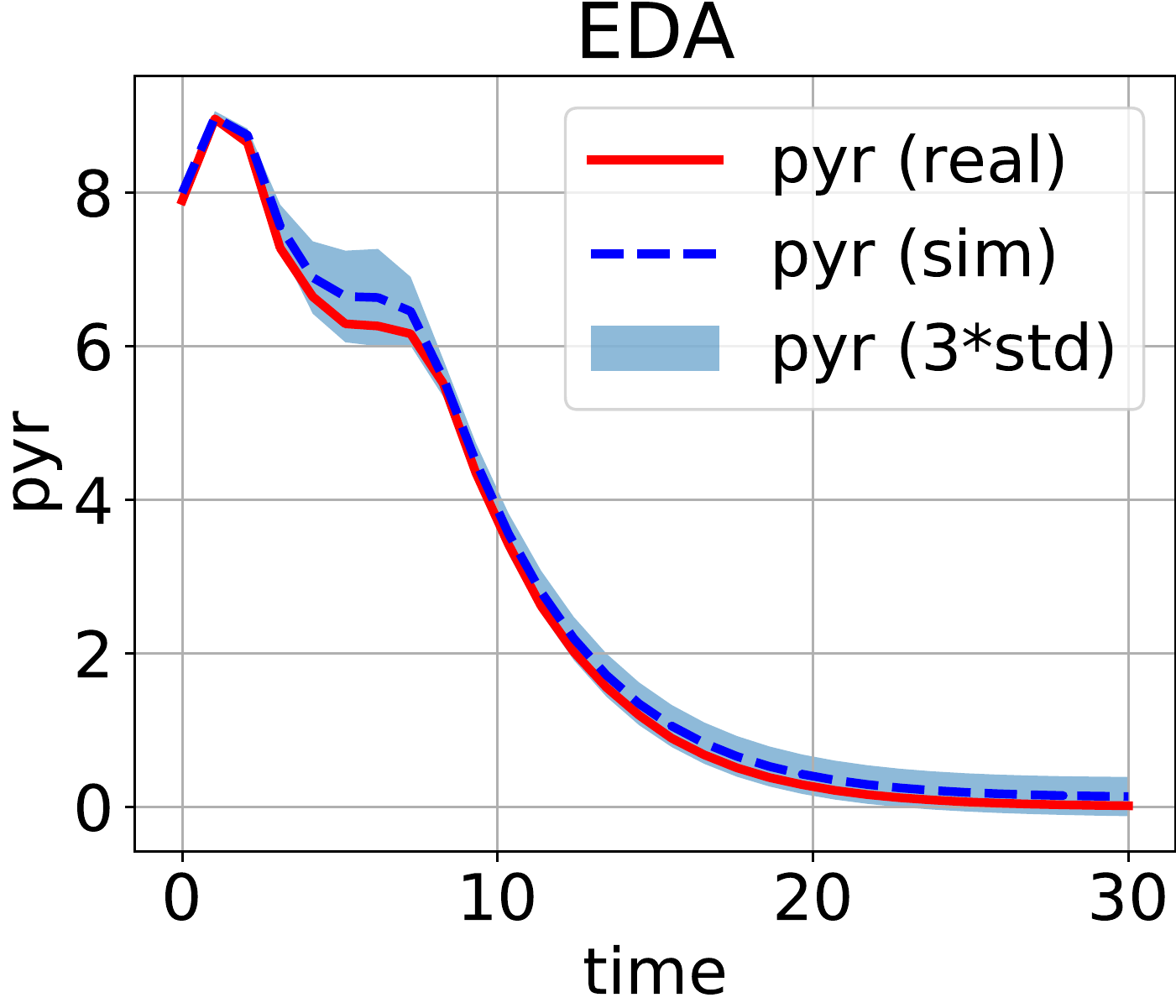} \\
        \vskip 3mm
    \includegraphics[width=110px,height=90px]{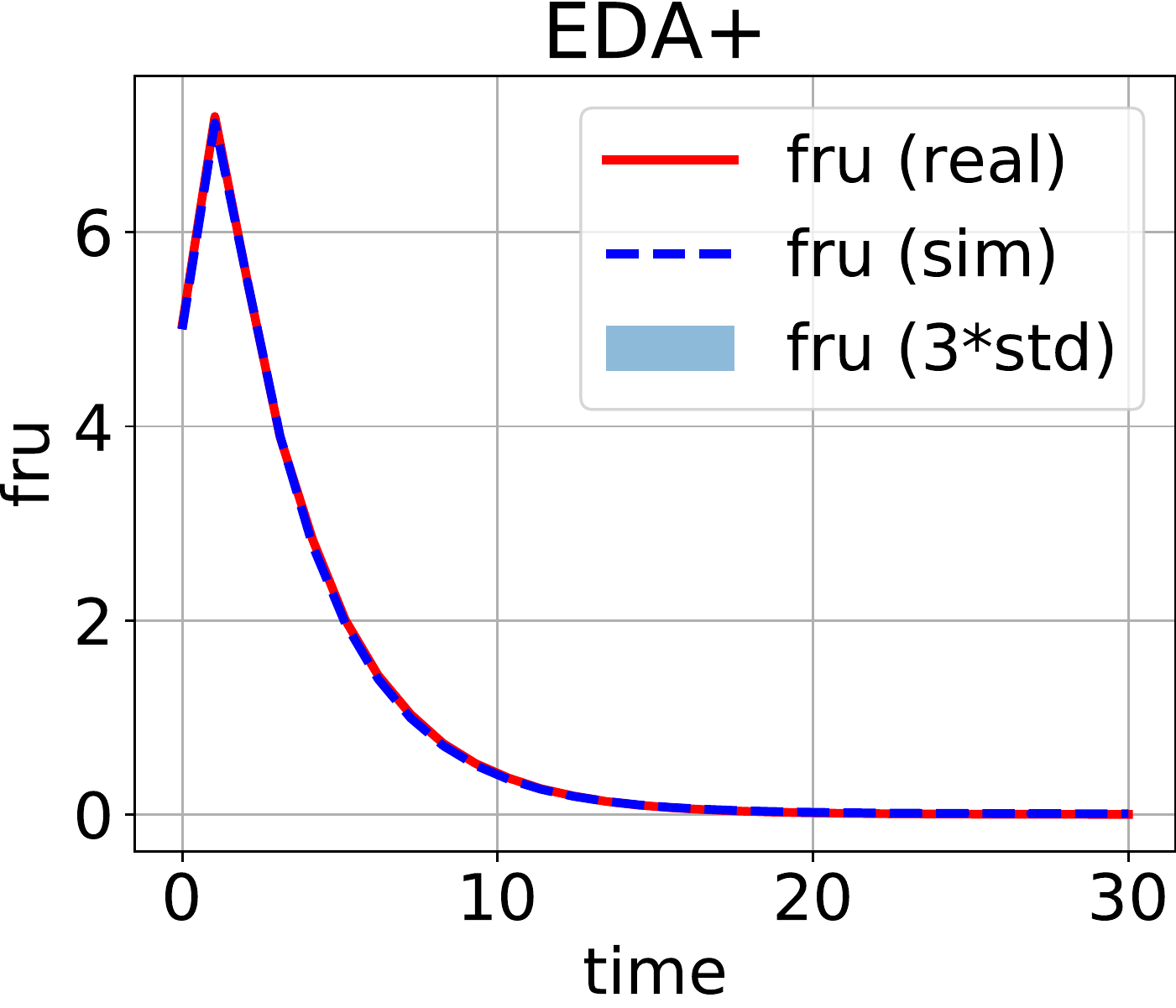}\quad
    \includegraphics[width=110px,height=90px]{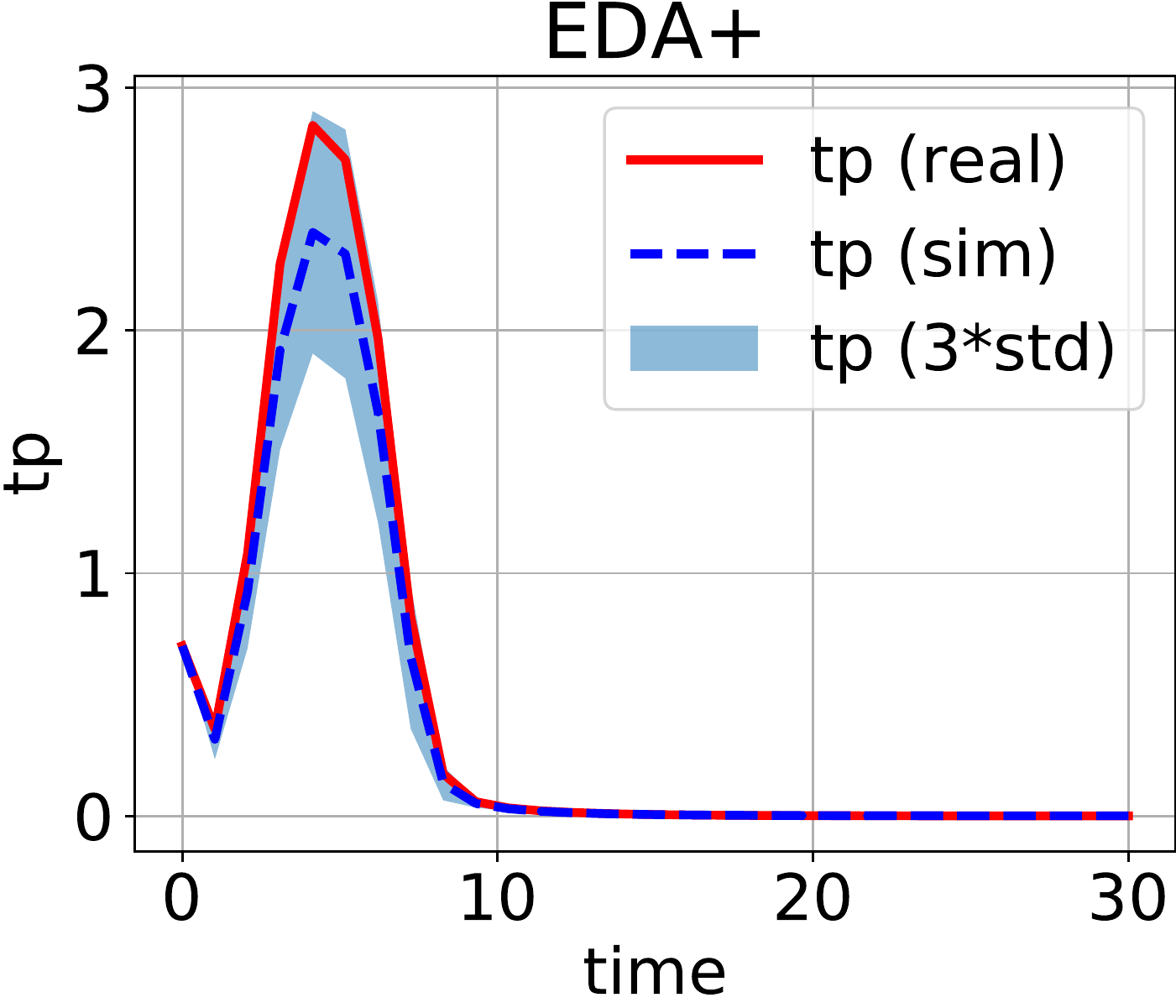}\quad
    \includegraphics[width=110px,height=90px]{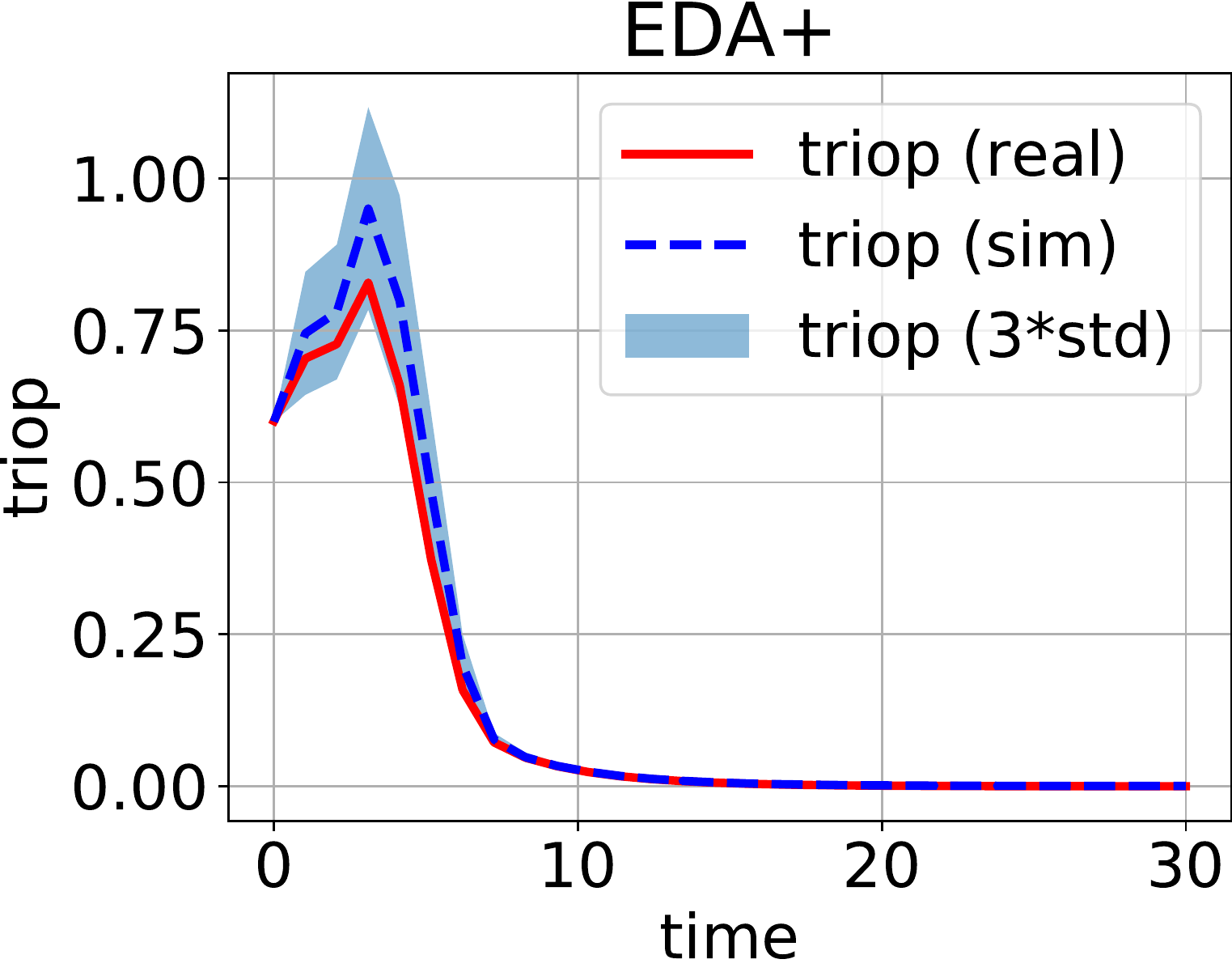}\quad
    \includegraphics[width=110px,height=90px]{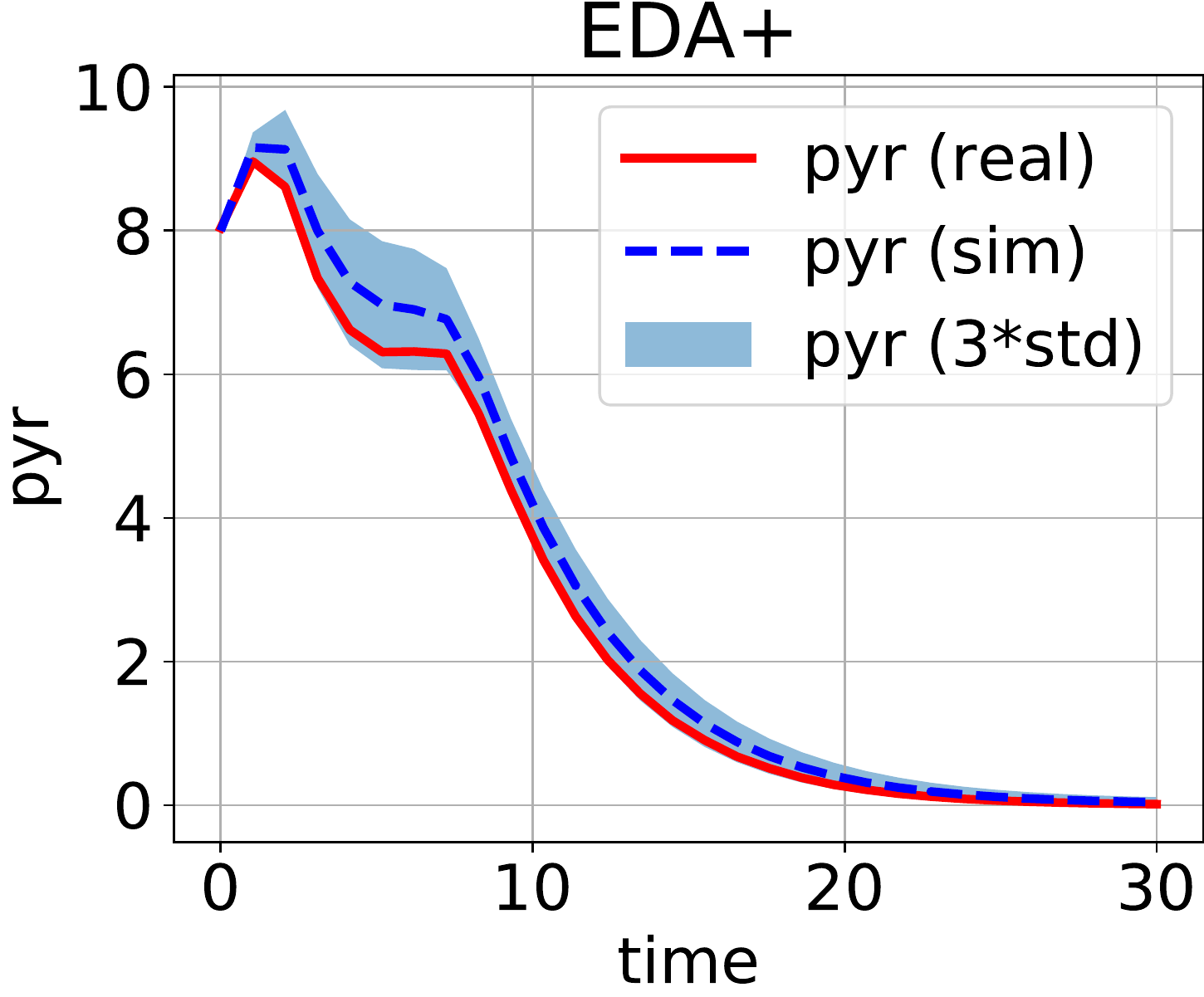} \\
        \vskip 3mm
    \includegraphics[width=110px,height=90px]{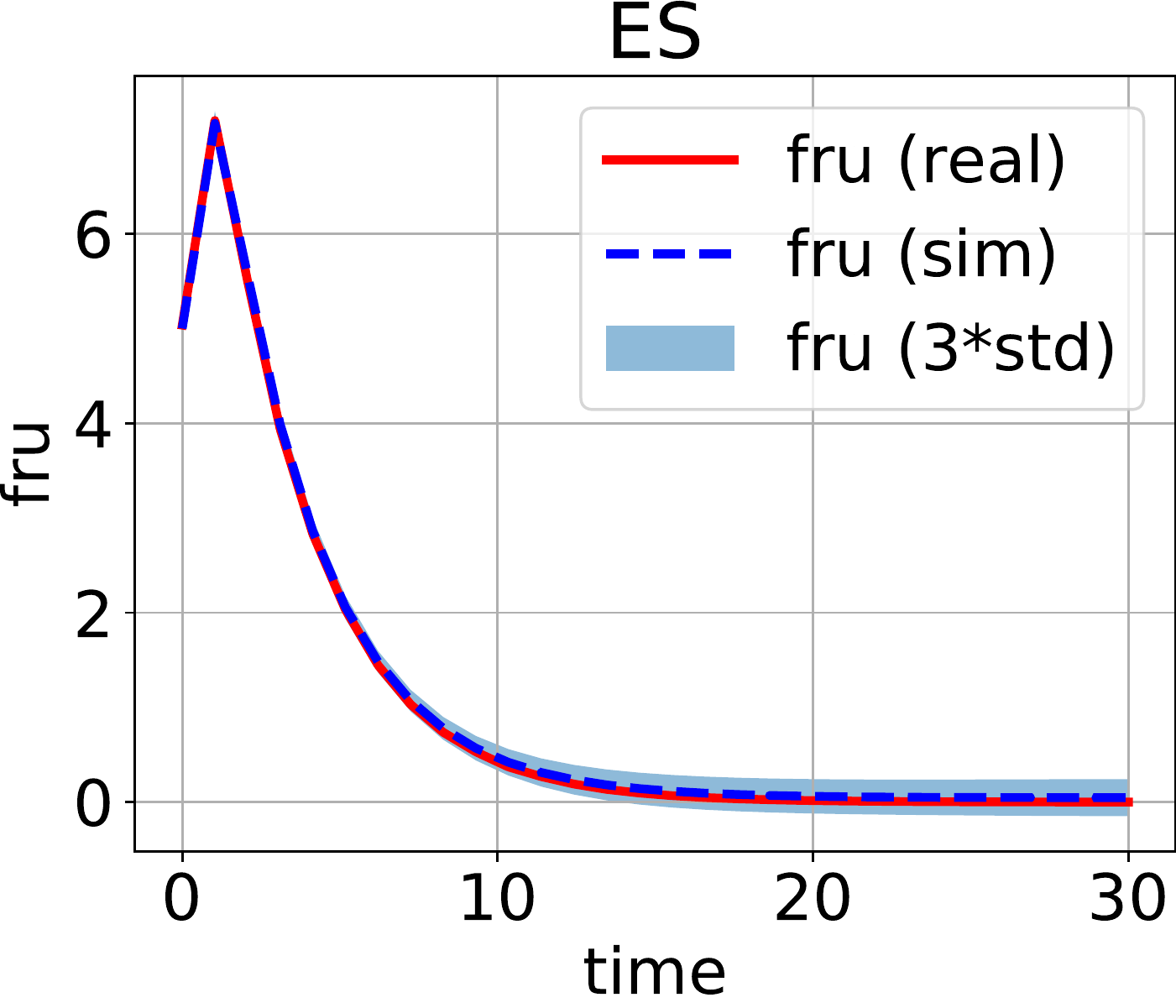}\quad
    \includegraphics[width=110px,height=90px]{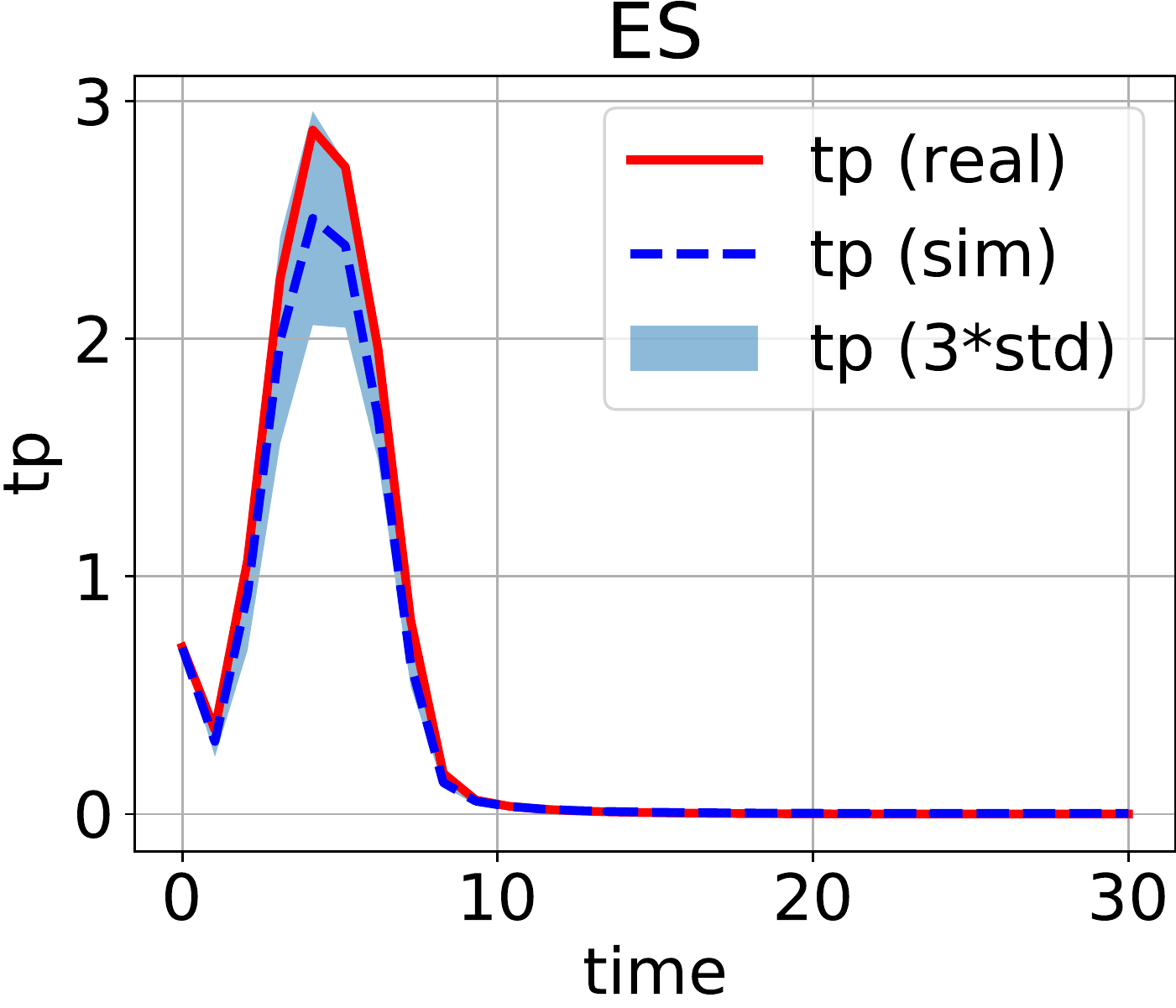}\quad
    \includegraphics[width=110px,height=90px]{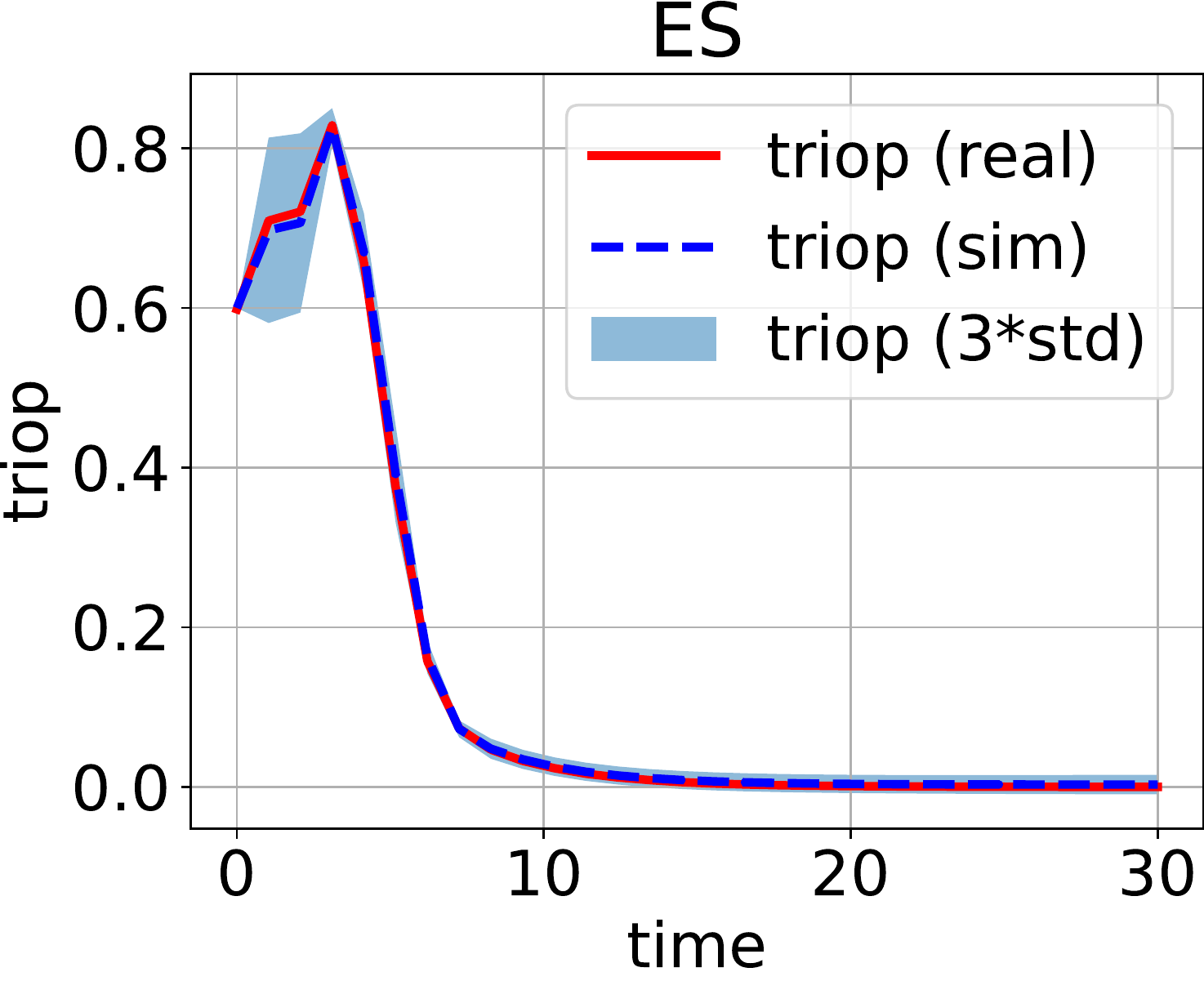}\quad
    \includegraphics[width=110px,height=90px]{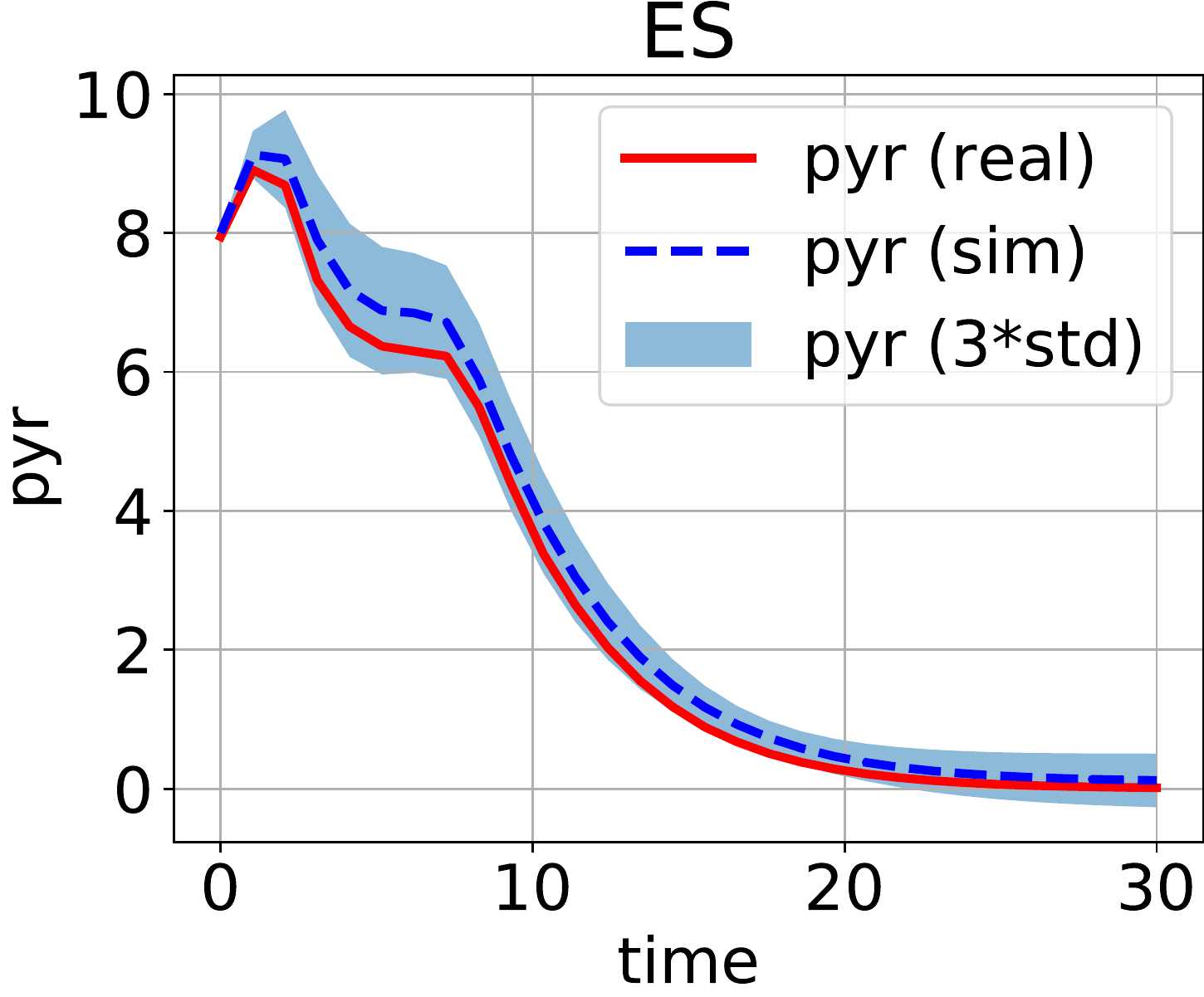}
    \caption{A comparison of the timecourses of the unobserved metabolites in the Case 1. Real timecourses are depicted in red, and the average value and a confidence interval ($3\times$ standard deviation) over $3$ runs of the simulator are in blue. The titles of the plots indicate optimization methods.}
    \label{fig:obs_1}
\end{figure*}

\begin{figure*}[!tbp]
    \centering
    \includegraphics[width=110px,height=90px]{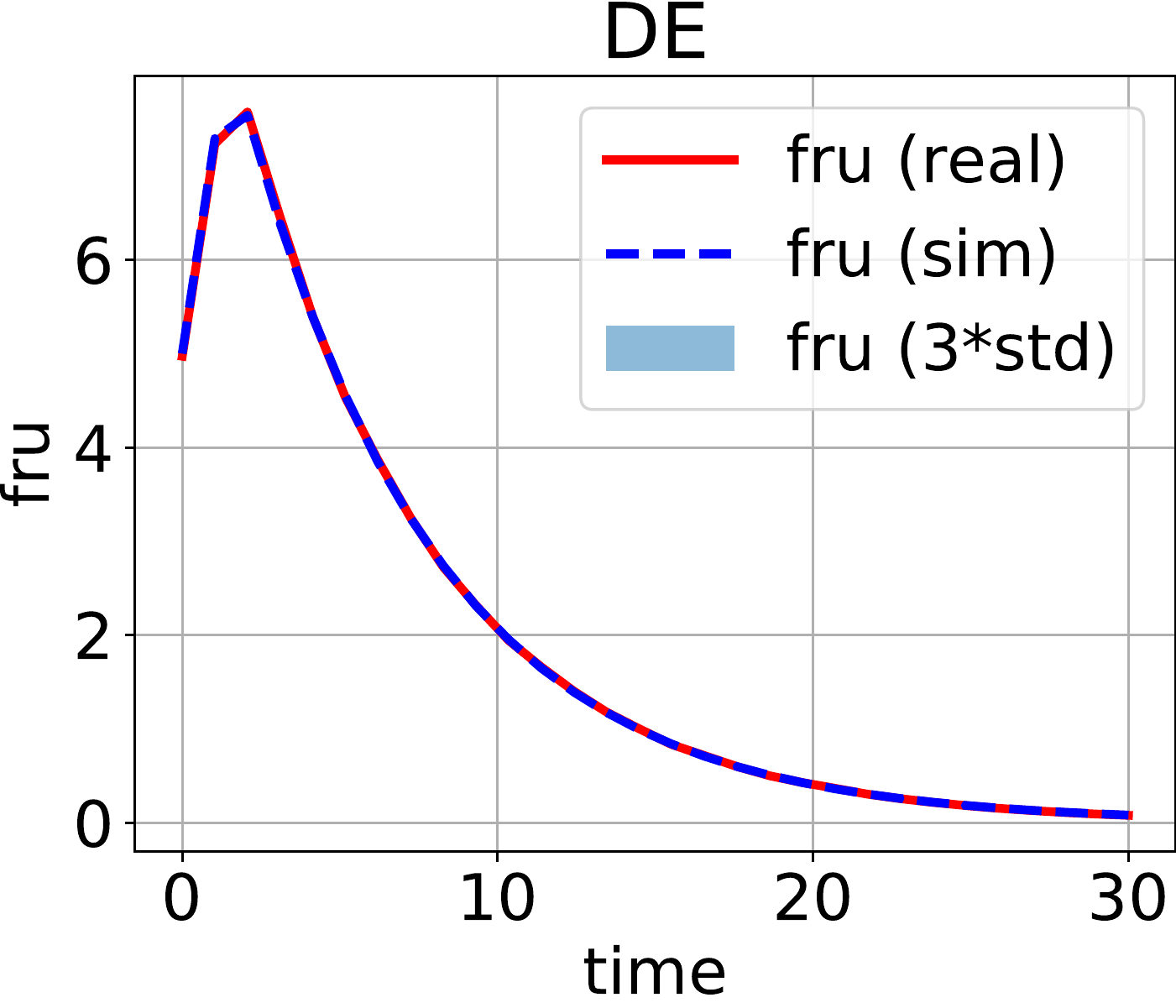}\quad
    \includegraphics[width=110px,height=90px]{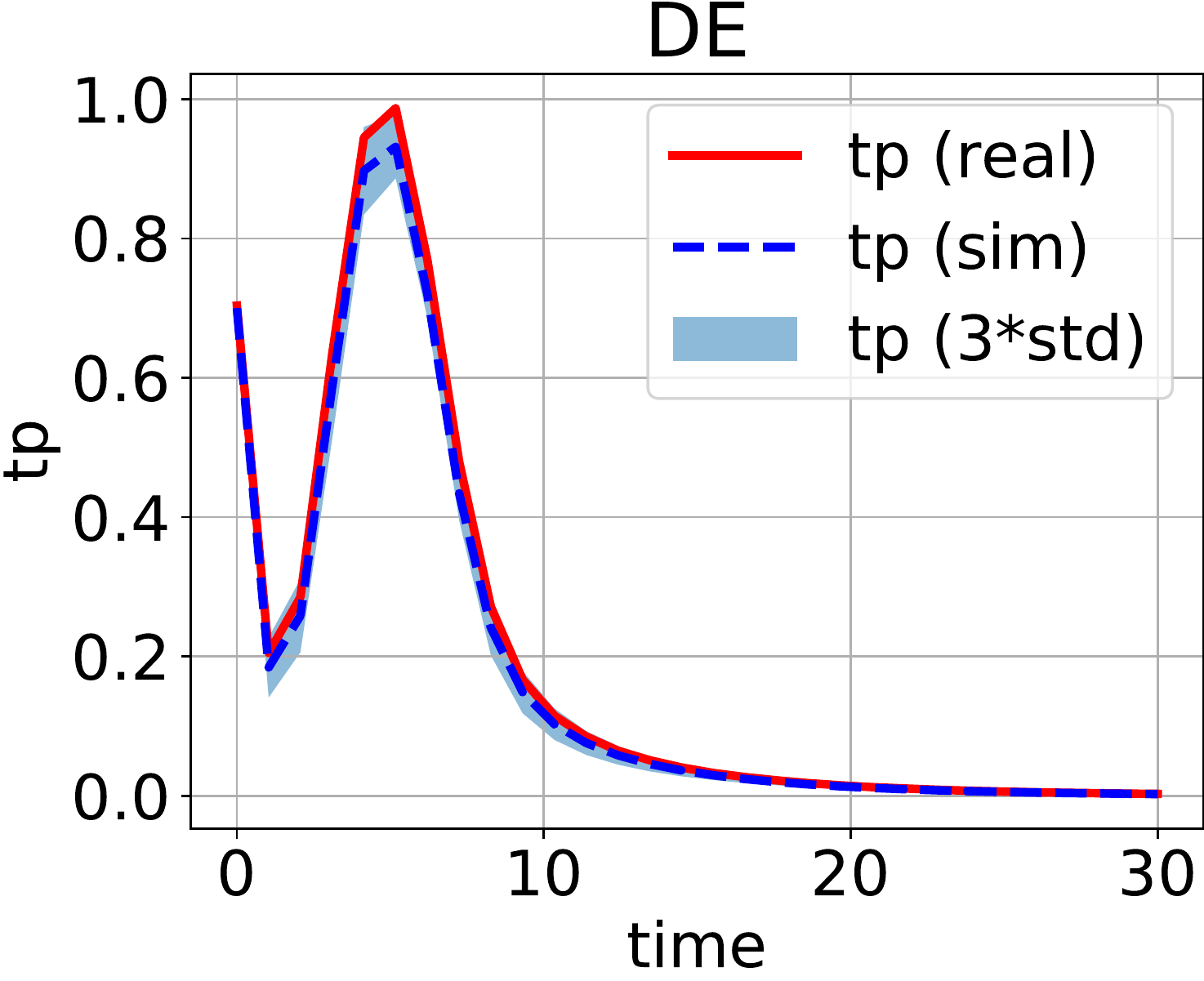}\quad
    \includegraphics[width=110px,height=90px]{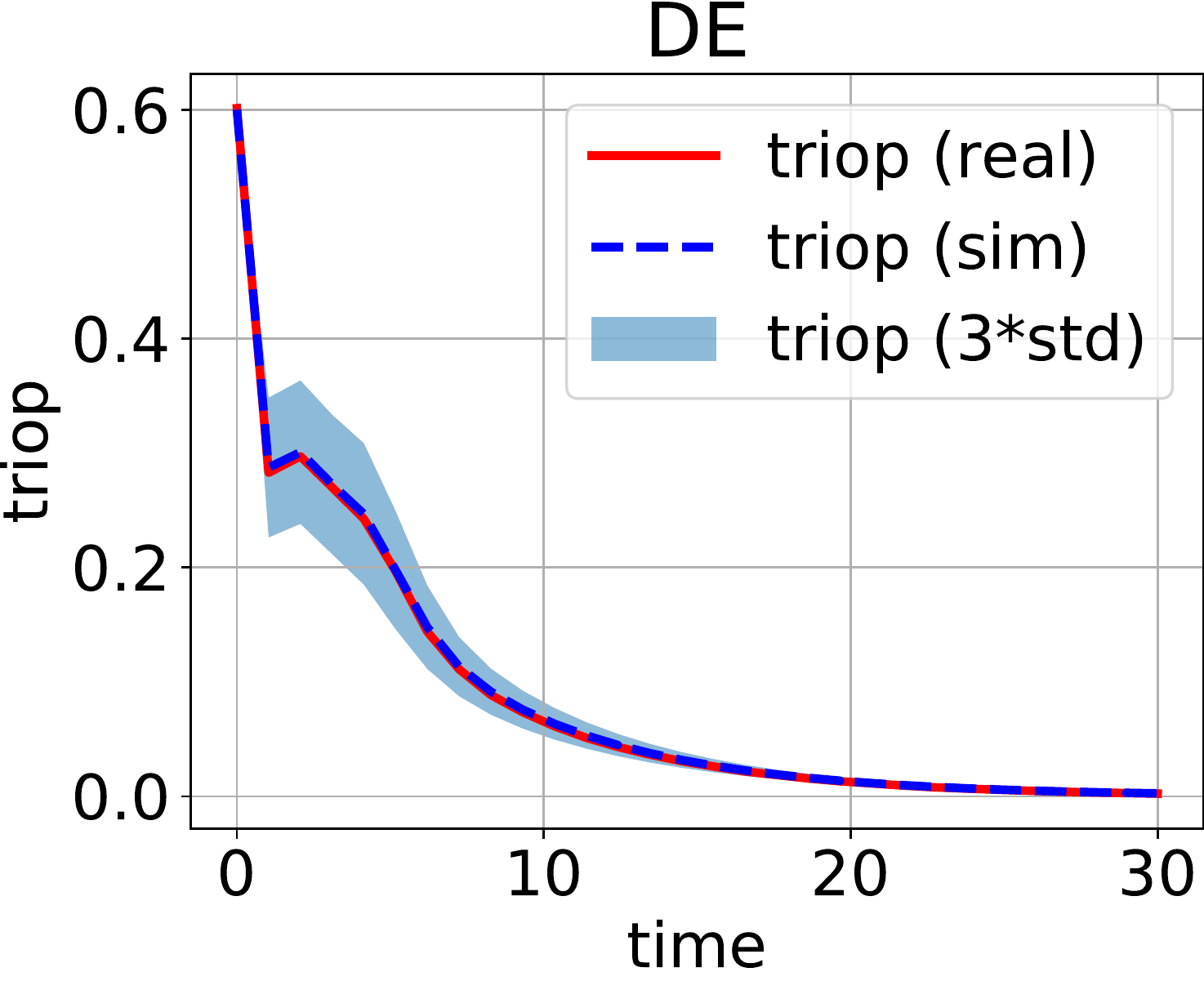}\quad
    \includegraphics[width=110px,height=90px]{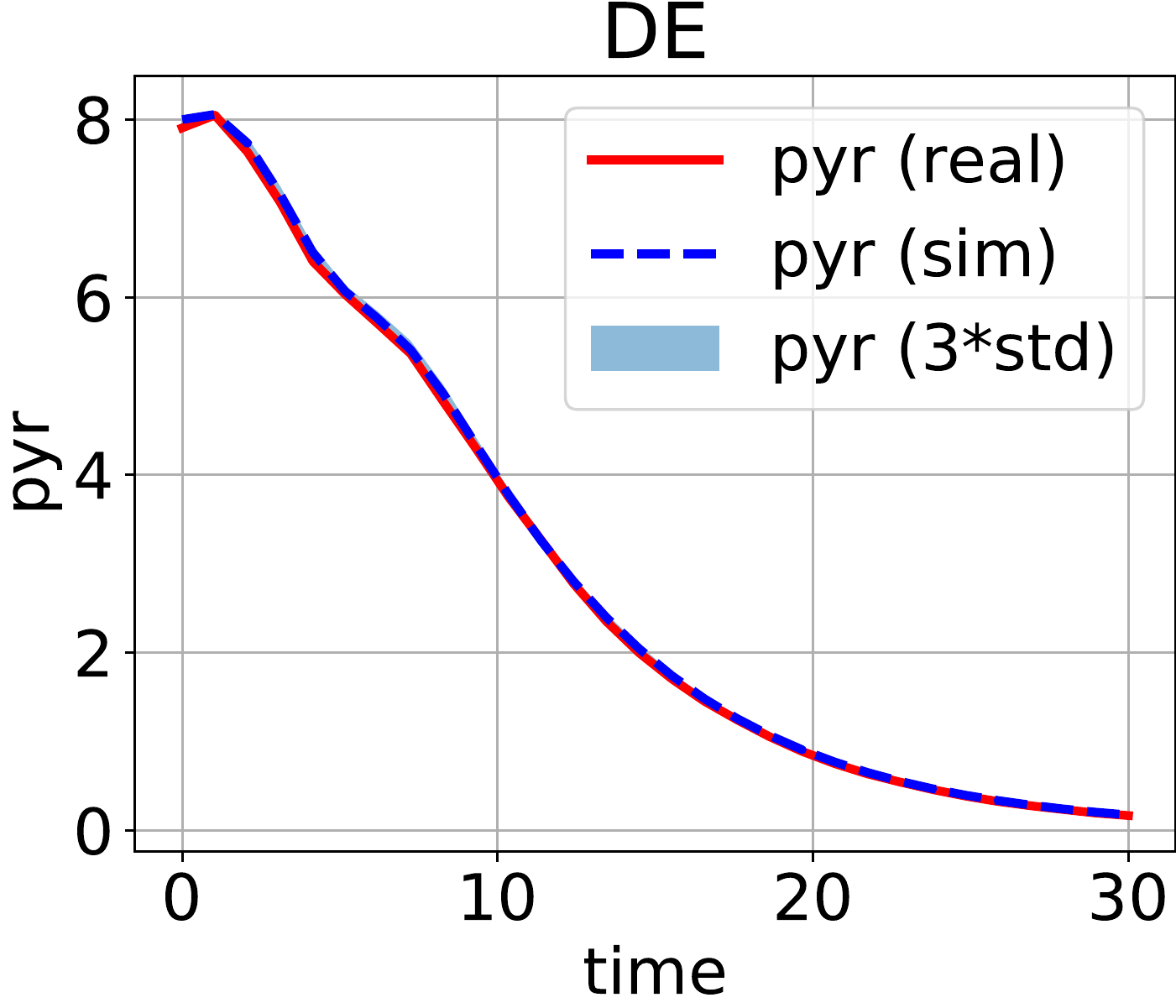} \\
        \vskip 3mm
    \includegraphics[width=110px,height=90px]{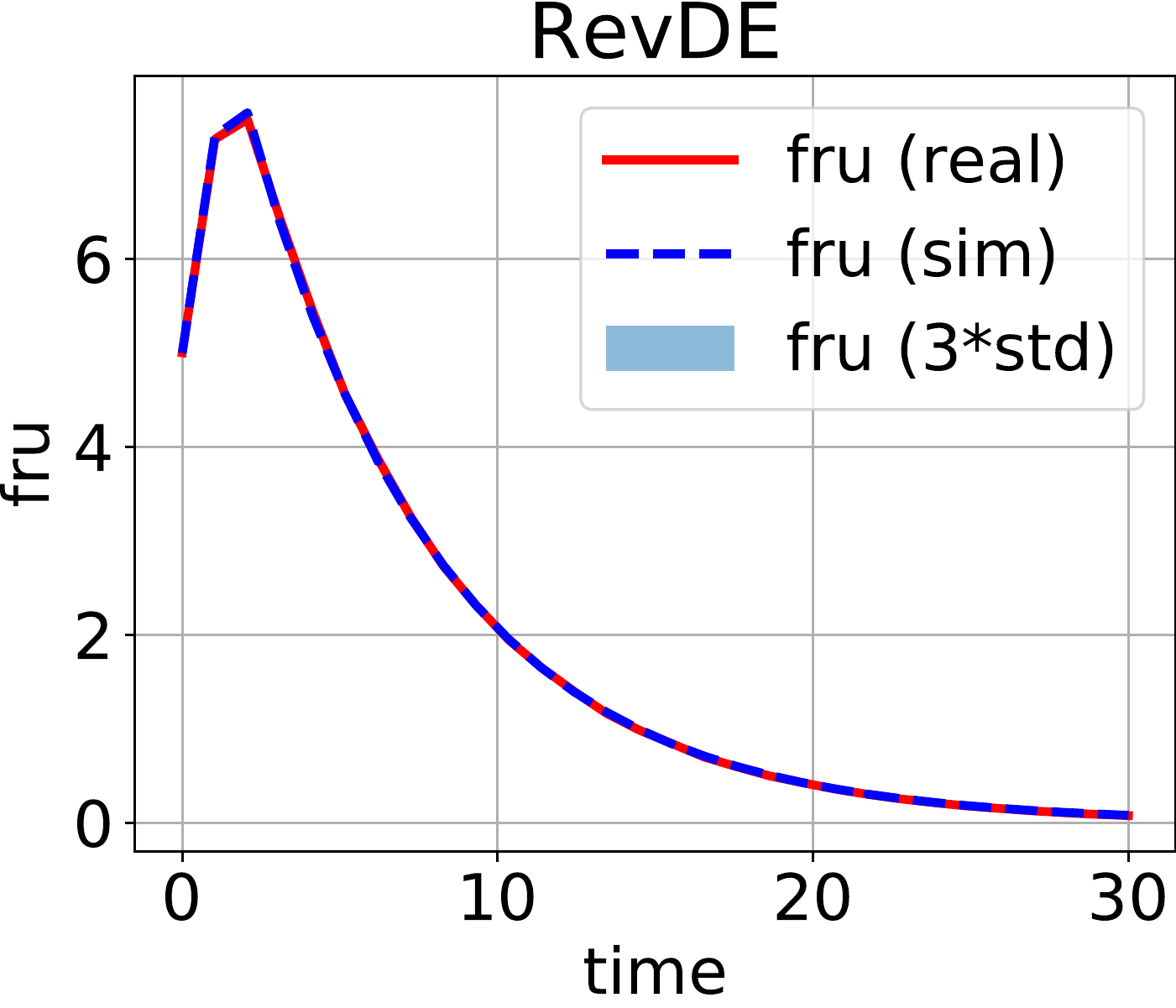}\quad
    \includegraphics[width=110px,height=90px]{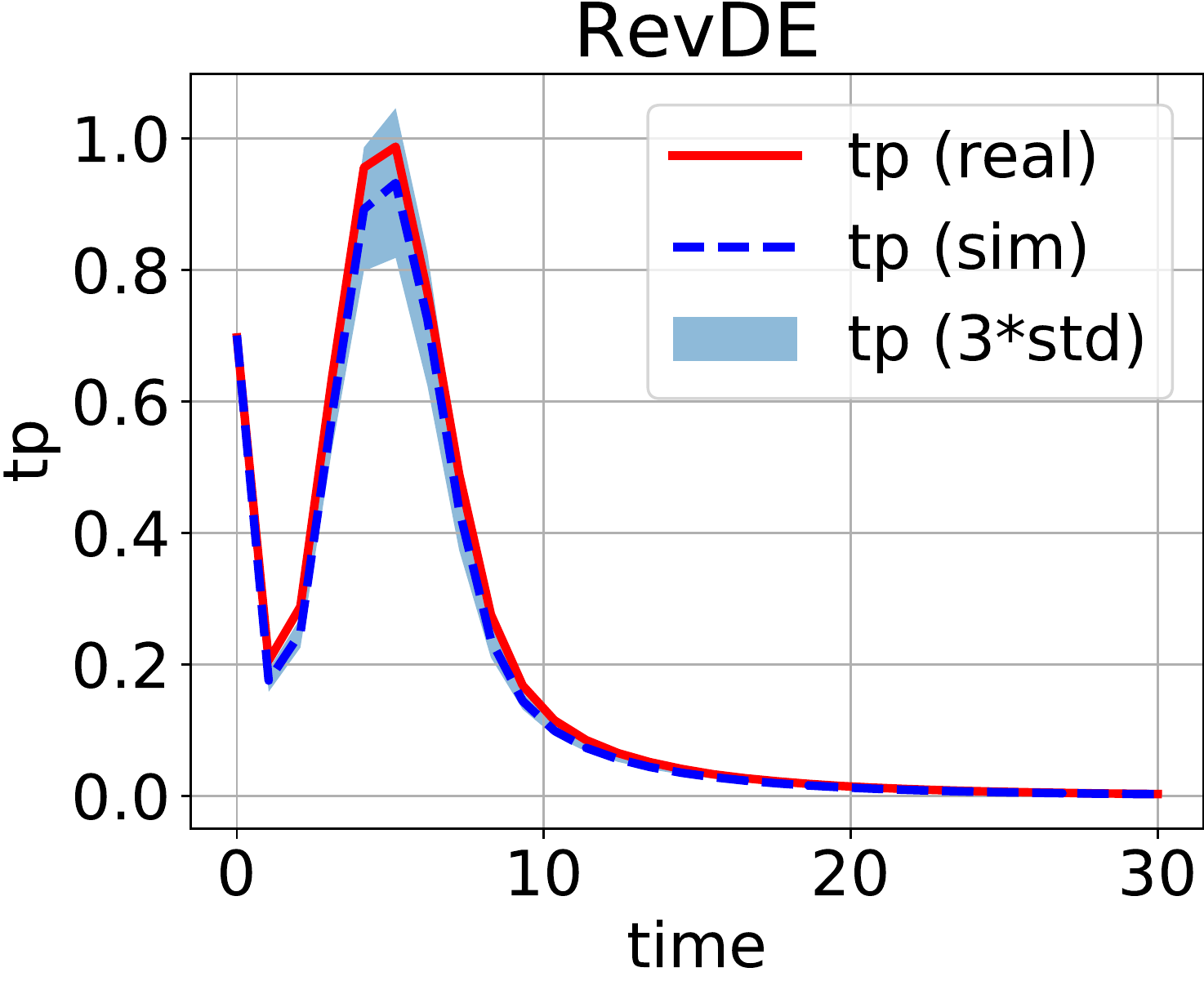}\quad
    \includegraphics[width=110px,height=90px]{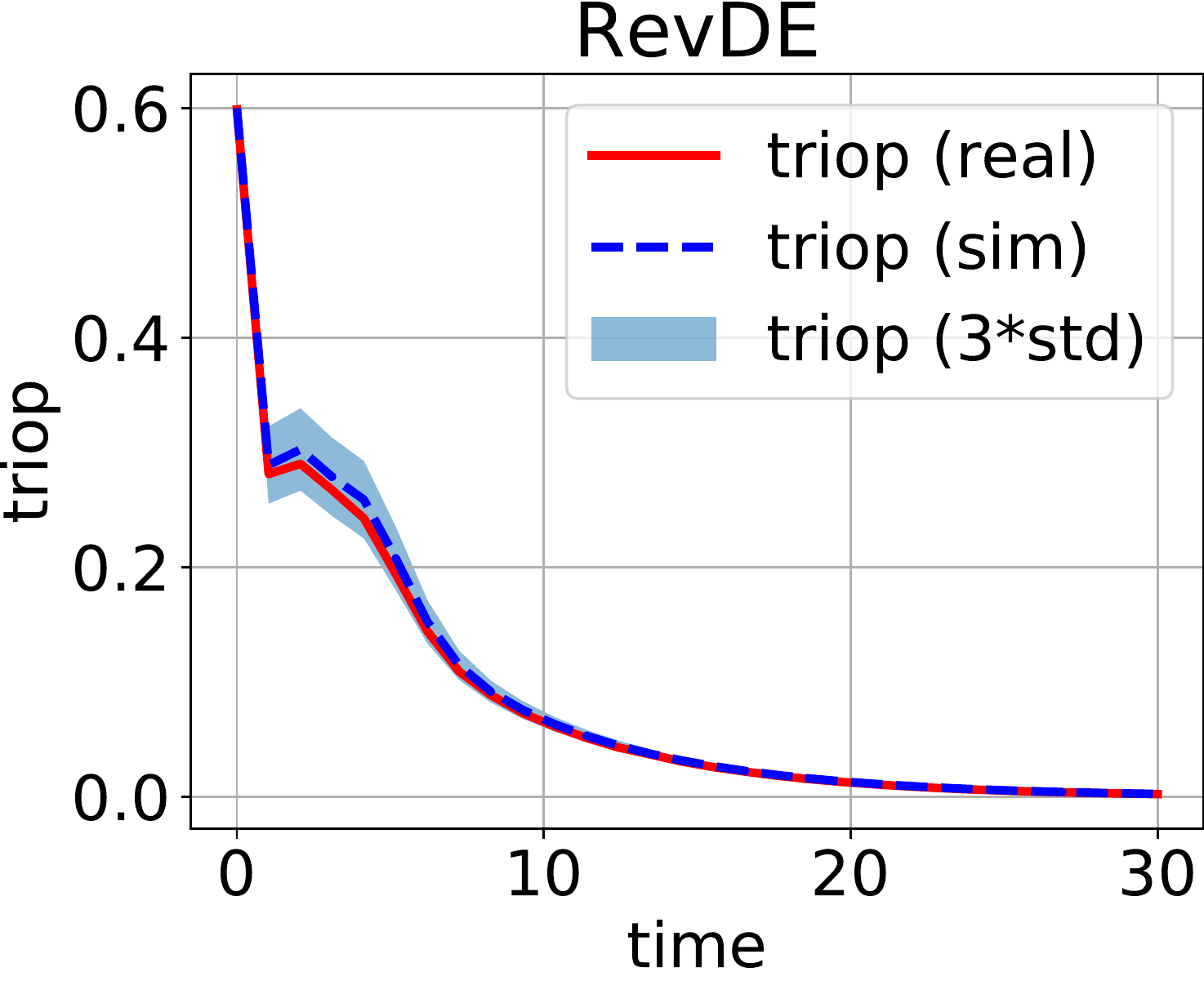}\quad
    \includegraphics[width=110px,height=90px]{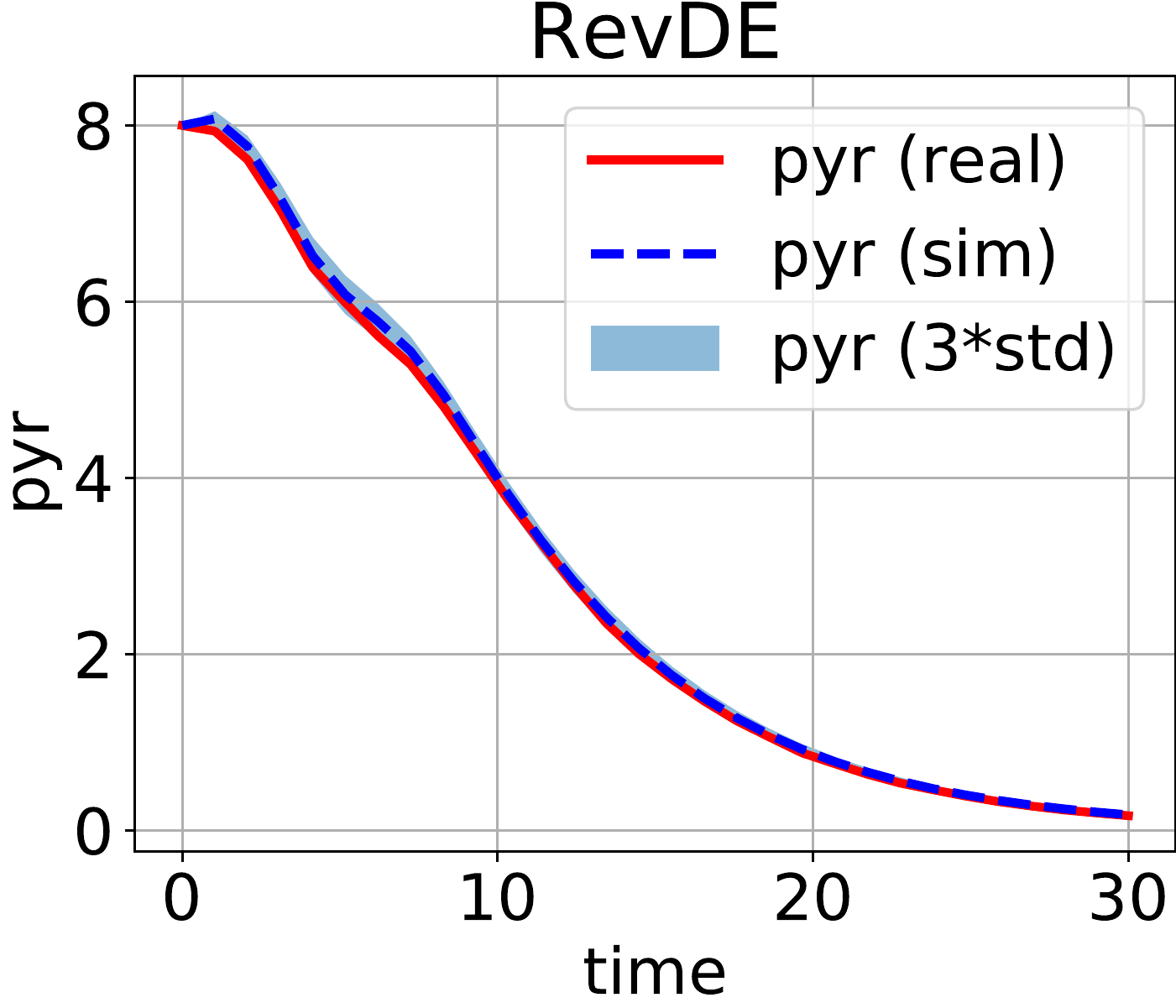} \\
        \vskip 3mm
    \includegraphics[width=110px,height=90px]{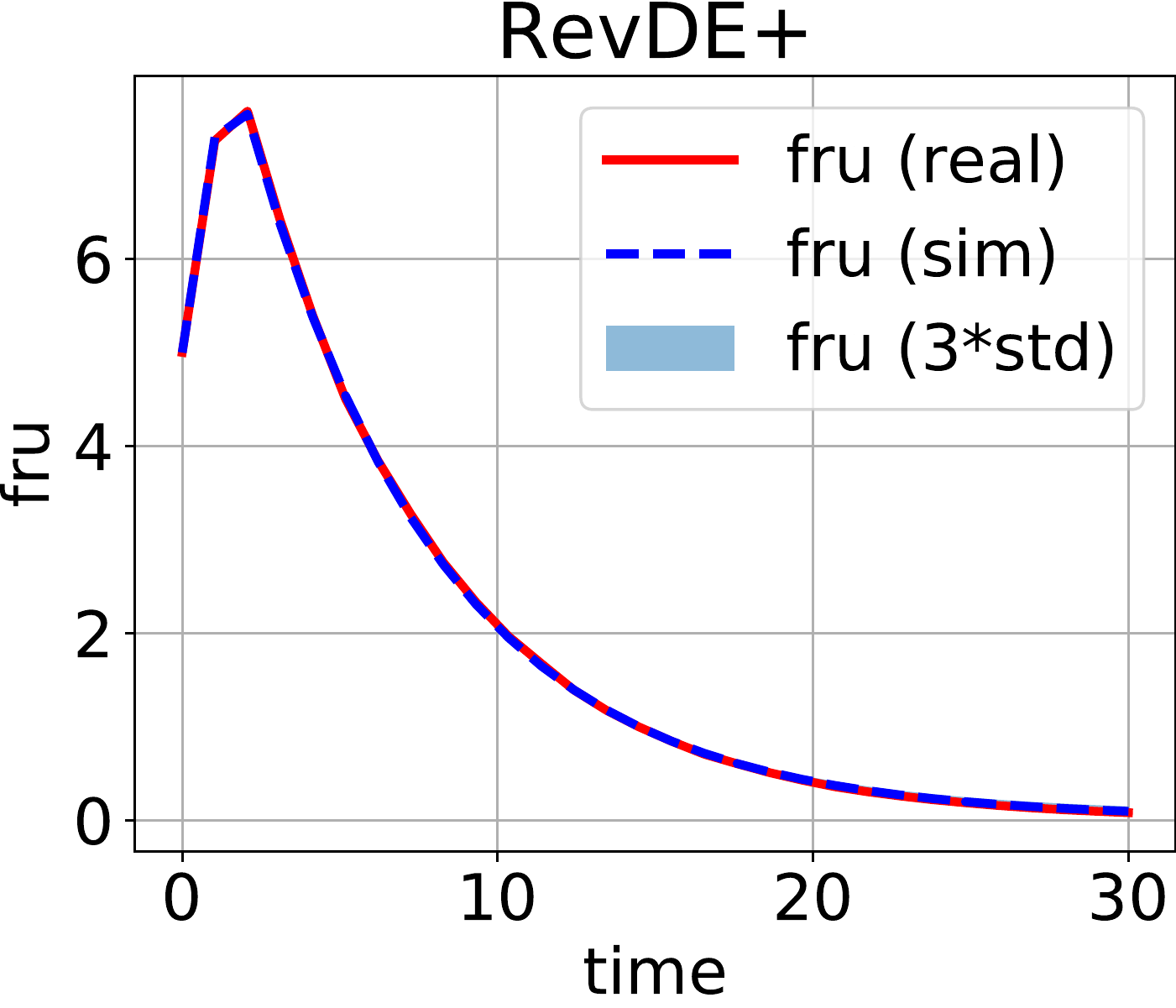}\quad
    \includegraphics[width=110px,height=90px]{figs/results_mutation/mutation__RevDEknn__obs_tp.pdf}\quad
    \includegraphics[width=110px,height=90px]{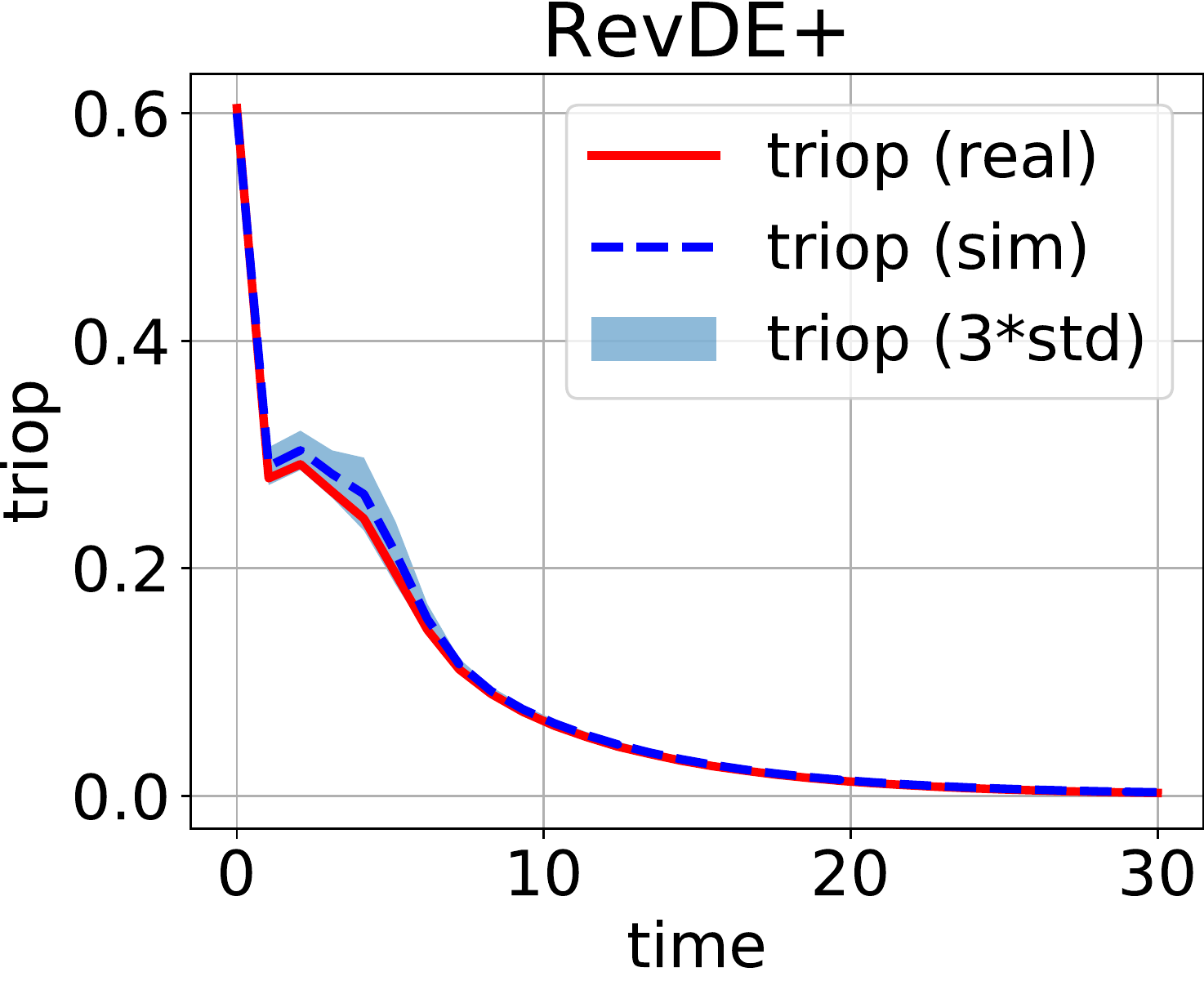}\quad
    \includegraphics[width=110px,height=90px]{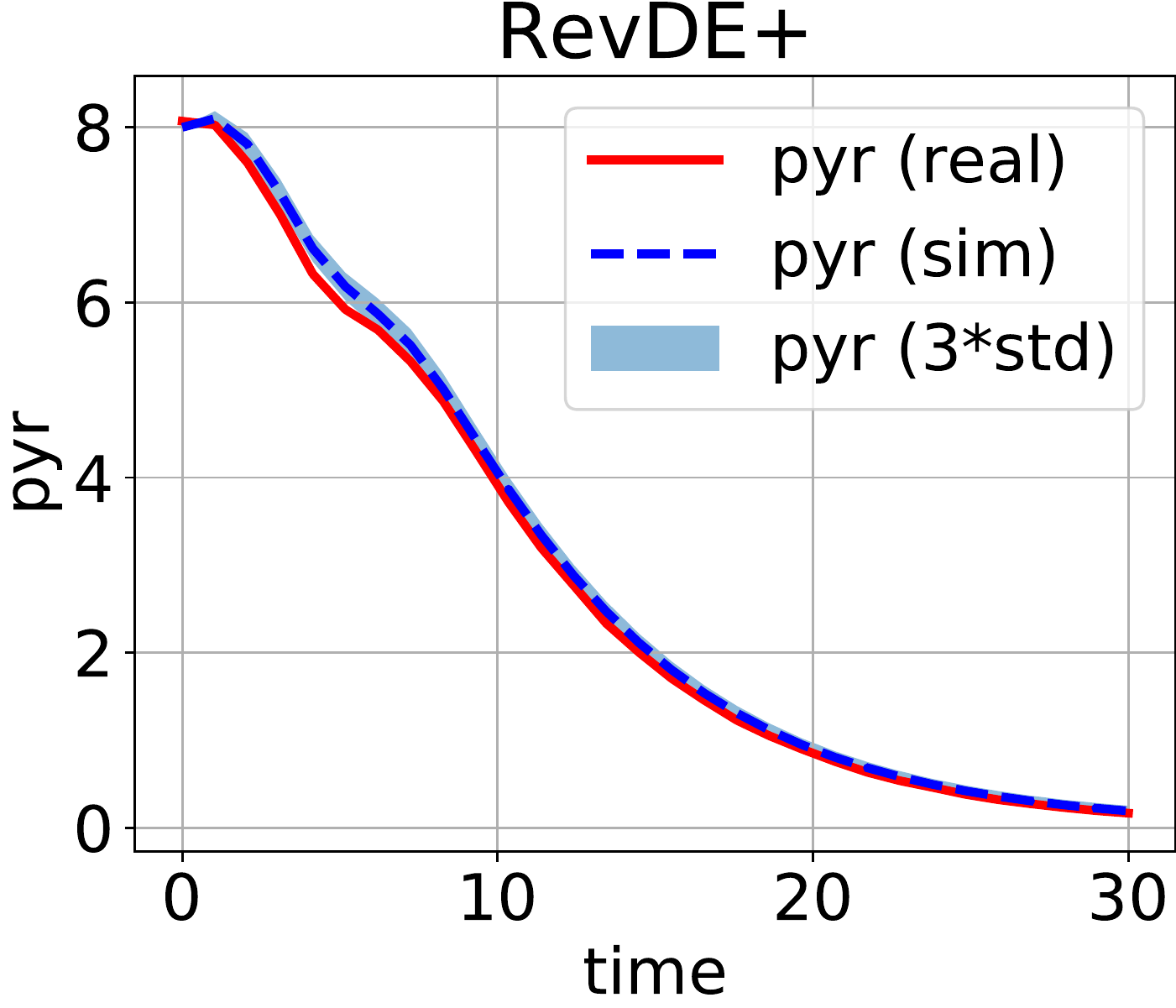} \\
        \vskip 3mm
    \includegraphics[width=110px,height=90px]{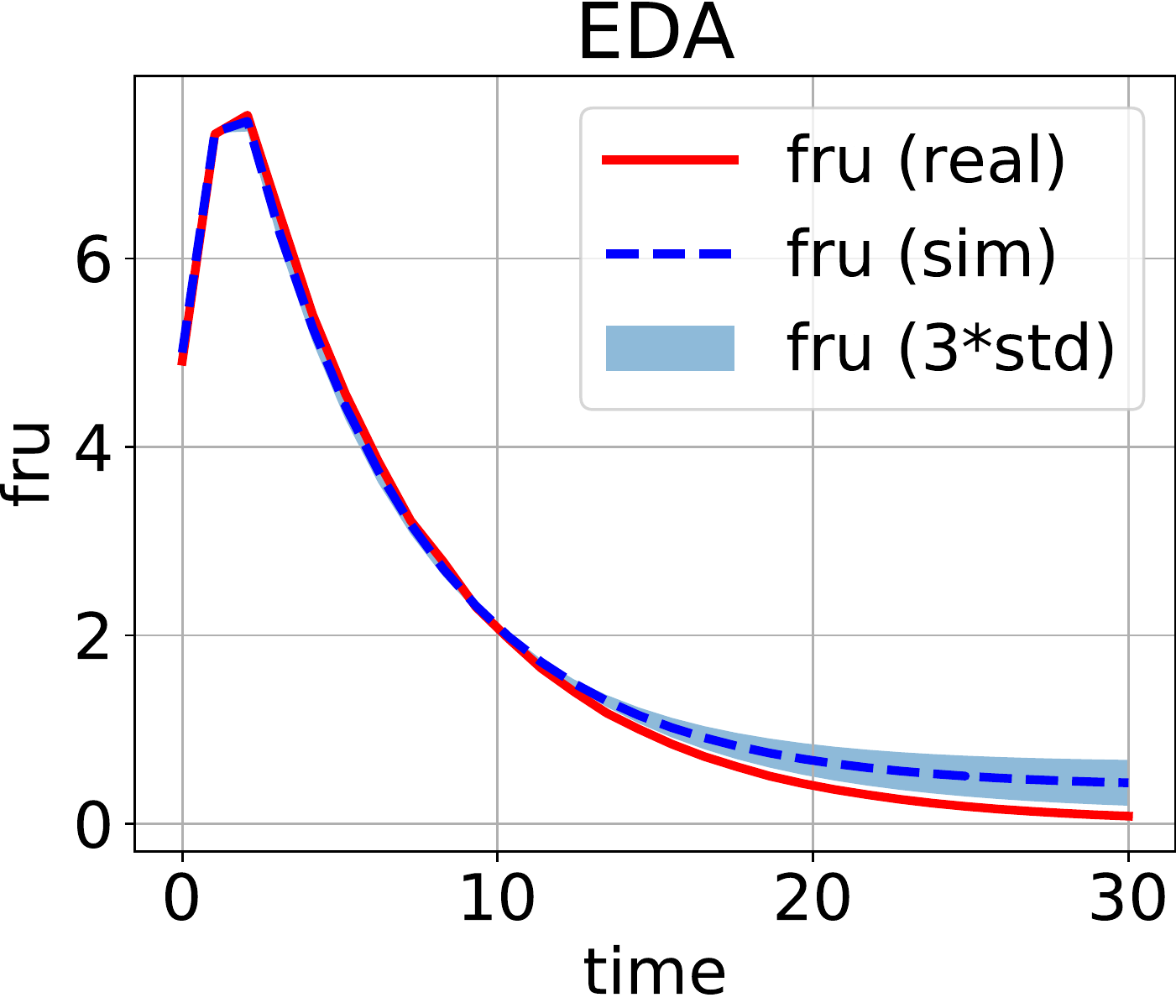}\quad
    \includegraphics[width=110px,height=90px]{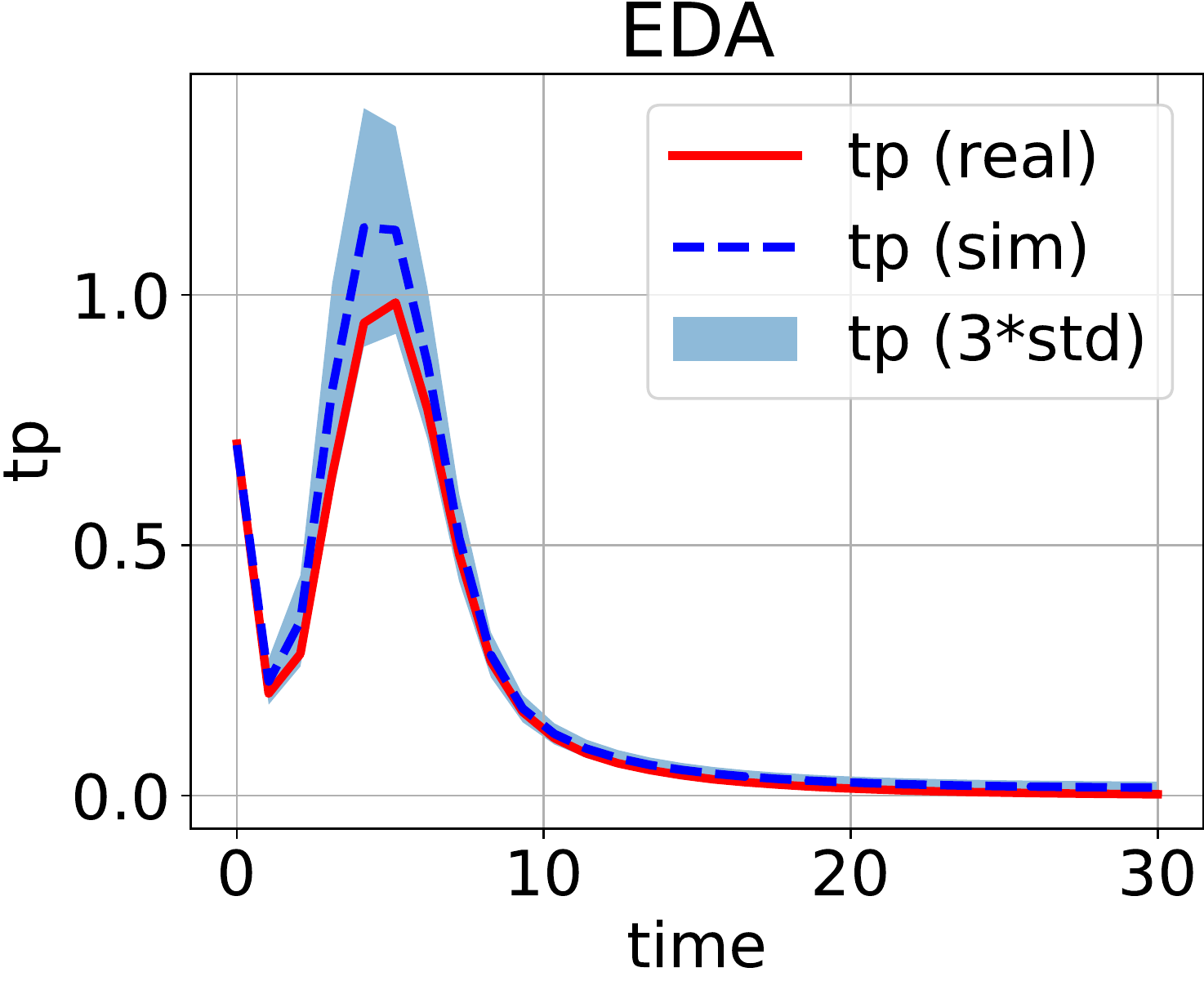}\quad
    \includegraphics[width=110px,height=90px]{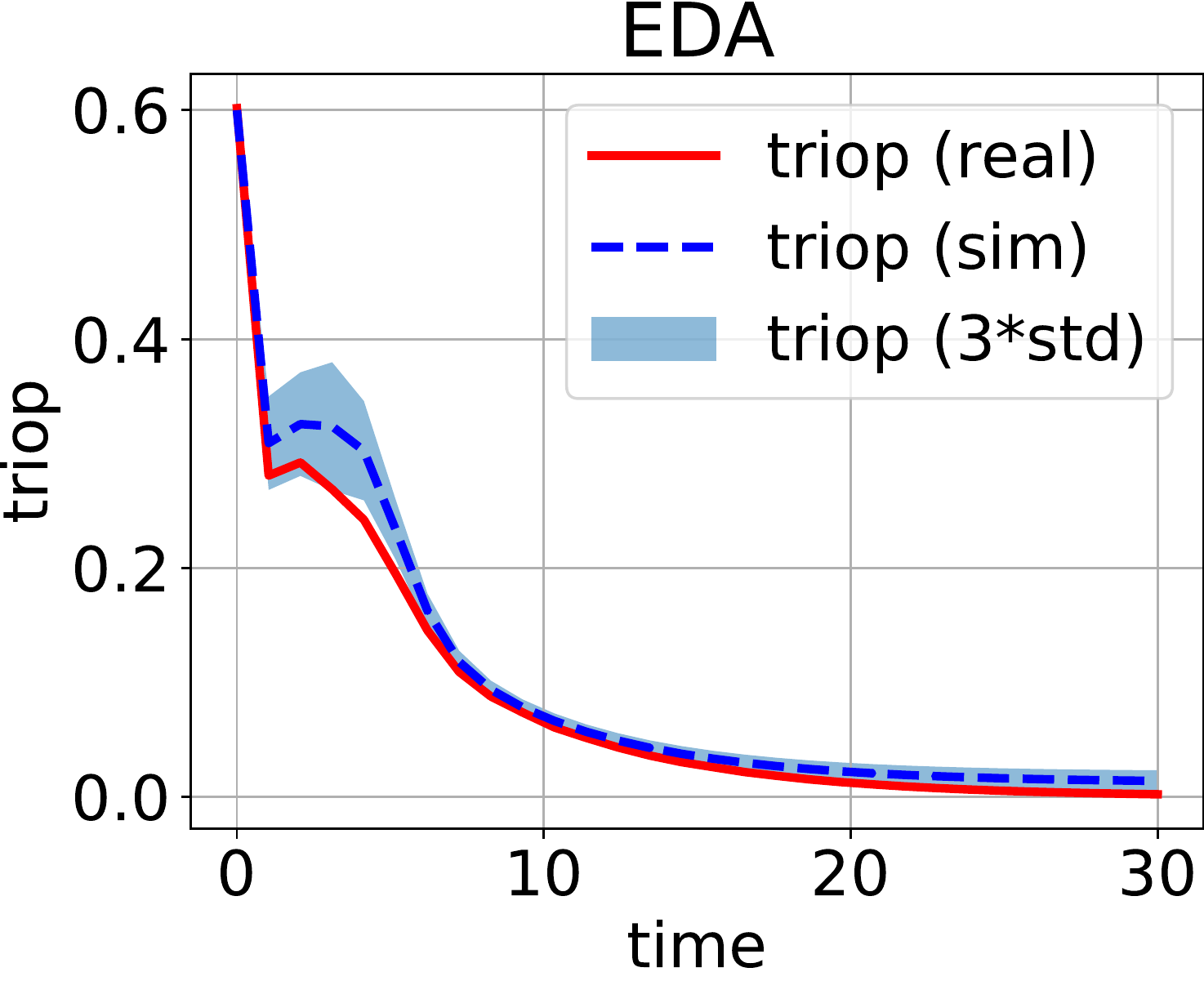}\quad
    \includegraphics[width=110px,height=90px]{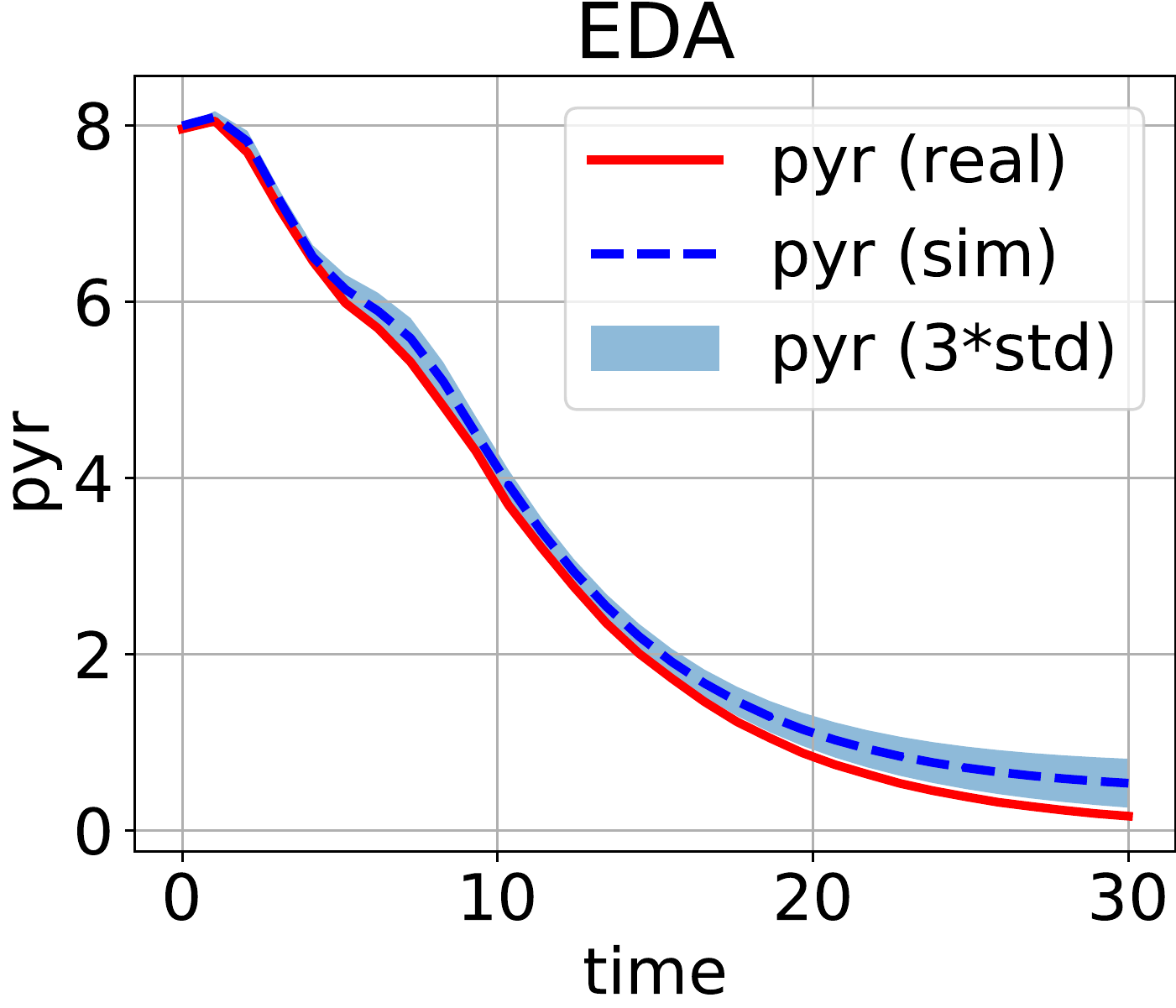} \\
        \vskip 3mm
    \includegraphics[width=110px,height=90px]{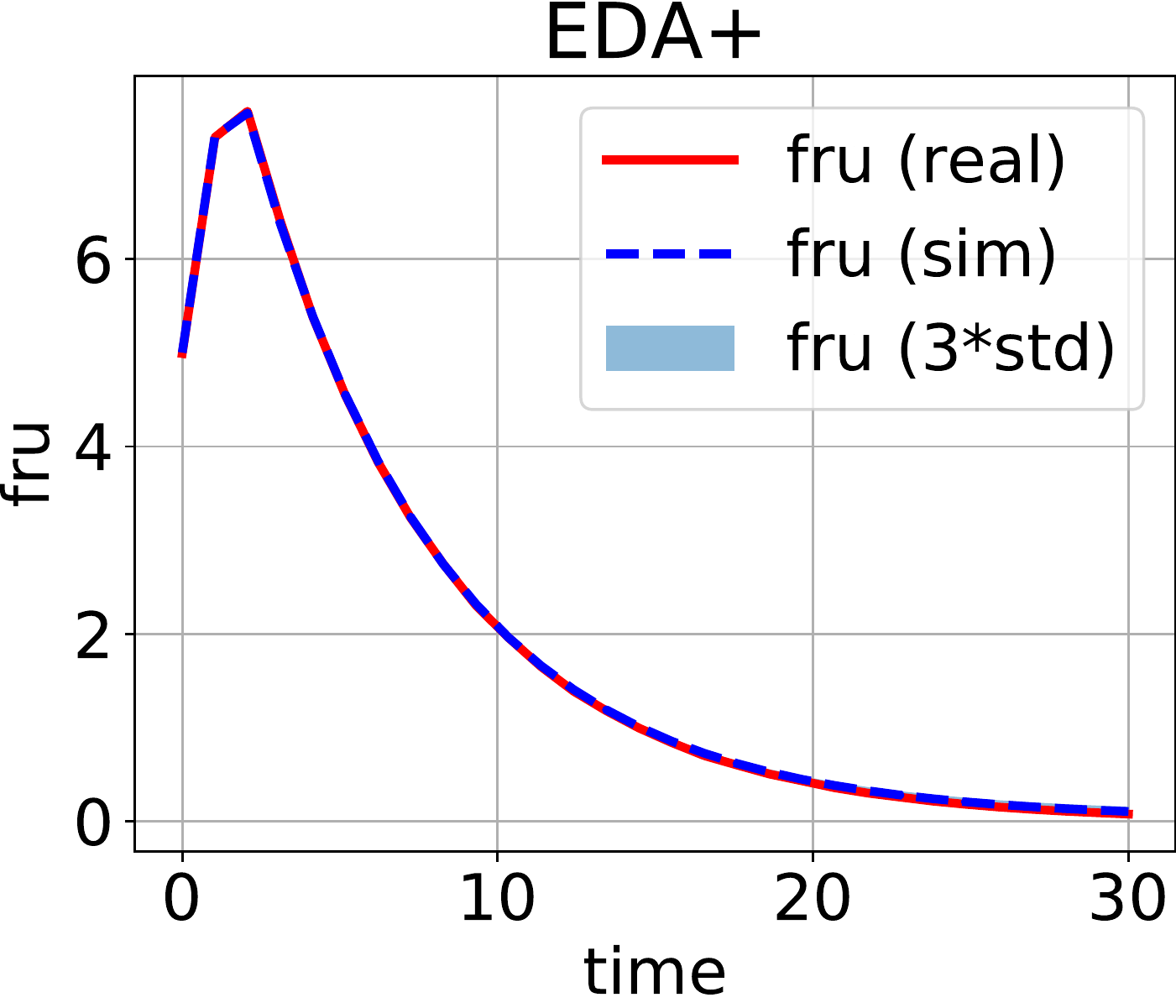}\quad
    \includegraphics[width=110px,height=90px]{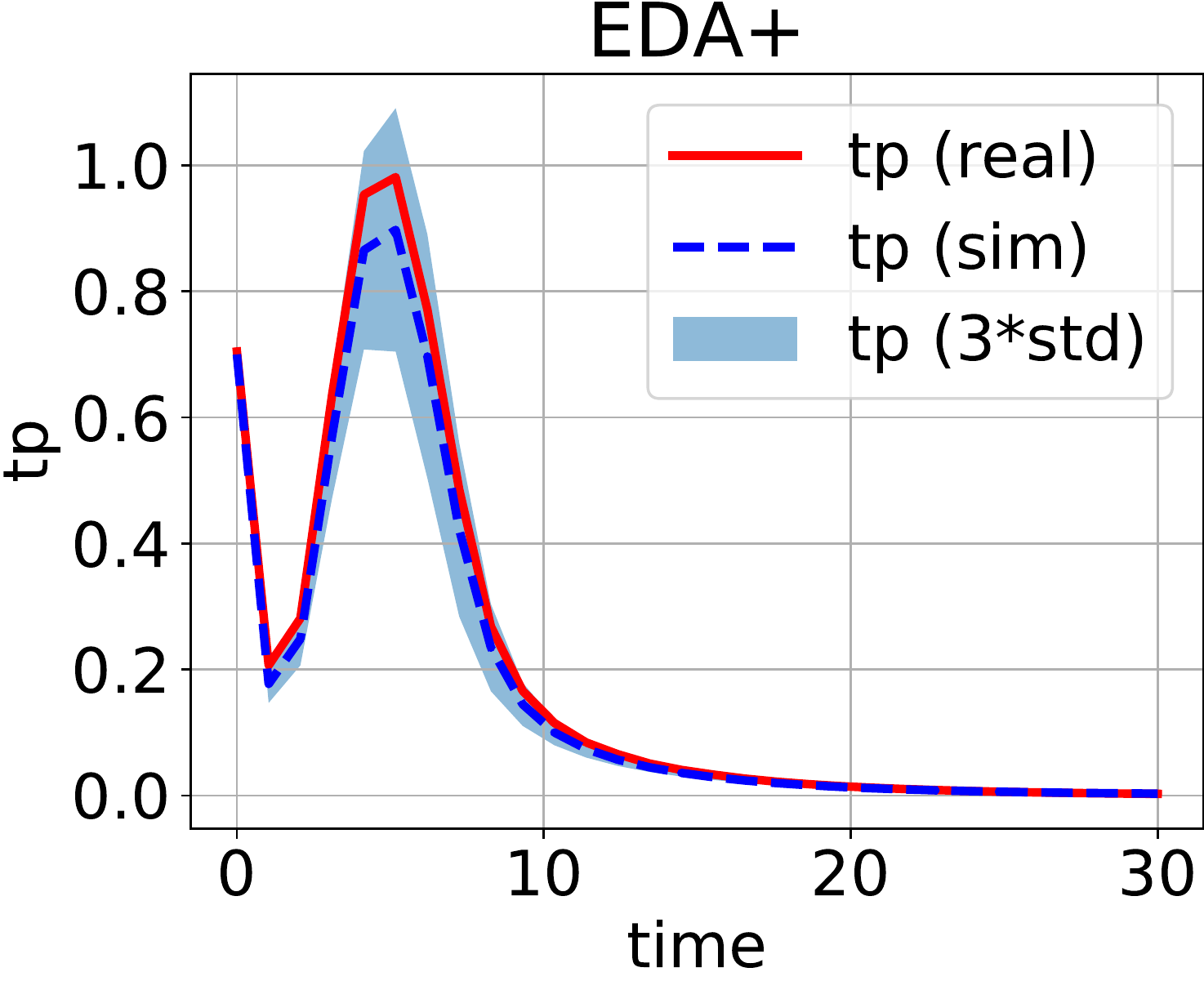}\quad
    \includegraphics[width=110px,height=90px]{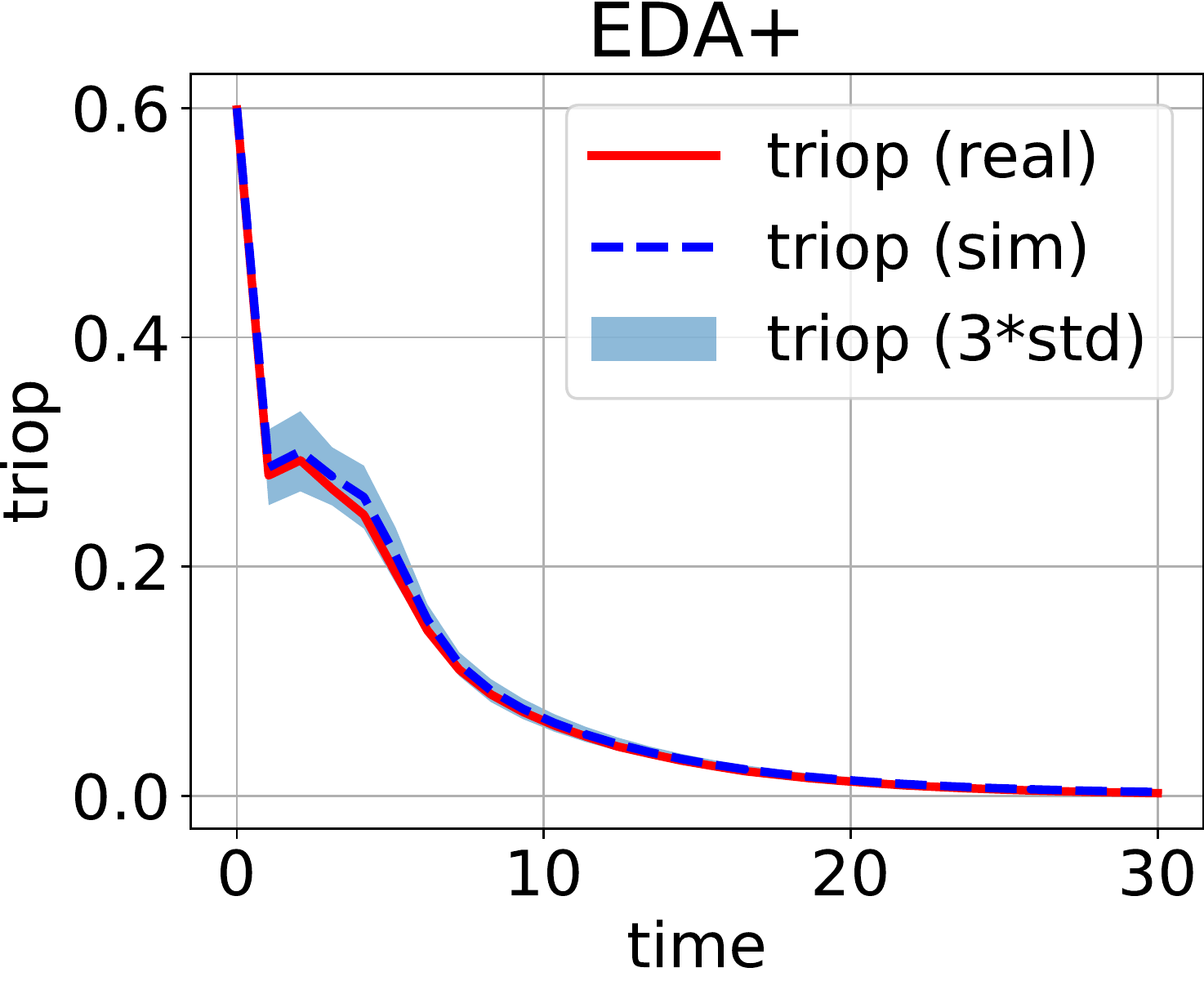}\quad
    \includegraphics[width=110px,height=90px]{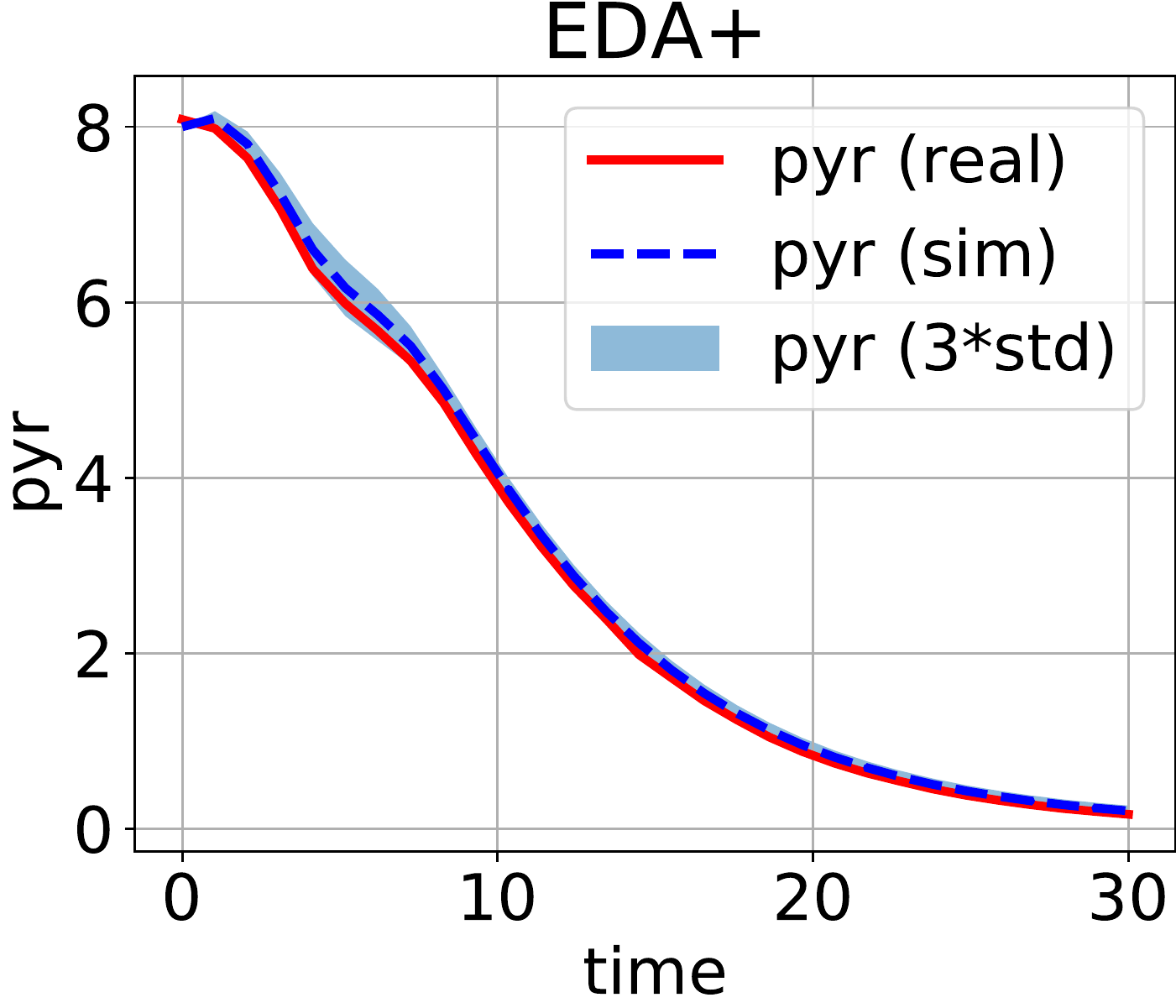} \\
        \vskip 3mm
    \includegraphics[width=110px,height=90px]{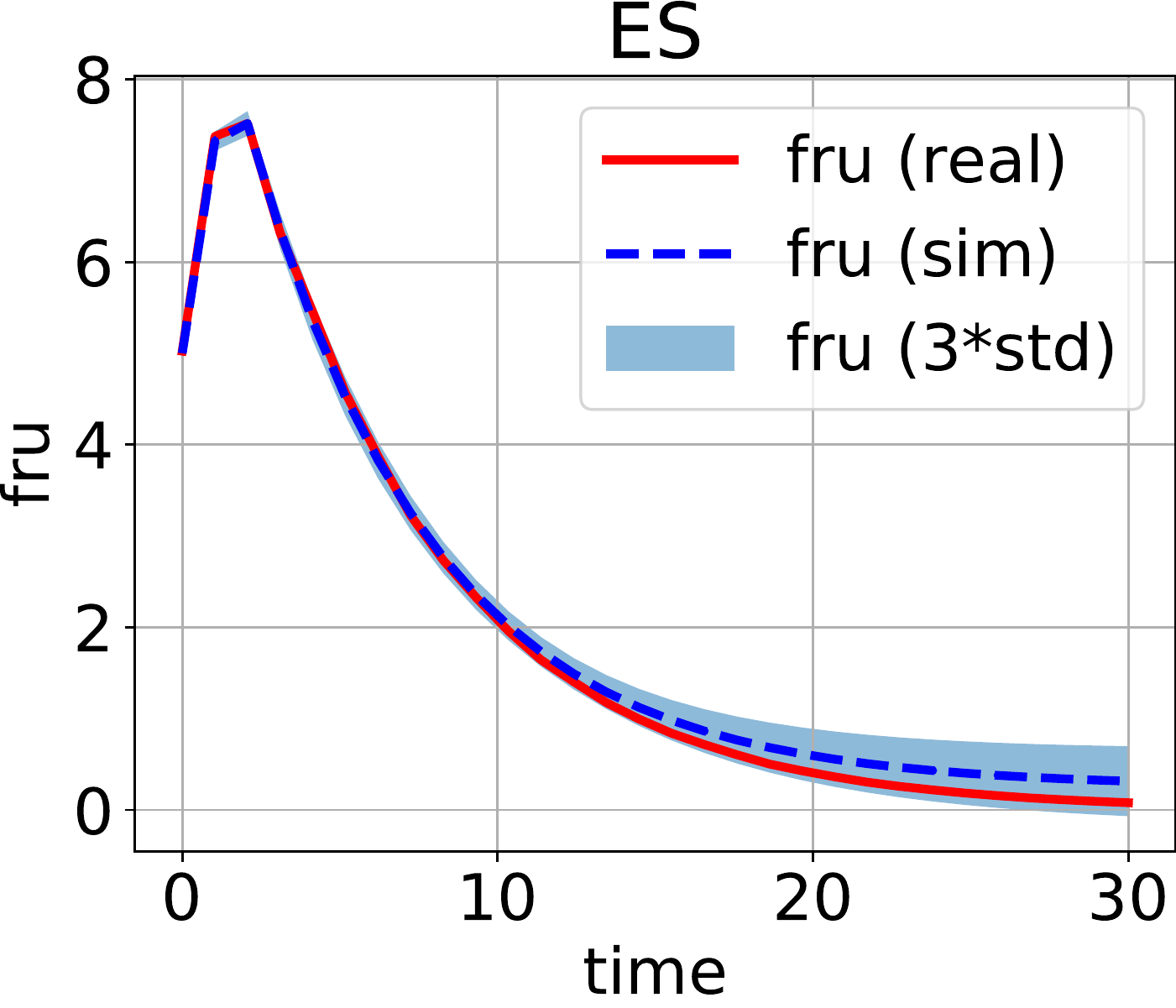}\quad
    \includegraphics[width=110px,height=90px]{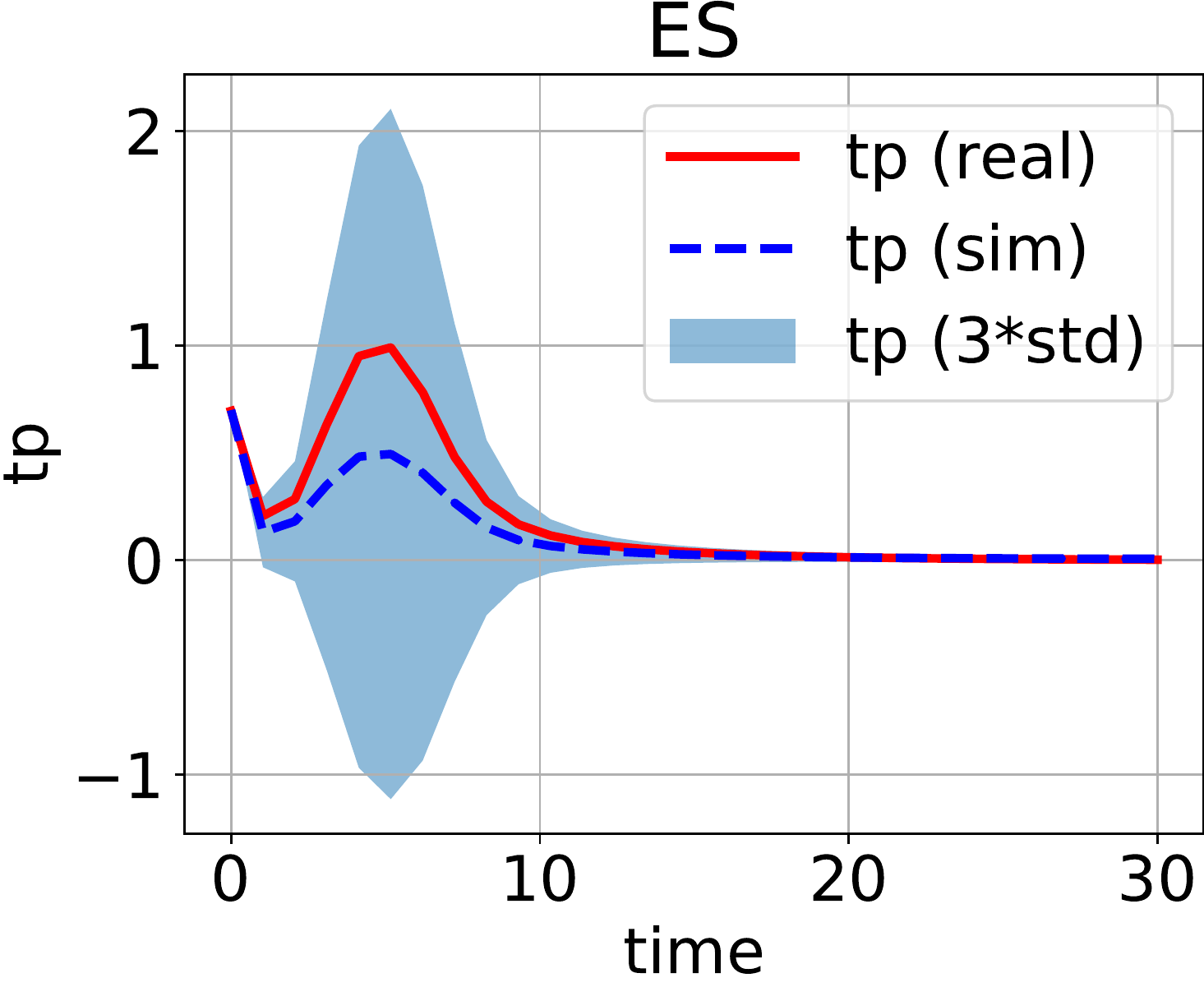}\quad
    \includegraphics[width=110px,height=90px]{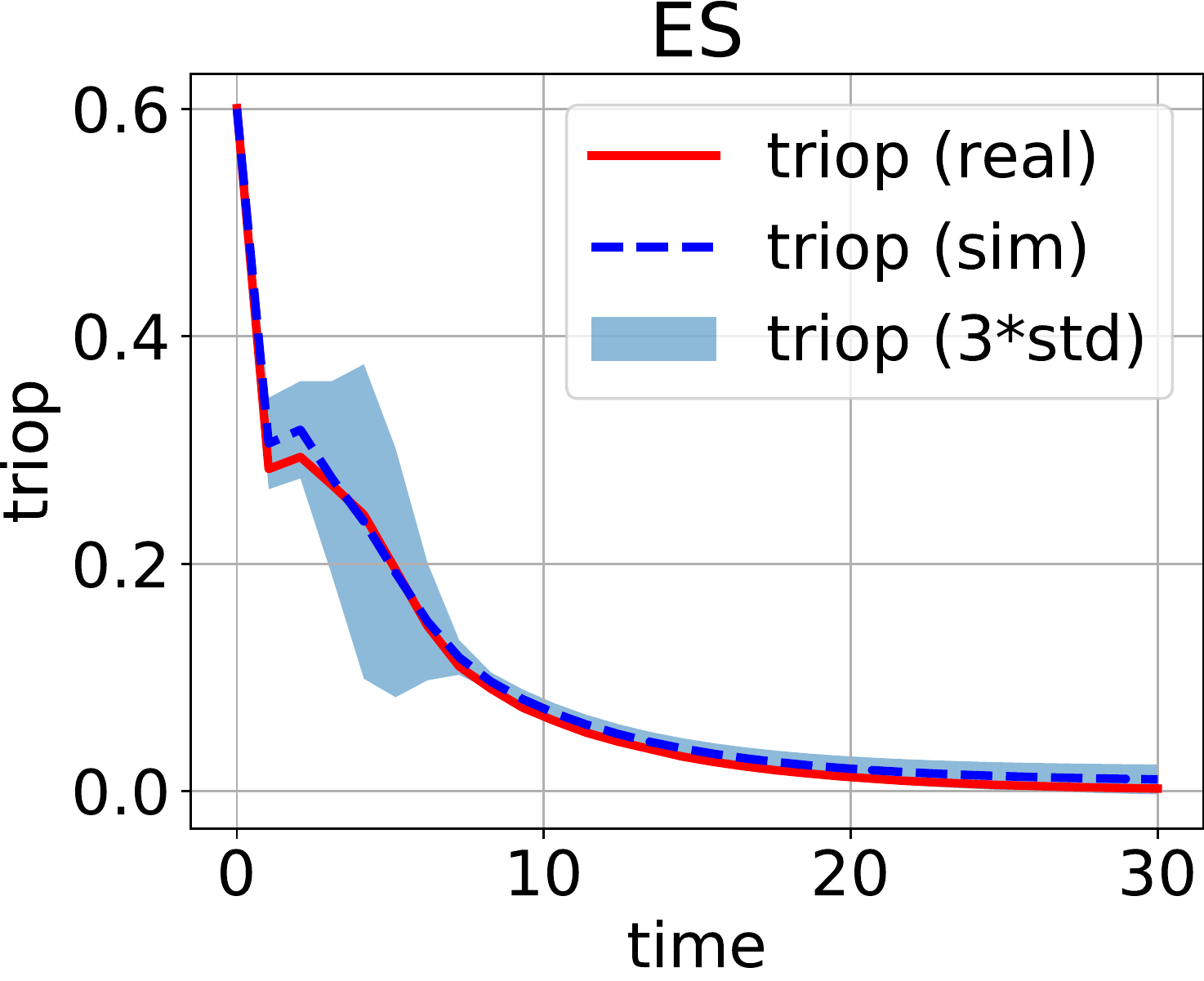}\quad
    \includegraphics[width=110px,height=90px]{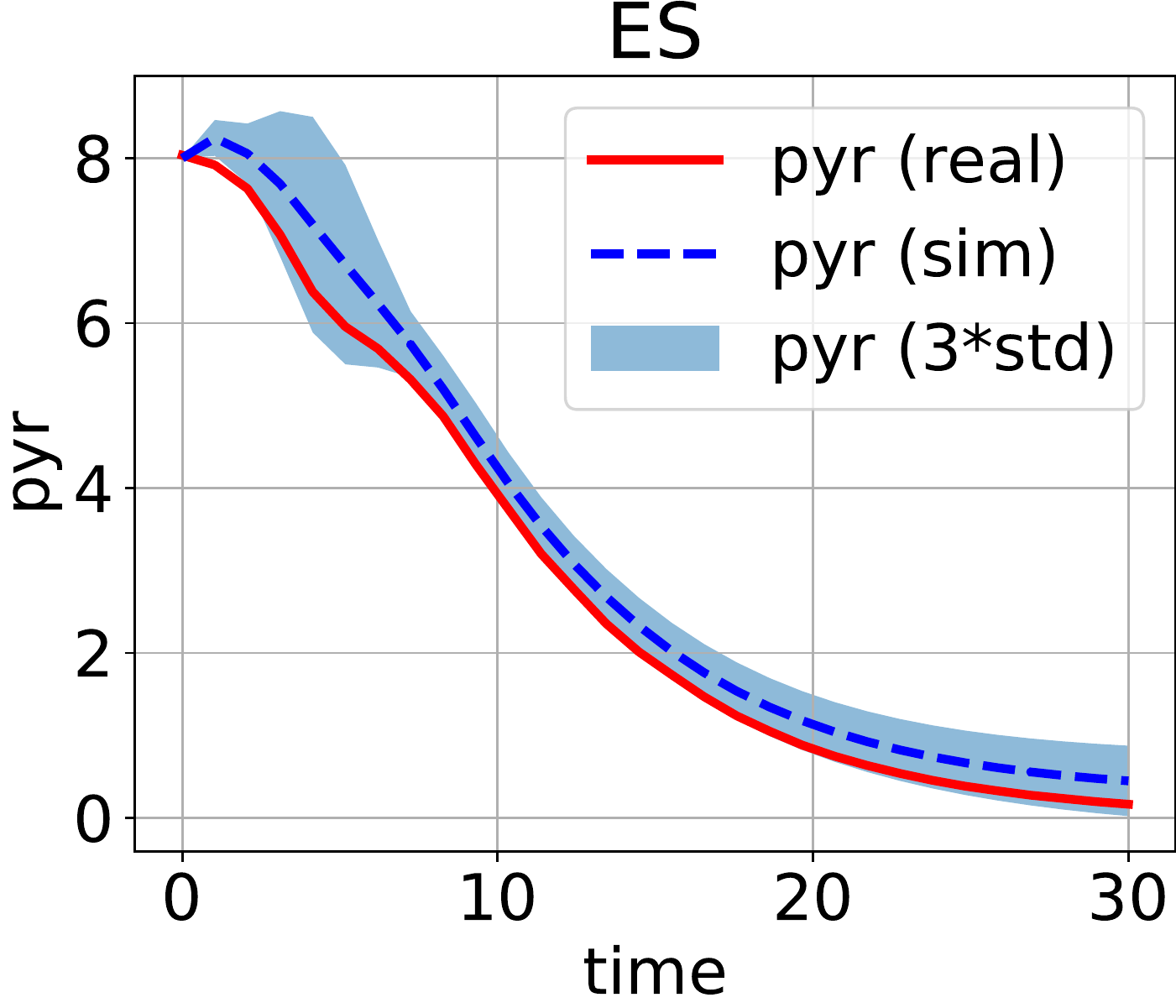}
    \caption{A comparison of the timecourses of the unobserved metabolites in the Case 2. Real timecourses are depicted in red, and the average value and a confidence interval ($3\times$ standard deviation) over $3$ runs of the simulator are in blue. The titles of the plots indicate optimization methods.}
    \label{fig:obs_2}
\end{figure*}

\begin{figure*}[!tbp]
    \centering
    \includegraphics[width=110px,height=90px]{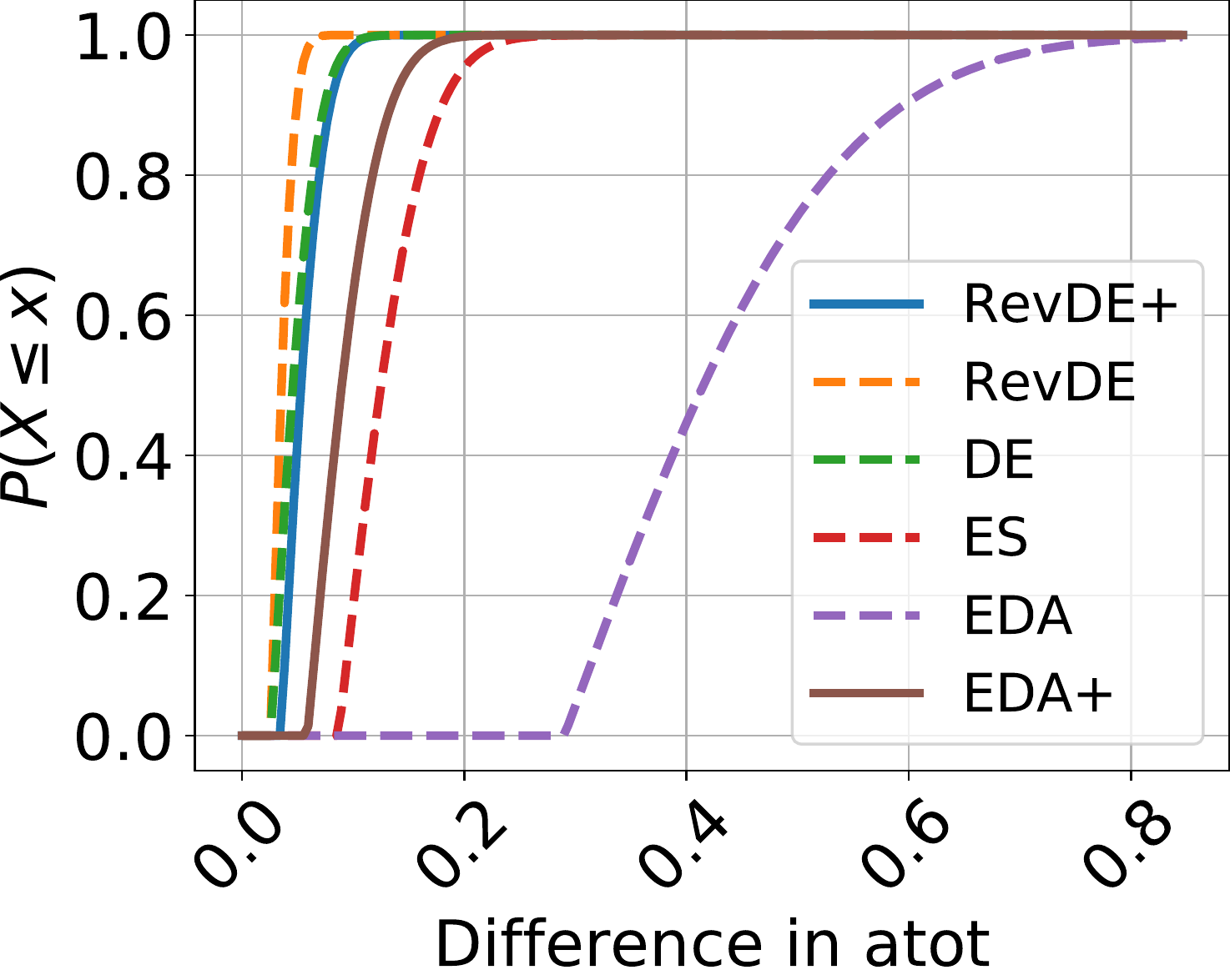}\quad
    \includegraphics[width=110px,height=90px]{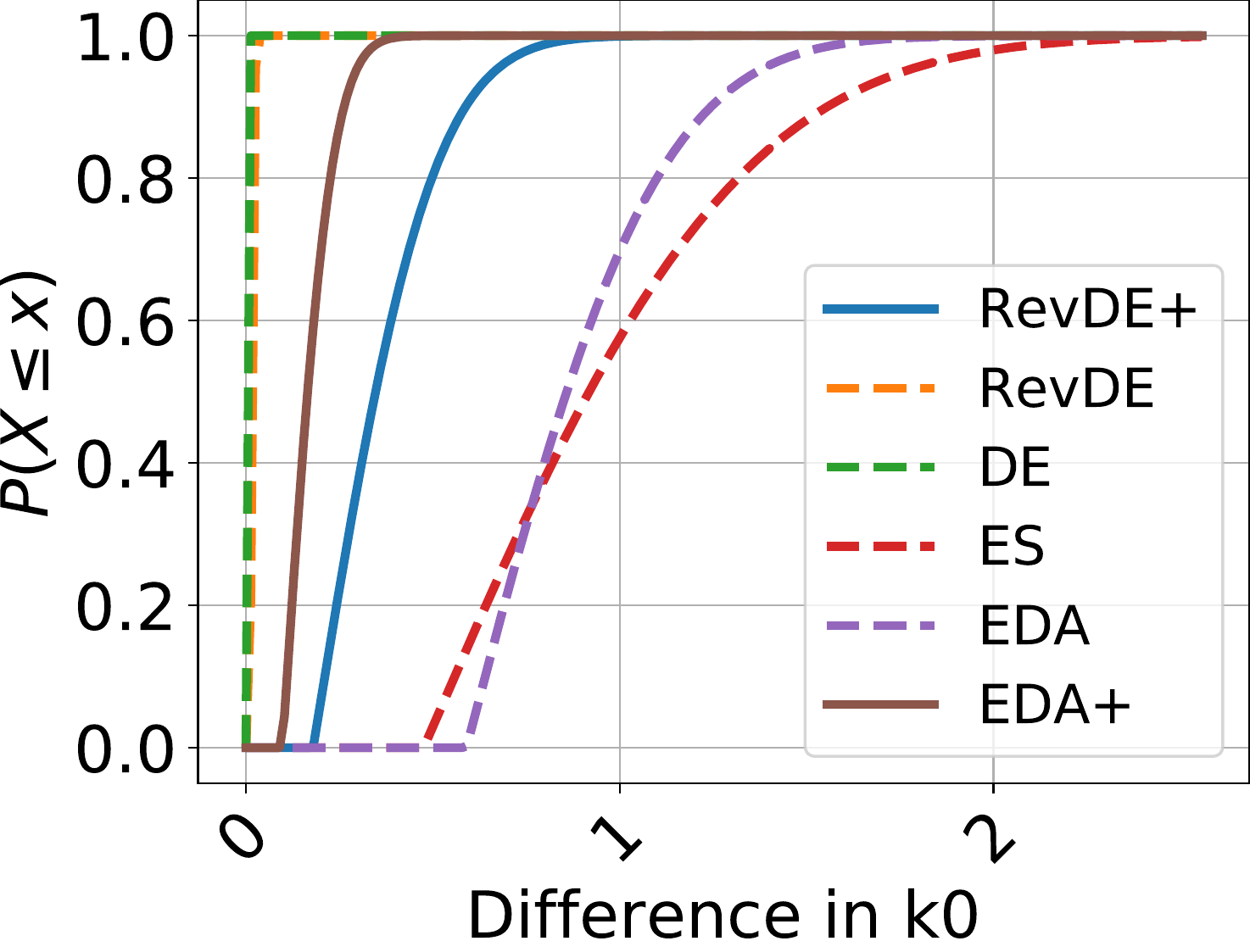}\quad
    \includegraphics[width=110px,height=90px]{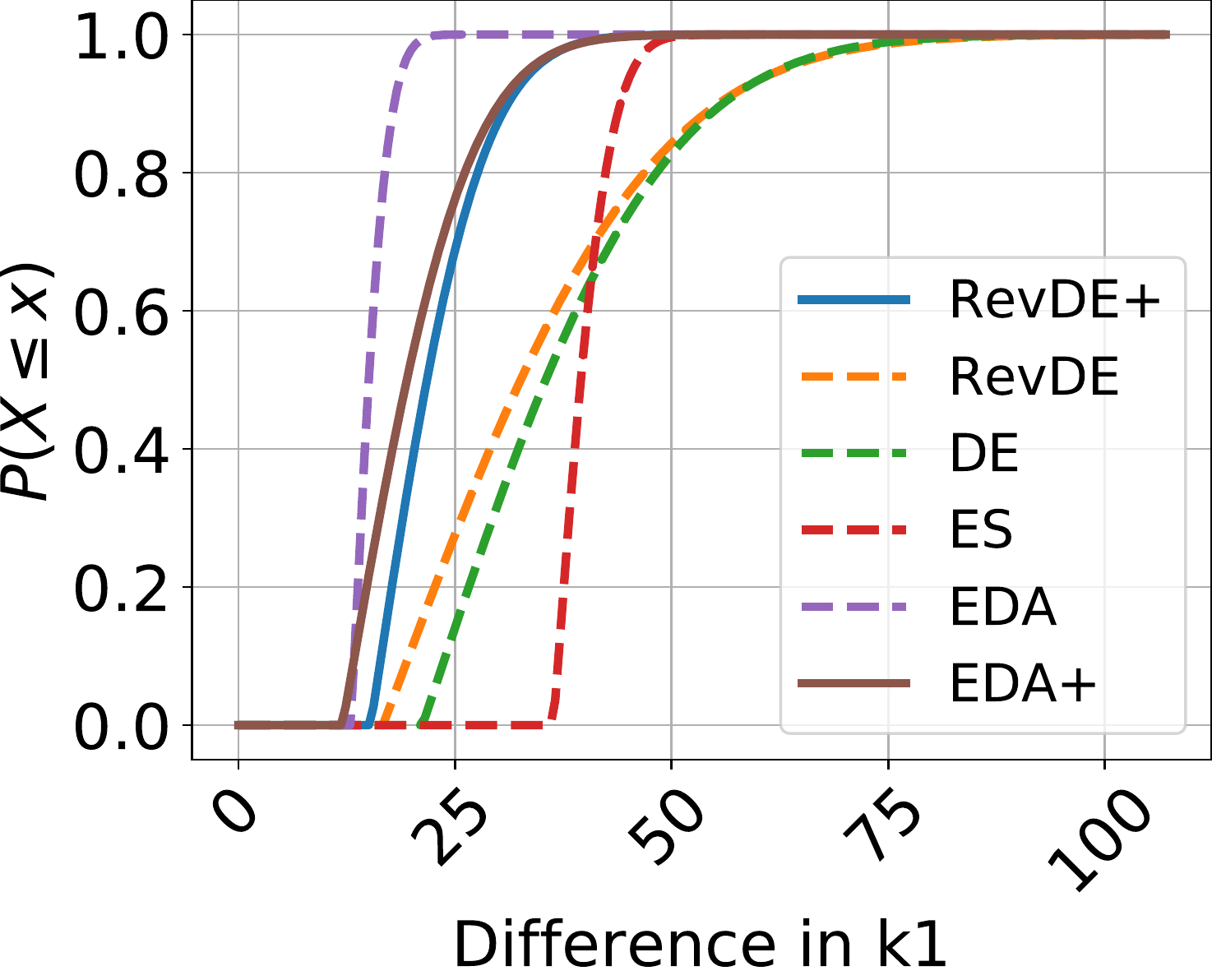}\quad
    \includegraphics[width=110px,height=90px]{figs/parameters_difference/wolf_differences_params_k2.pdf} \\
        \vskip 3mm
    \includegraphics[width=110px,height=90px]{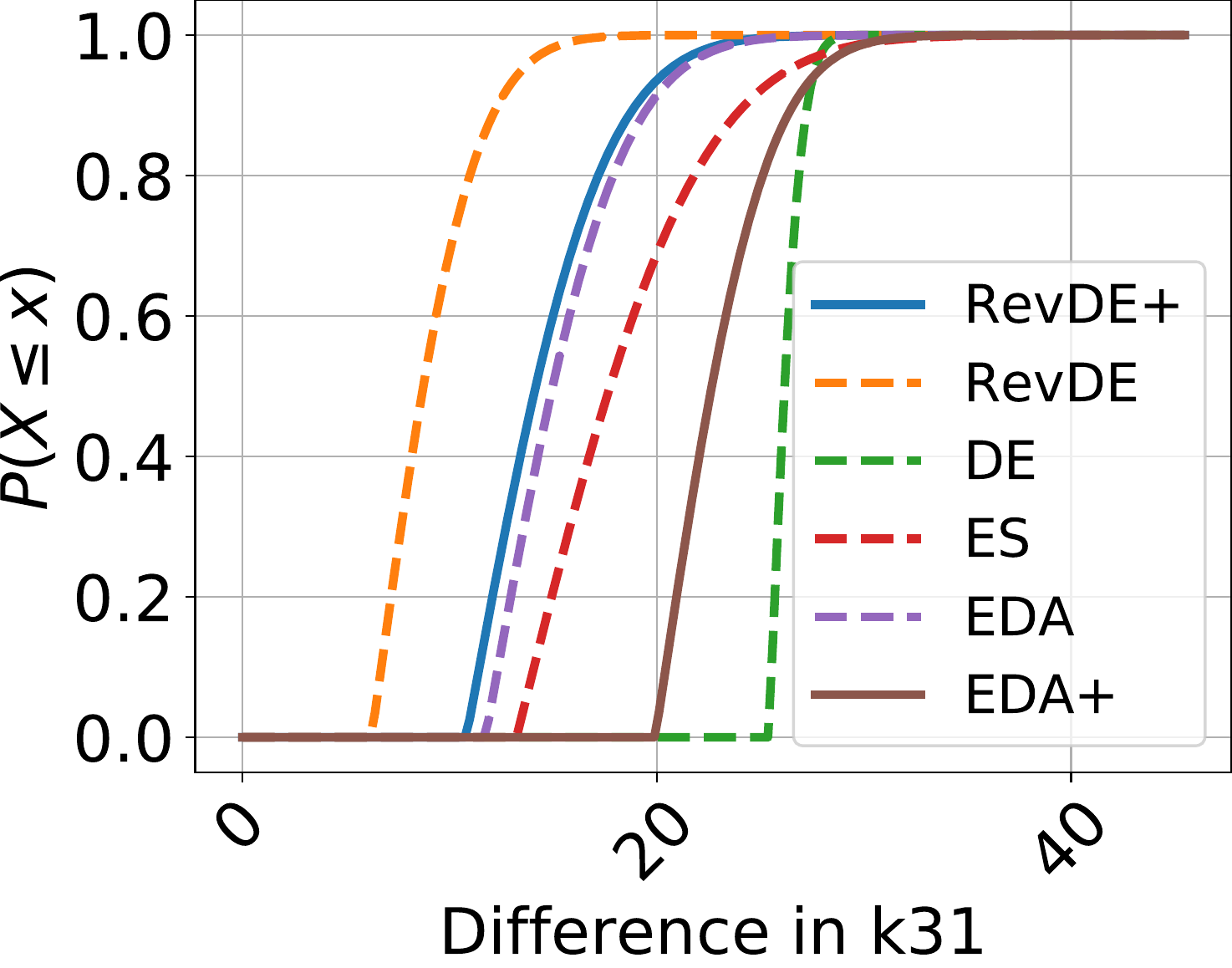}\quad
    \includegraphics[width=110px,height=90px]{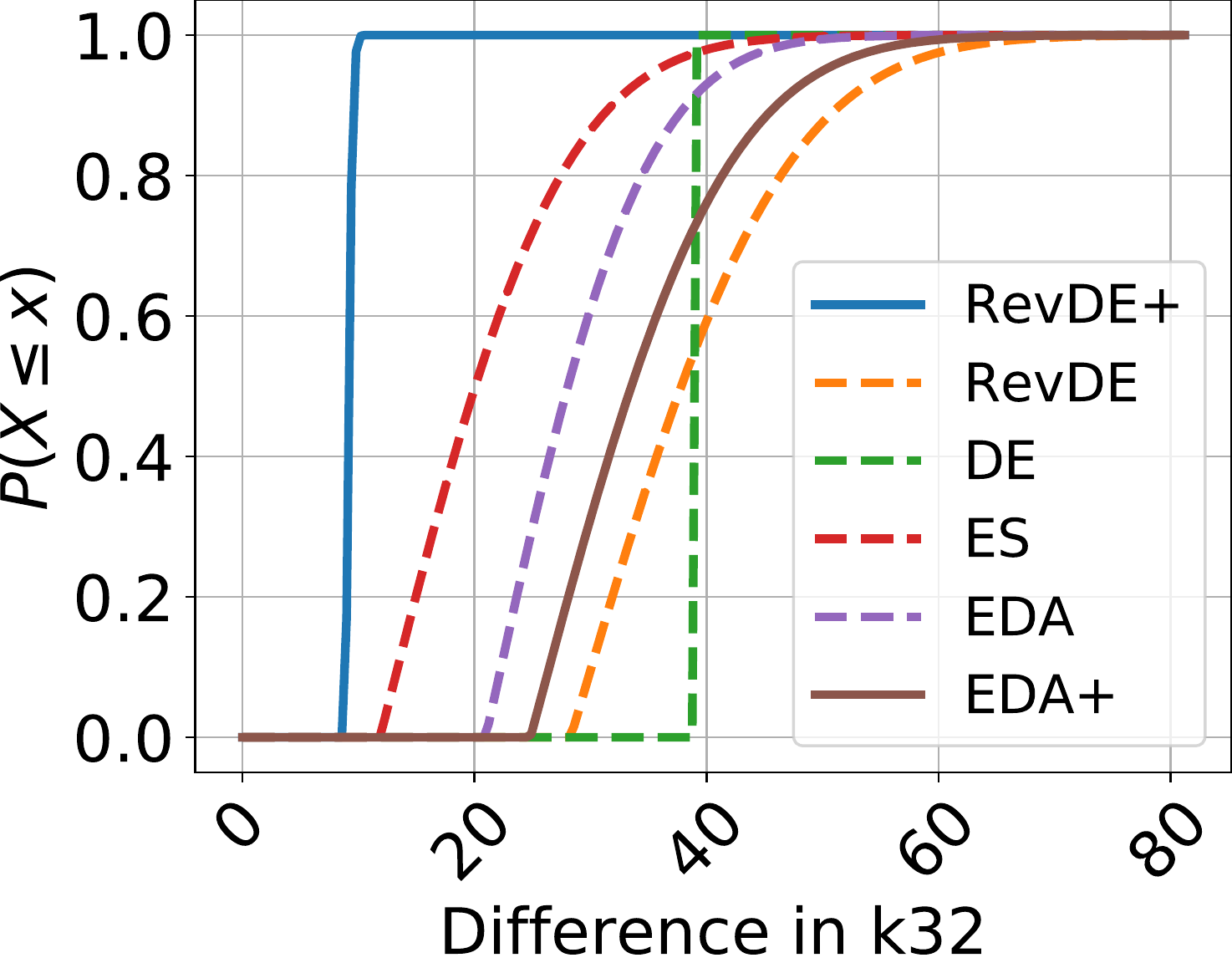}\quad
    \includegraphics[width=110px,height=90px]{figs/parameters_difference/wolf_differences_params_k33.pdf}\quad
    \includegraphics[width=110px,height=90px]{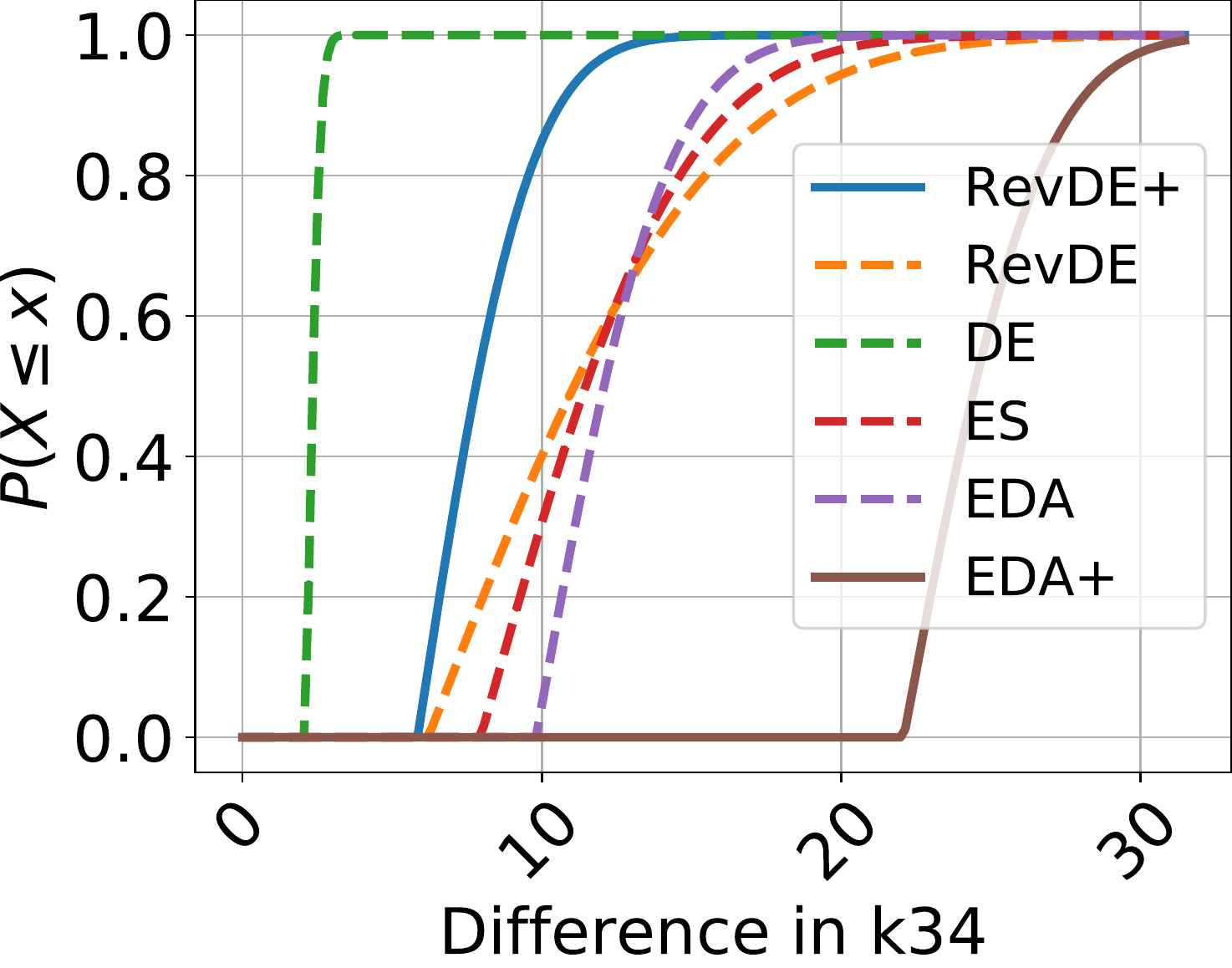} \\
        \vskip 3mm
    \includegraphics[width=110px,height=90px]{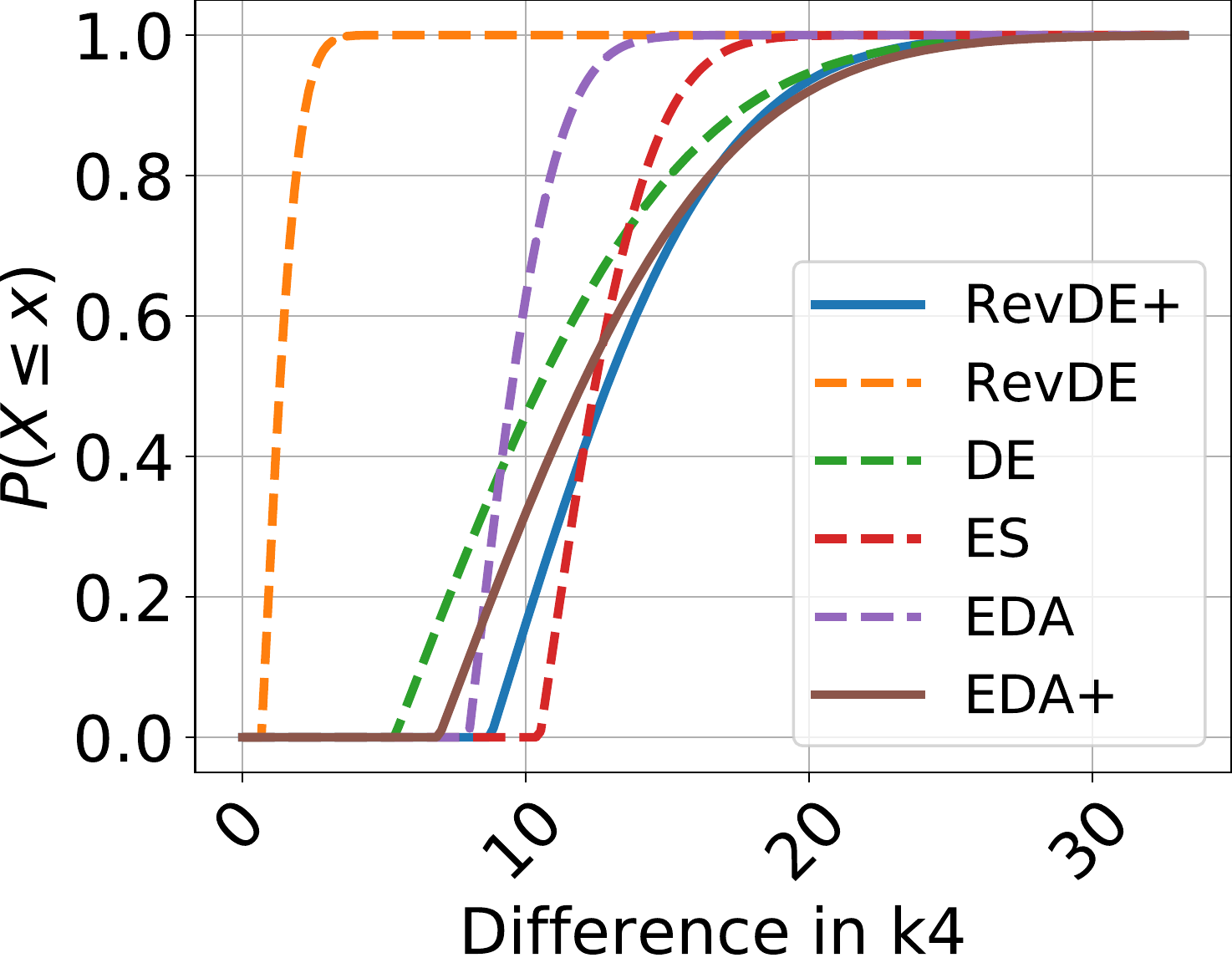}\quad
    \includegraphics[width=110px,height=90px]{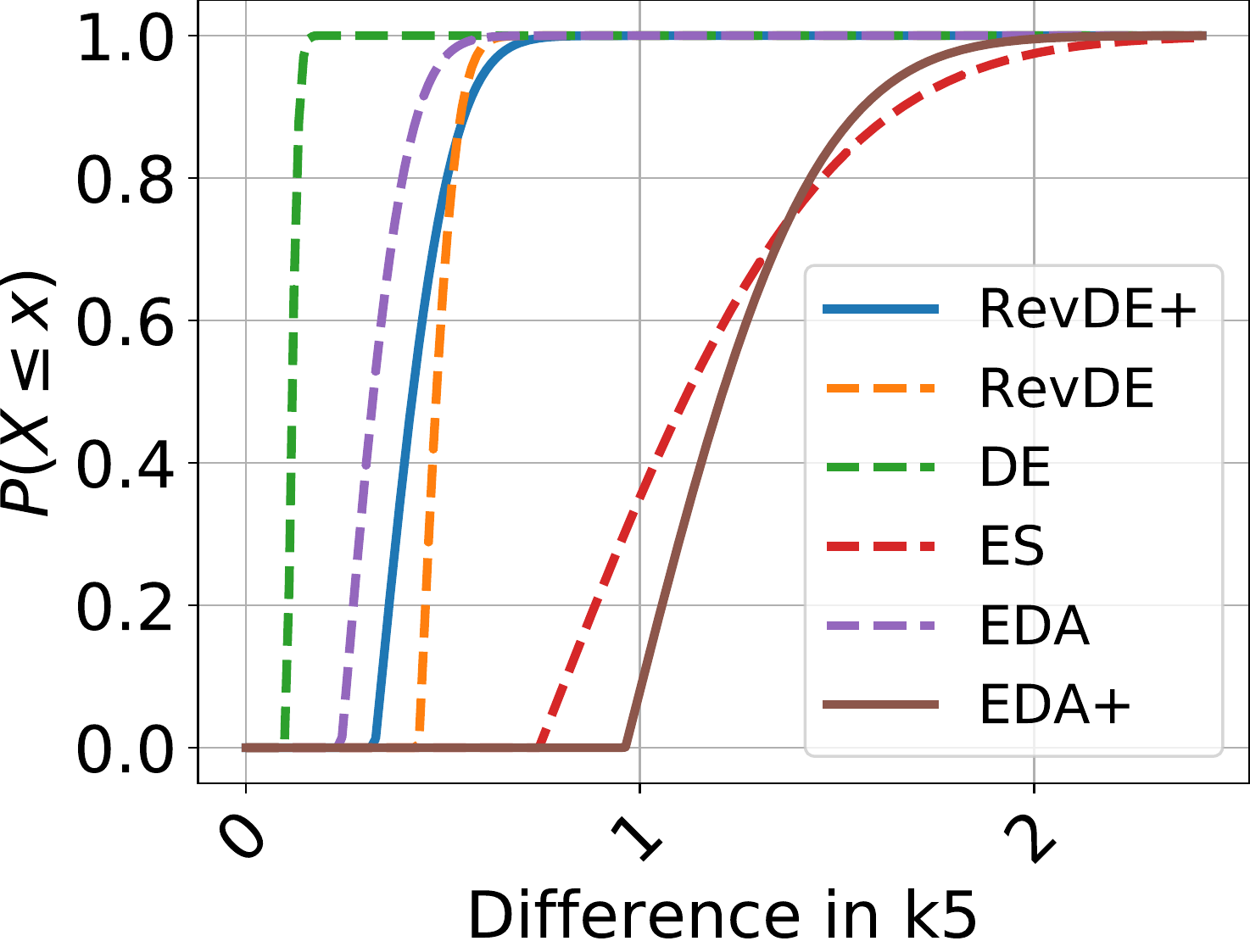}\quad
    \includegraphics[width=110px,height=90px]{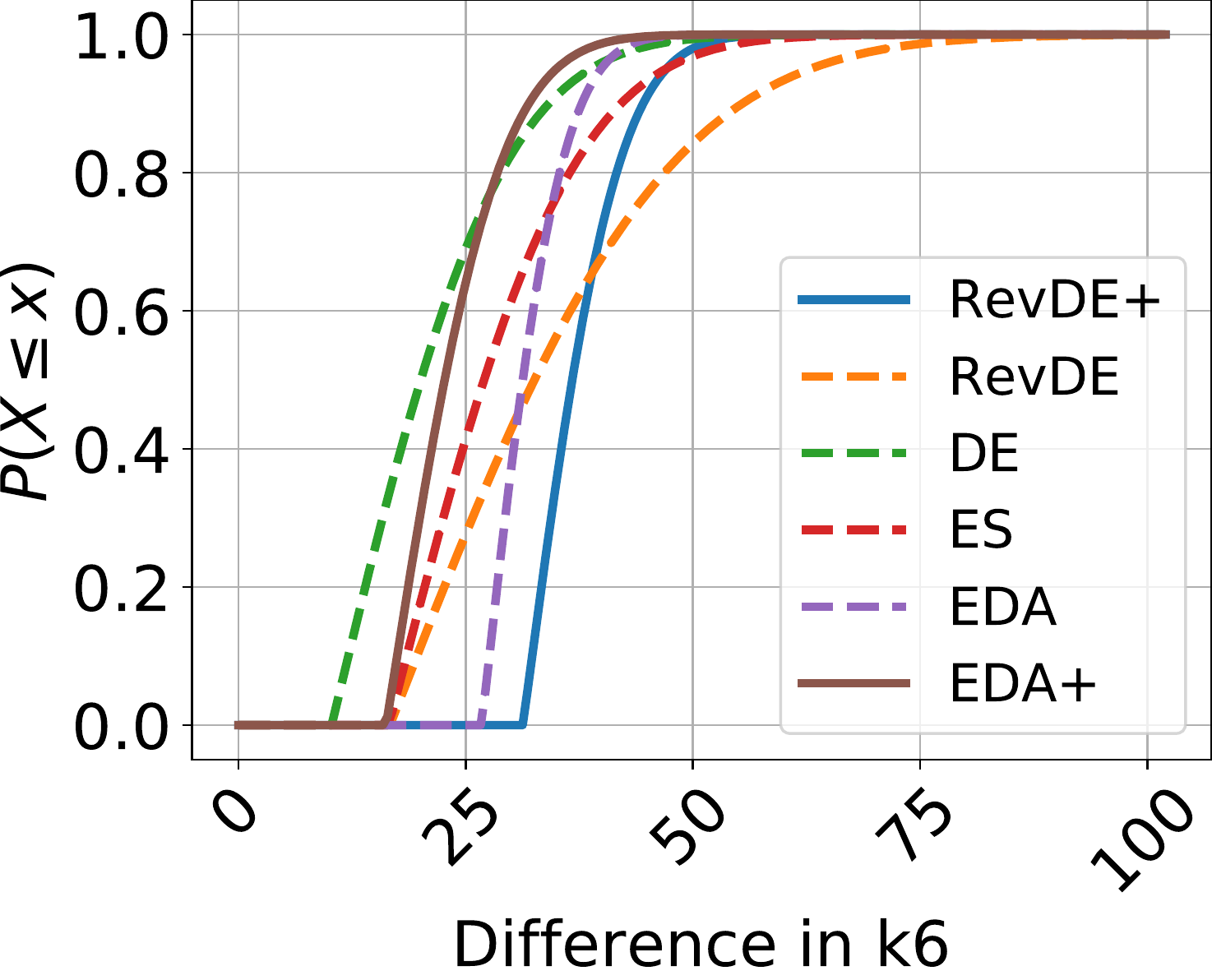}\quad
    \includegraphics[width=110px,height=90px]{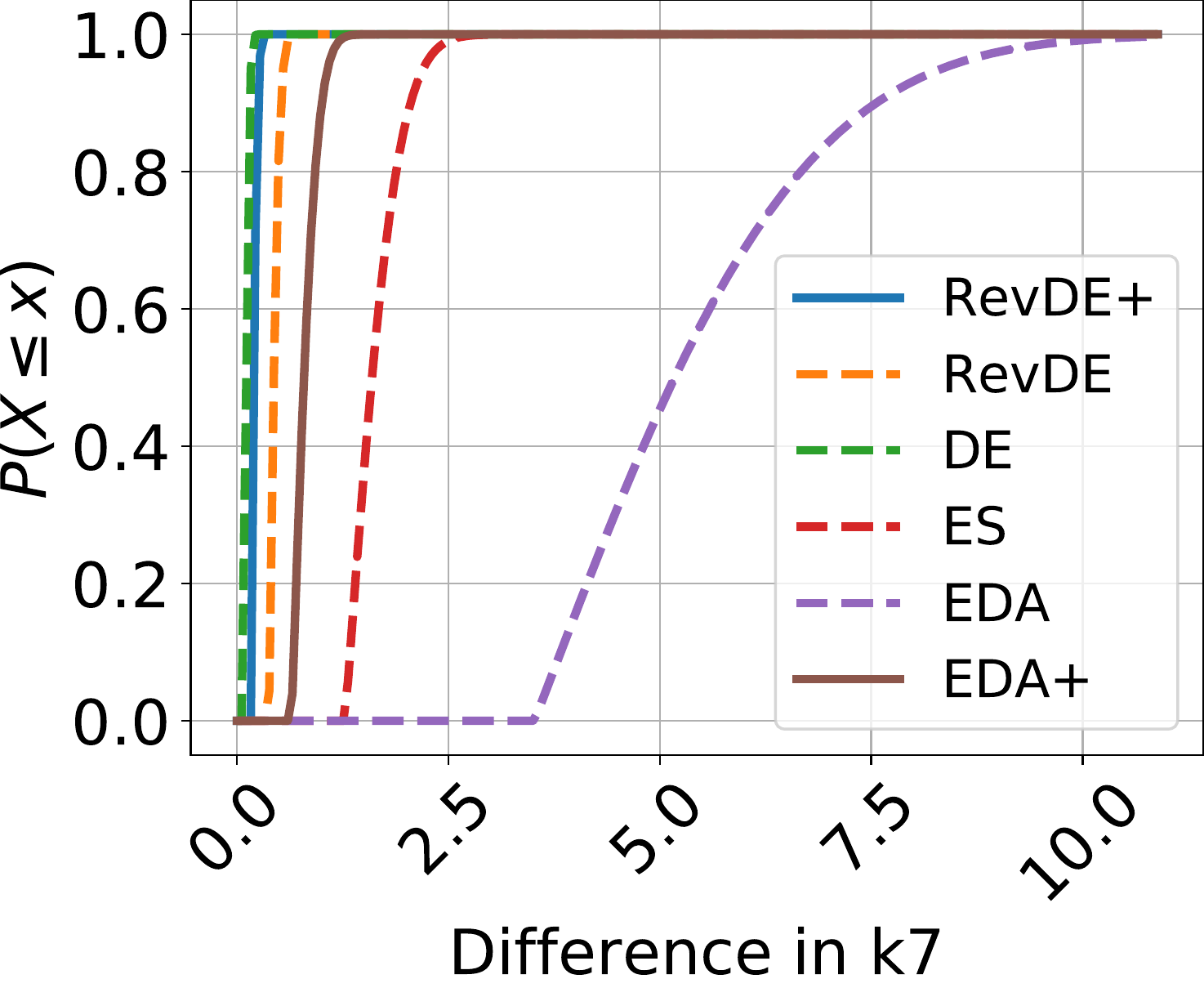} \\
        \vskip 3mm
    \includegraphics[width=110px,height=90px]{figs/parameters_difference/wolf_differences_params_k8.pdf}\quad
    \includegraphics[width=110px,height=90px]{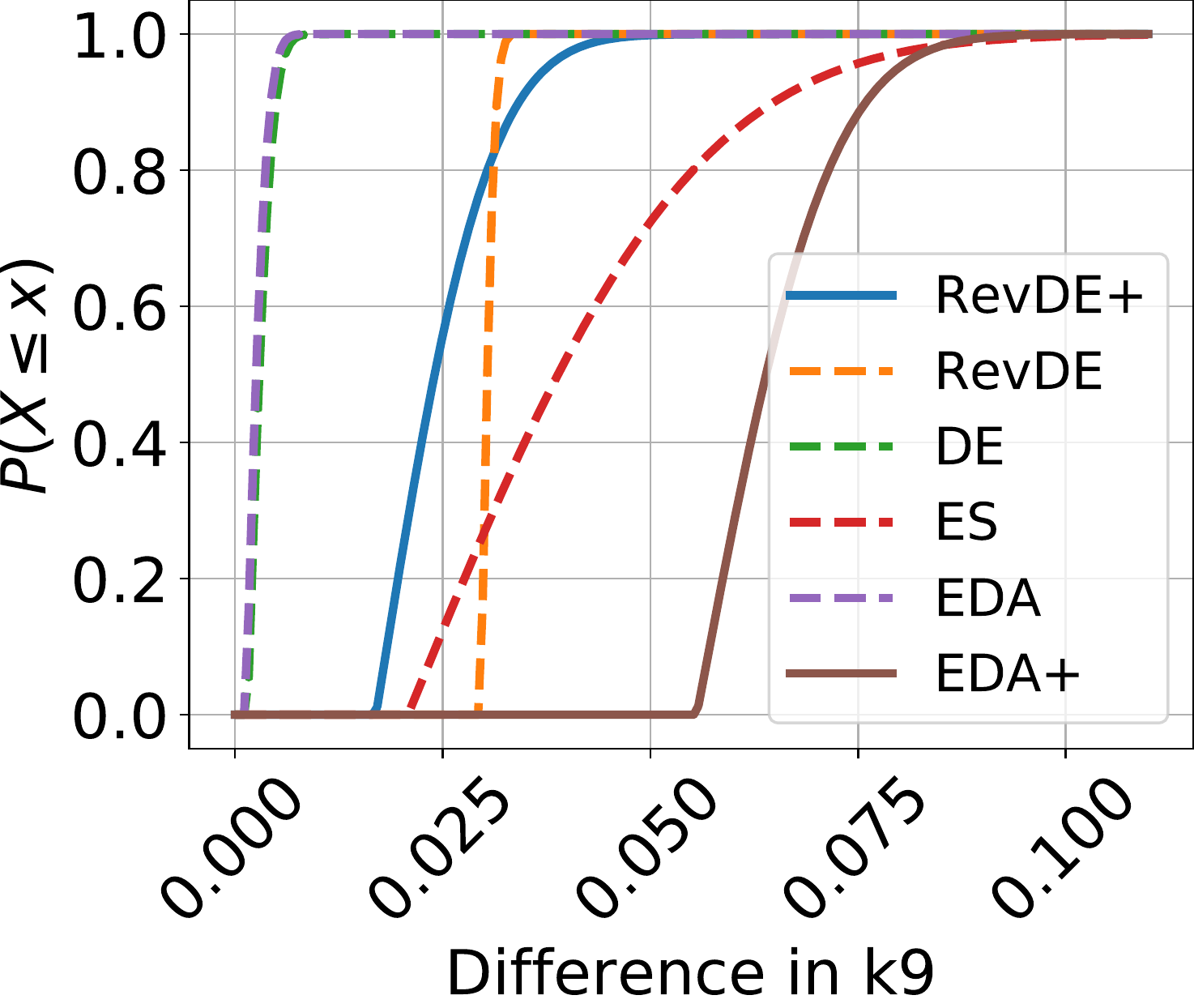}\quad
    \includegraphics[width=110px,height=90px]{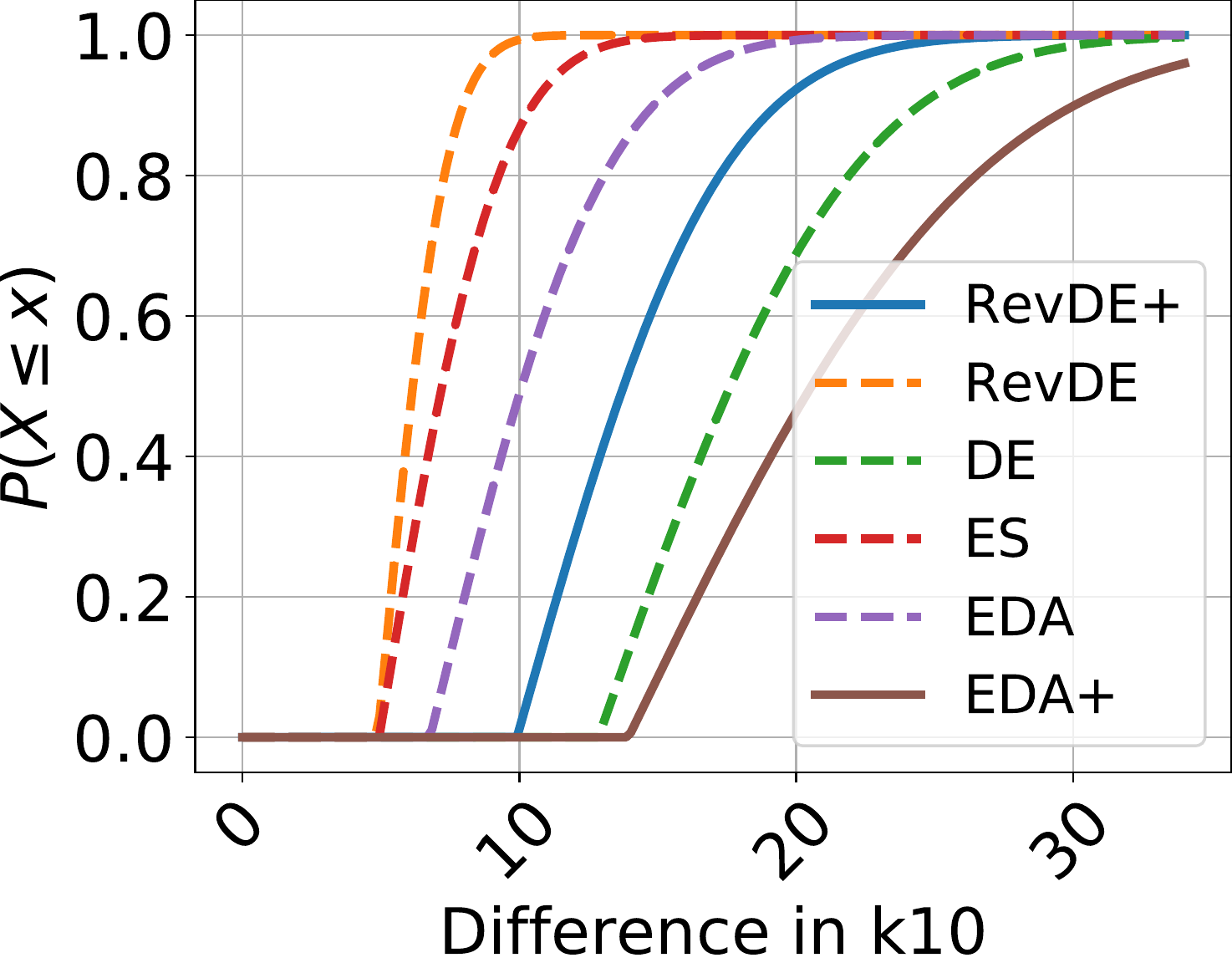}\quad
    \includegraphics[width=110px,height=90px]{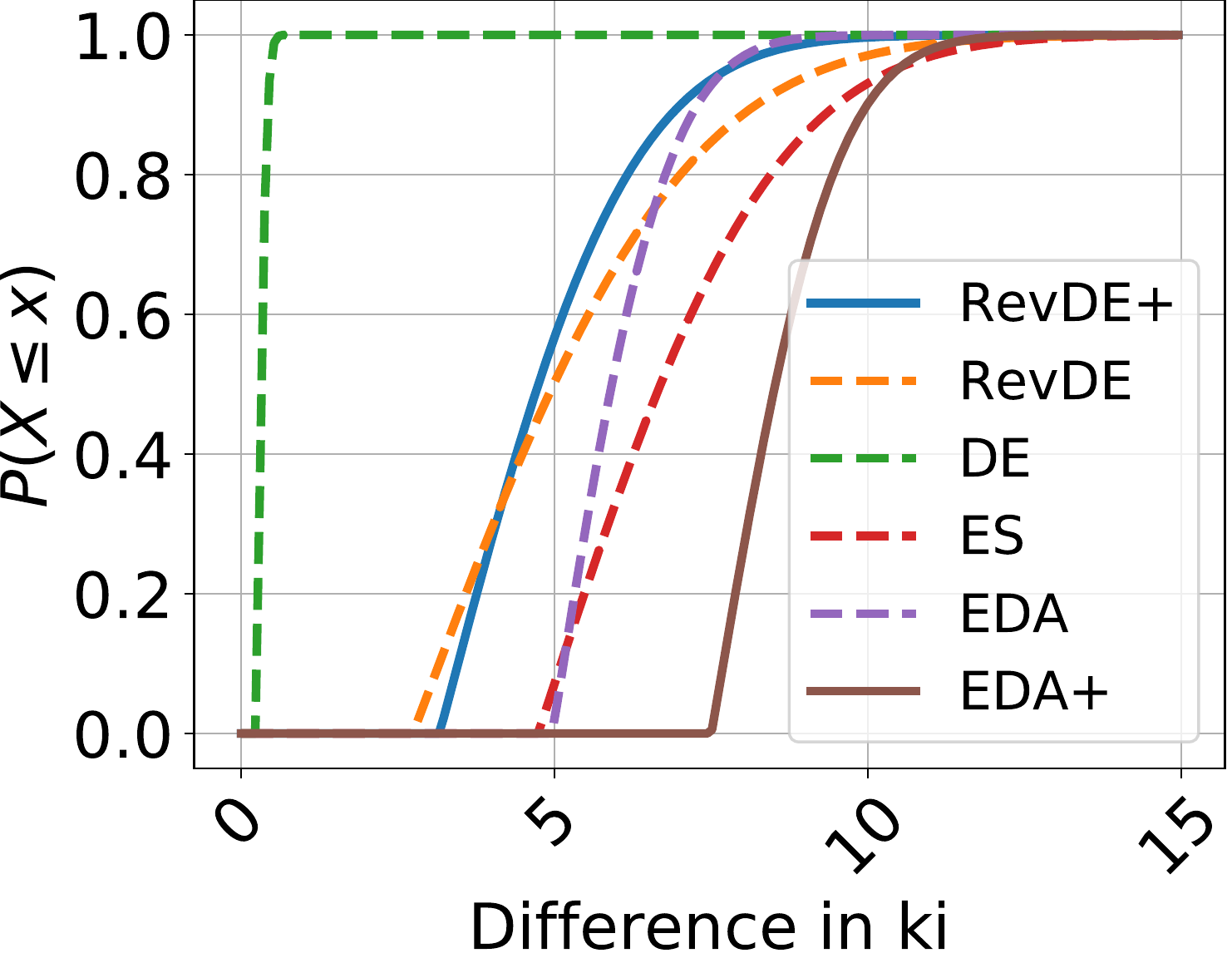} \\
        \vskip 3mm
    \includegraphics[width=110px,height=90px]{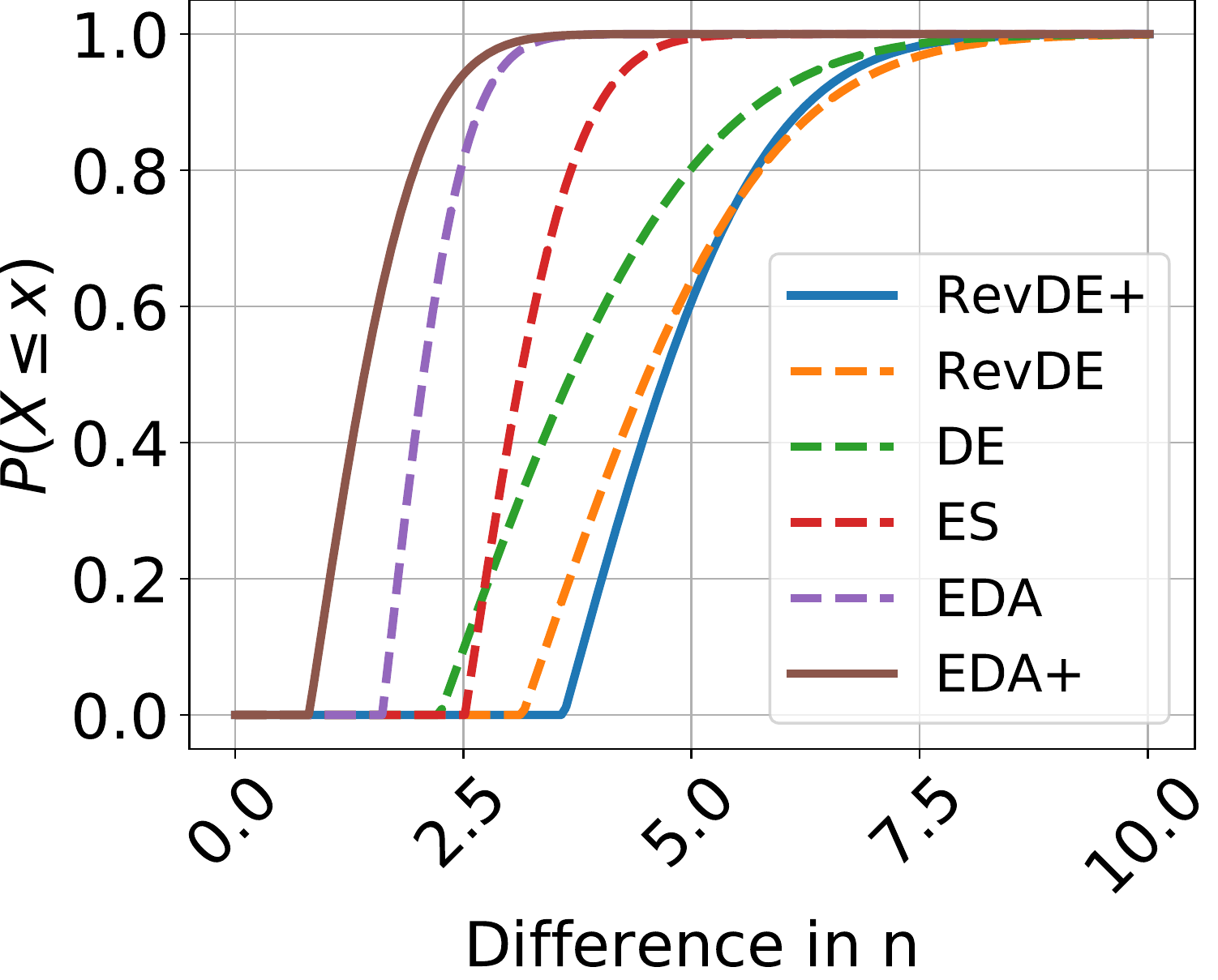}\quad
    \includegraphics[width=110px,height=90px]{figs/parameters_difference/wolf_differences_params_ntot.pdf}
    \caption{The cumulative distribution functions (cdfs) of the differences for all parameters. Ideally, a cdf of an optimization method should resemble a step-function centered at $0$. The averages and the scales are calculated over $3$ repetitions of the experiment in the Case 1.}
    \label{fig:differences_1}
\end{figure*}

\begin{figure*}[!tbp]
    \centering
    \includegraphics[width=110px,height=90px]{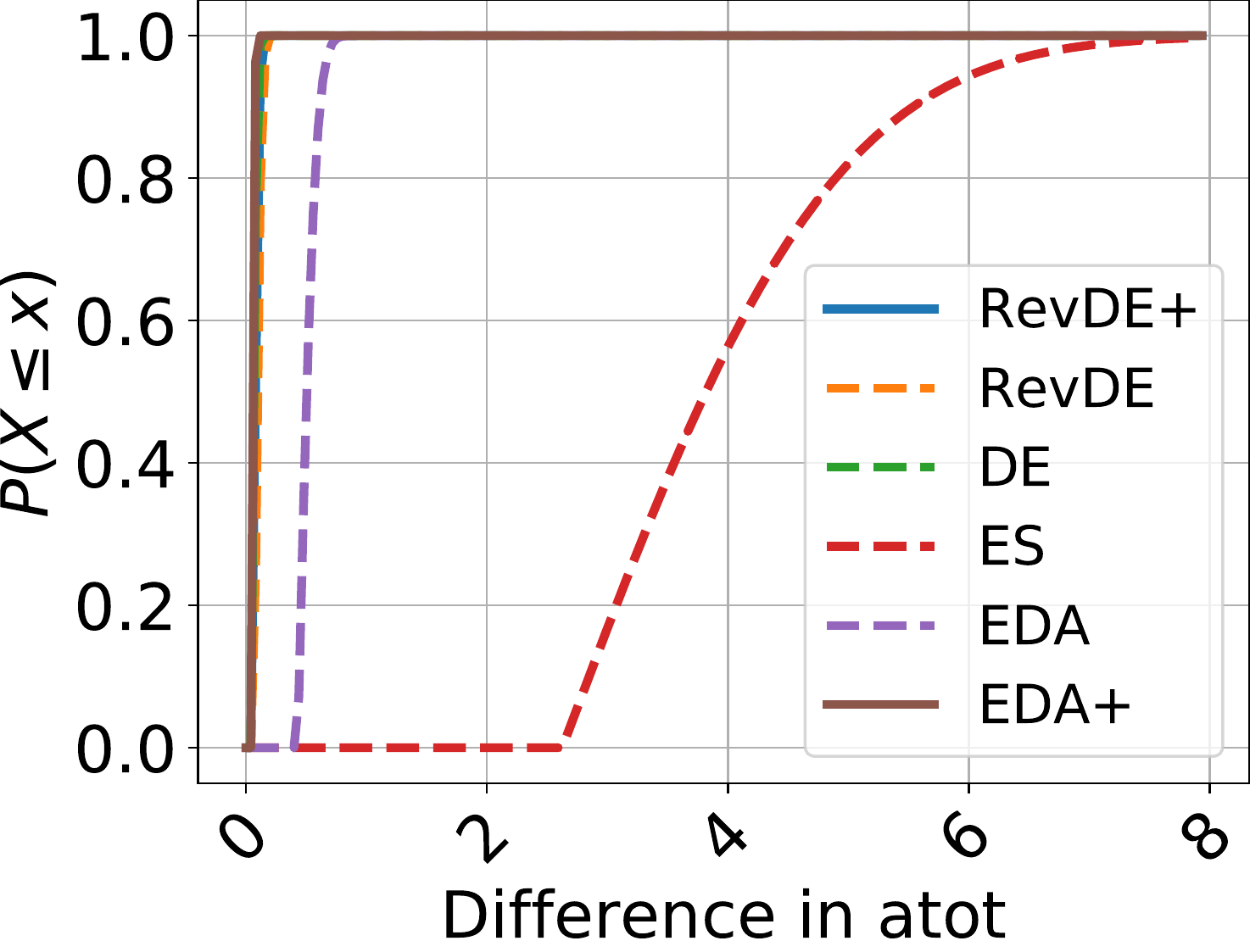}\quad
    \includegraphics[width=110px,height=90px]{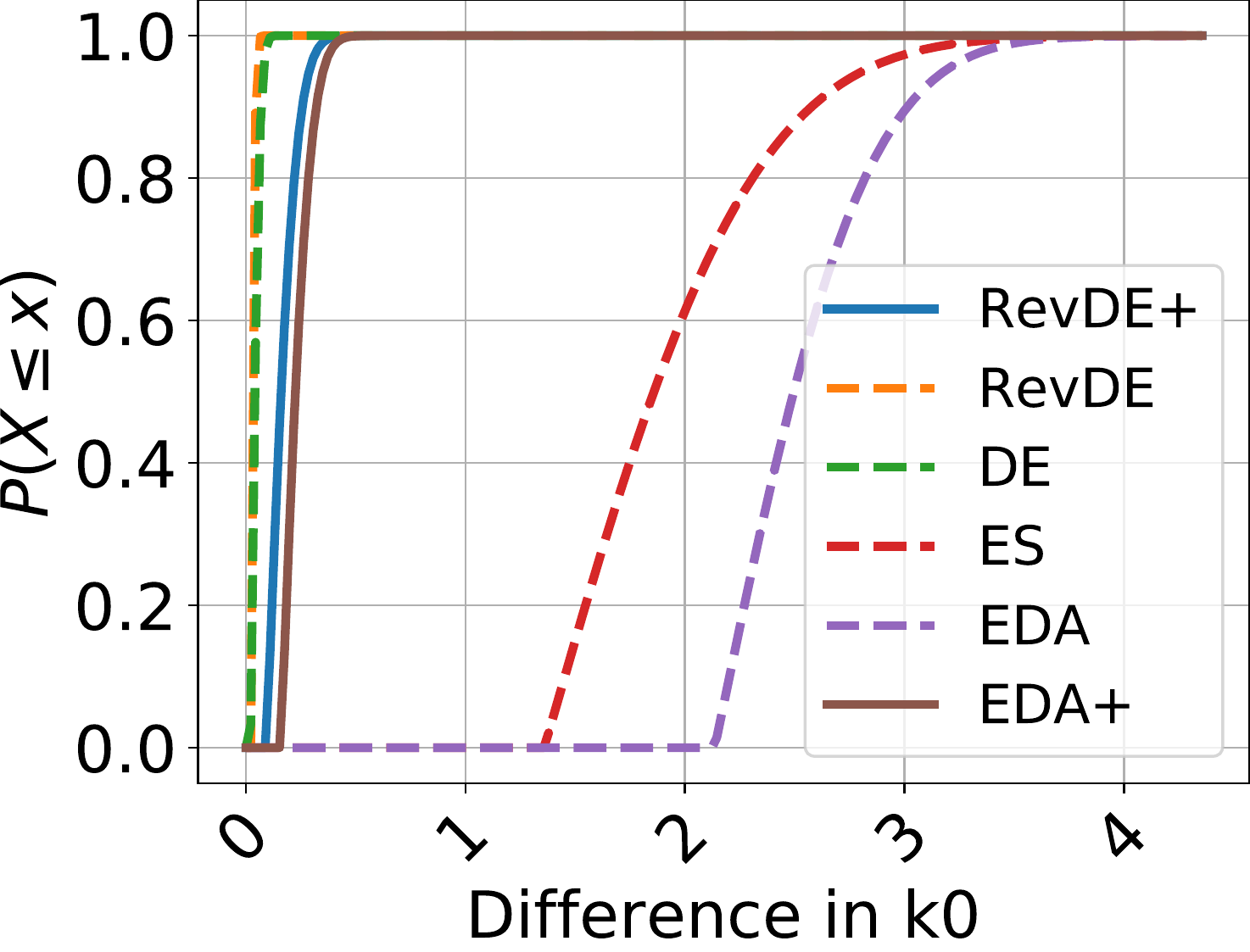}\quad
    \includegraphics[width=110px,height=90px]{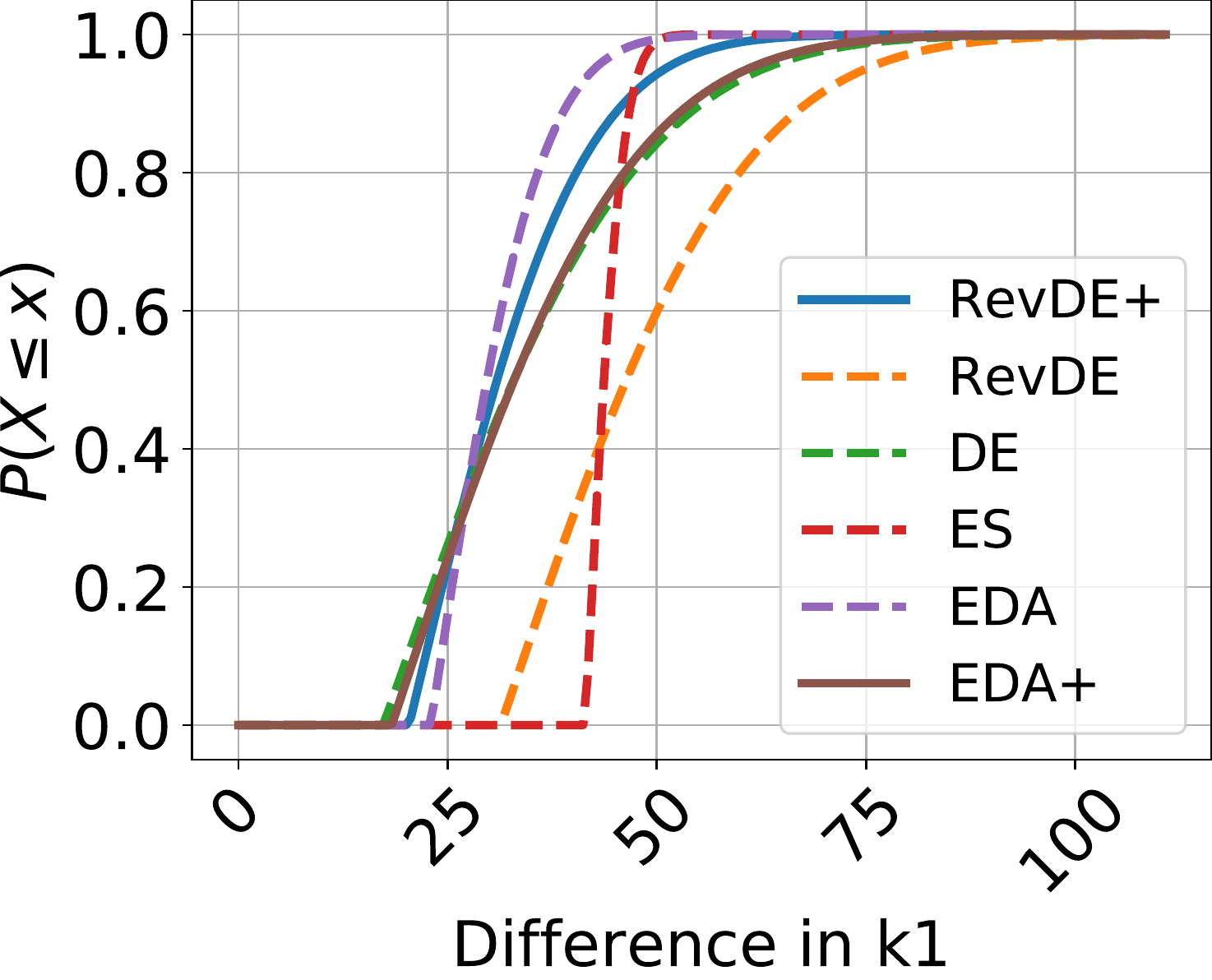}\quad
    \includegraphics[width=110px,height=90px]{figs/parameters_difference/mutation_differences_params_k2.pdf} \\
        \vskip 3mm
    \includegraphics[width=110px,height=90px]{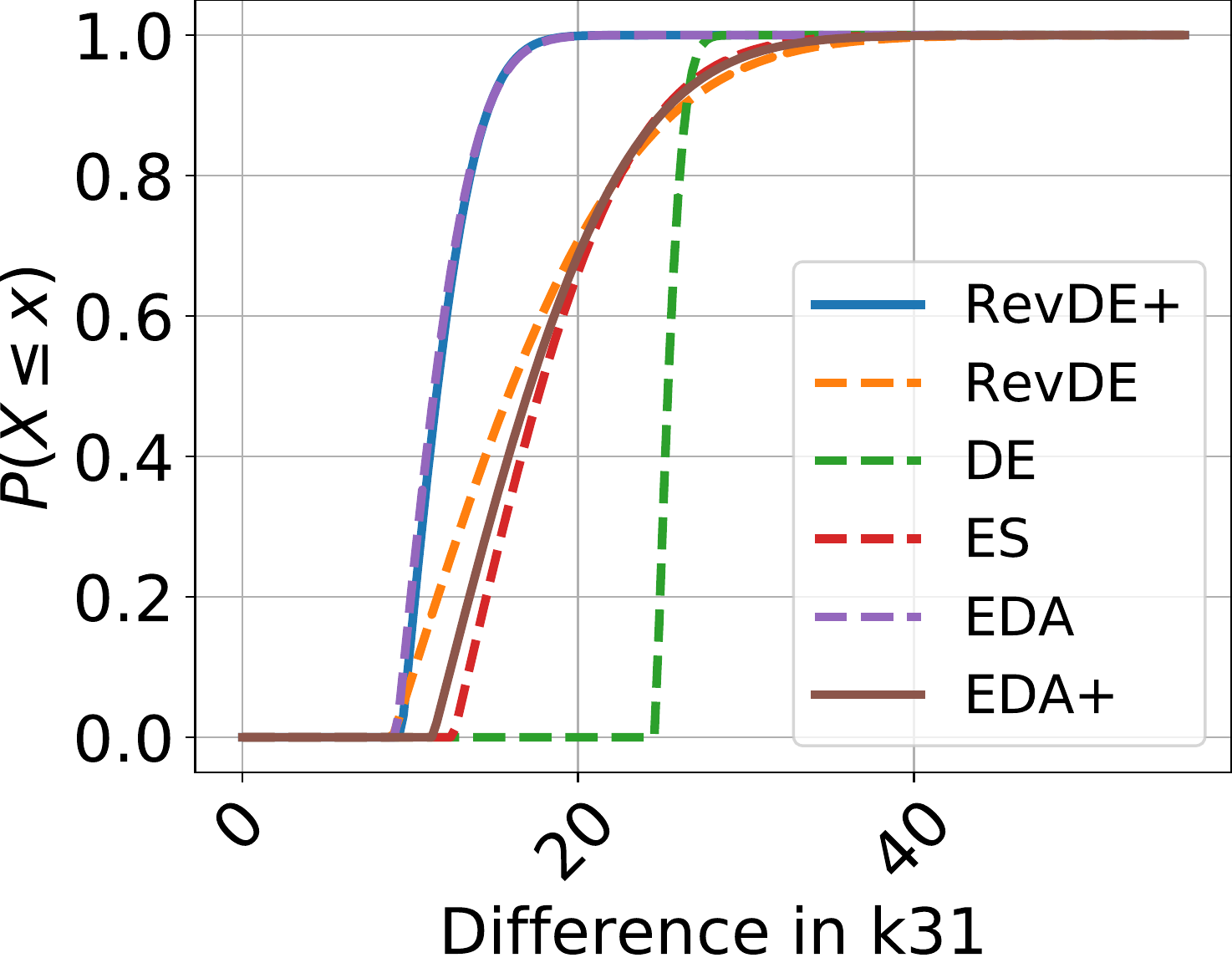}\quad
    \includegraphics[width=110px,height=90px]{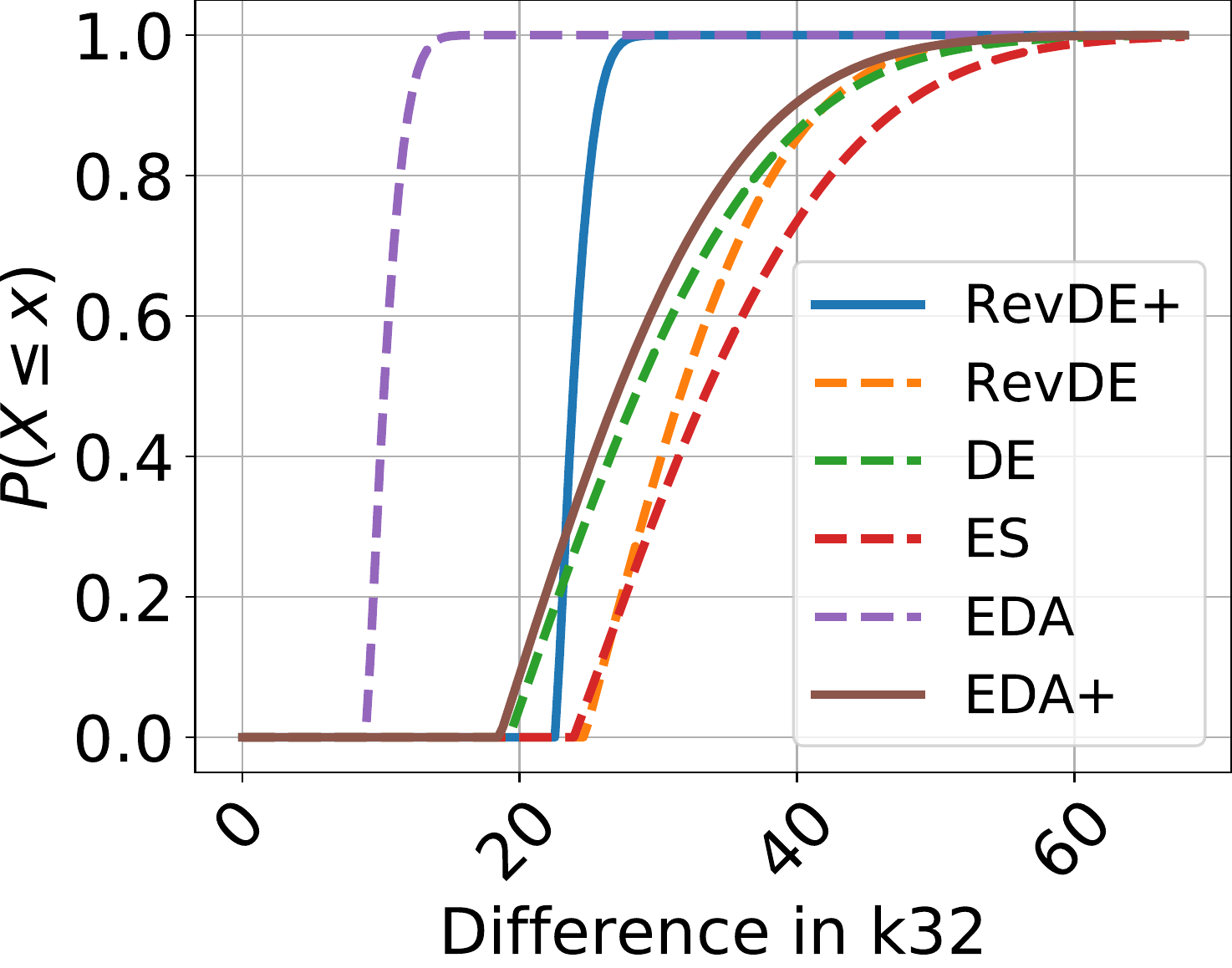}\quad
    \includegraphics[width=110px,height=90px]{figs/parameters_difference/mutation_differences_params_k33.pdf}\quad
    \includegraphics[width=110px,height=90px]{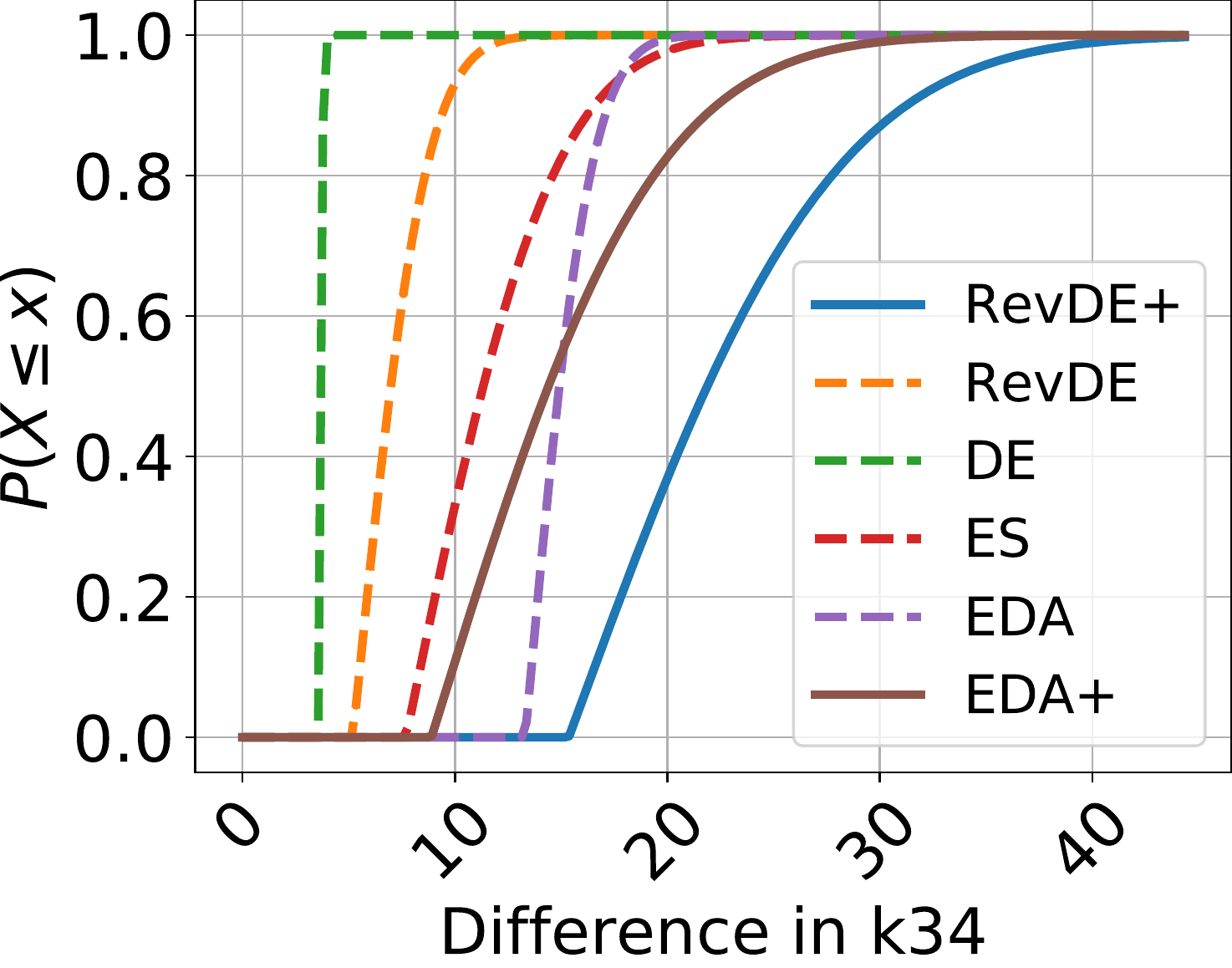} \\
        \vskip 3mm
    \includegraphics[width=110px,height=90px]{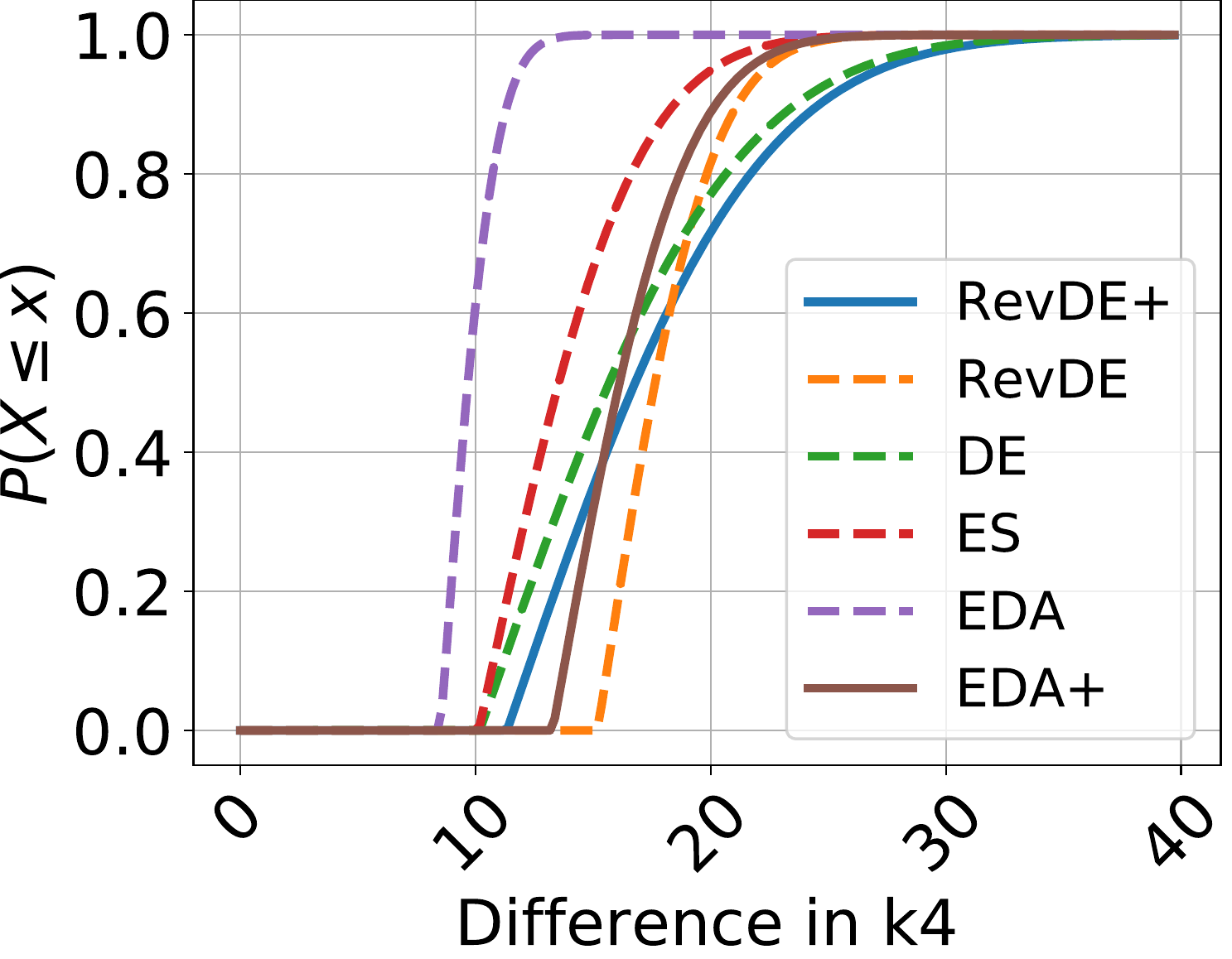}\quad
    \includegraphics[width=110px,height=90px]{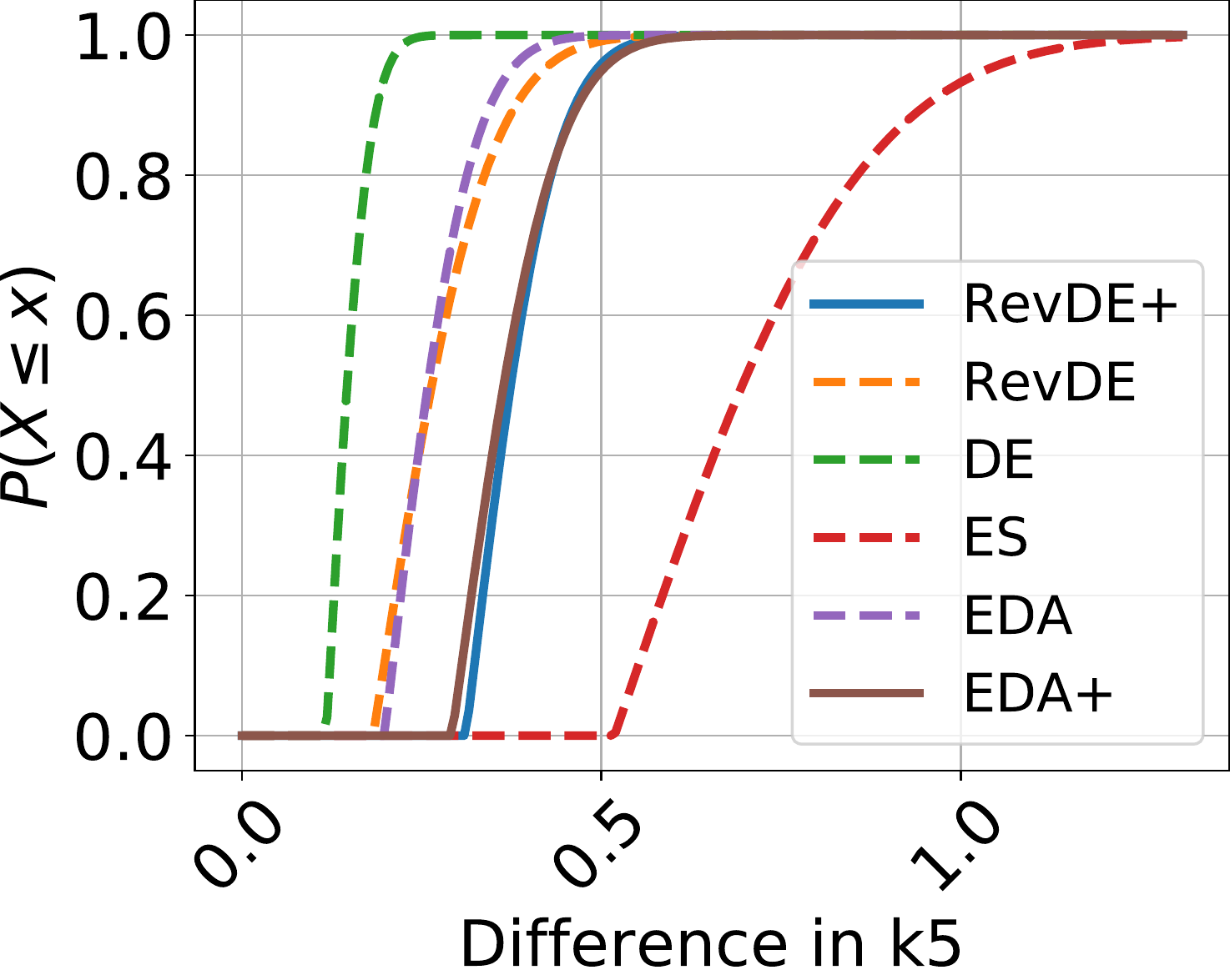}\quad
    \includegraphics[width=110px,height=90px]{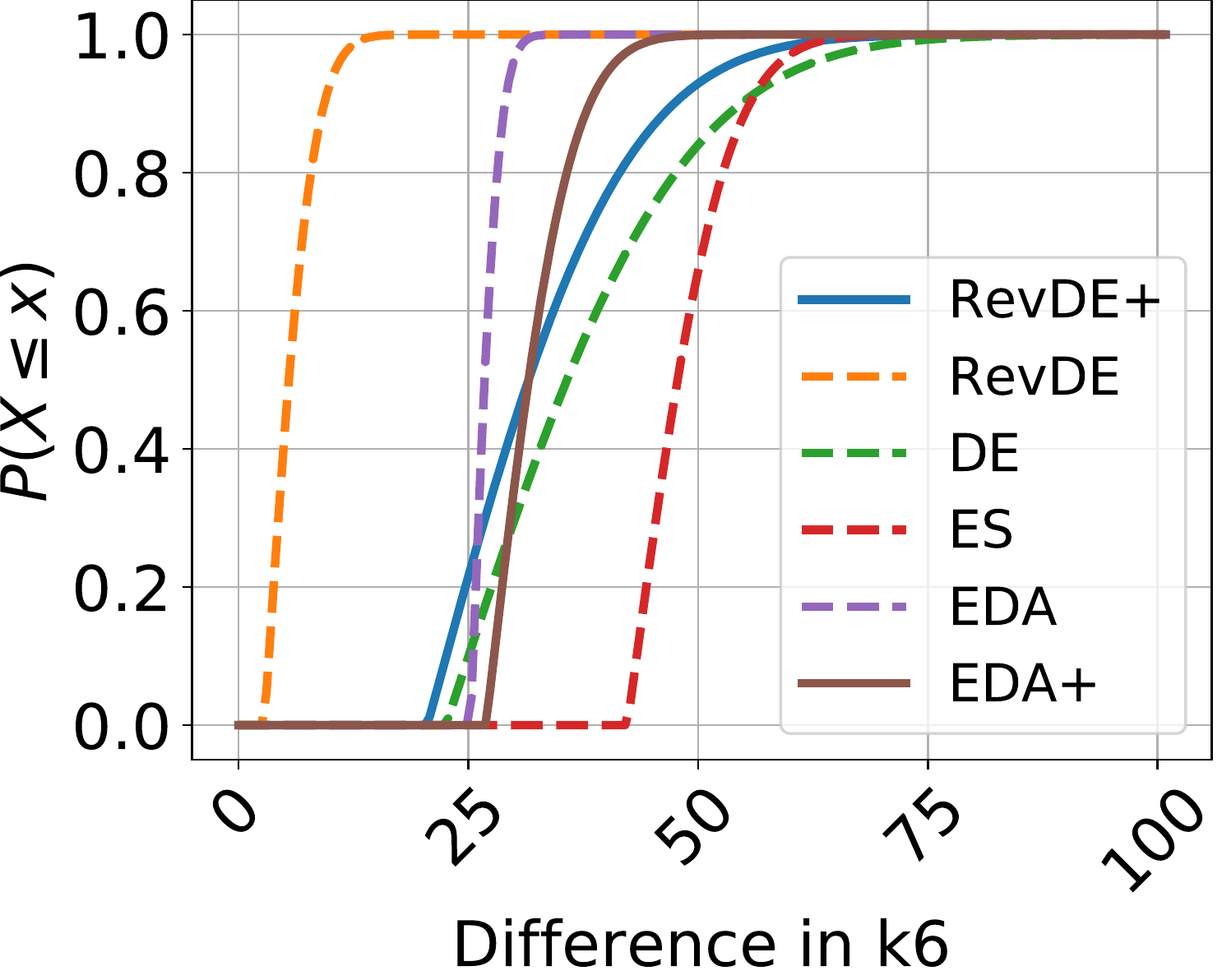}\quad
    \includegraphics[width=110px,height=90px]{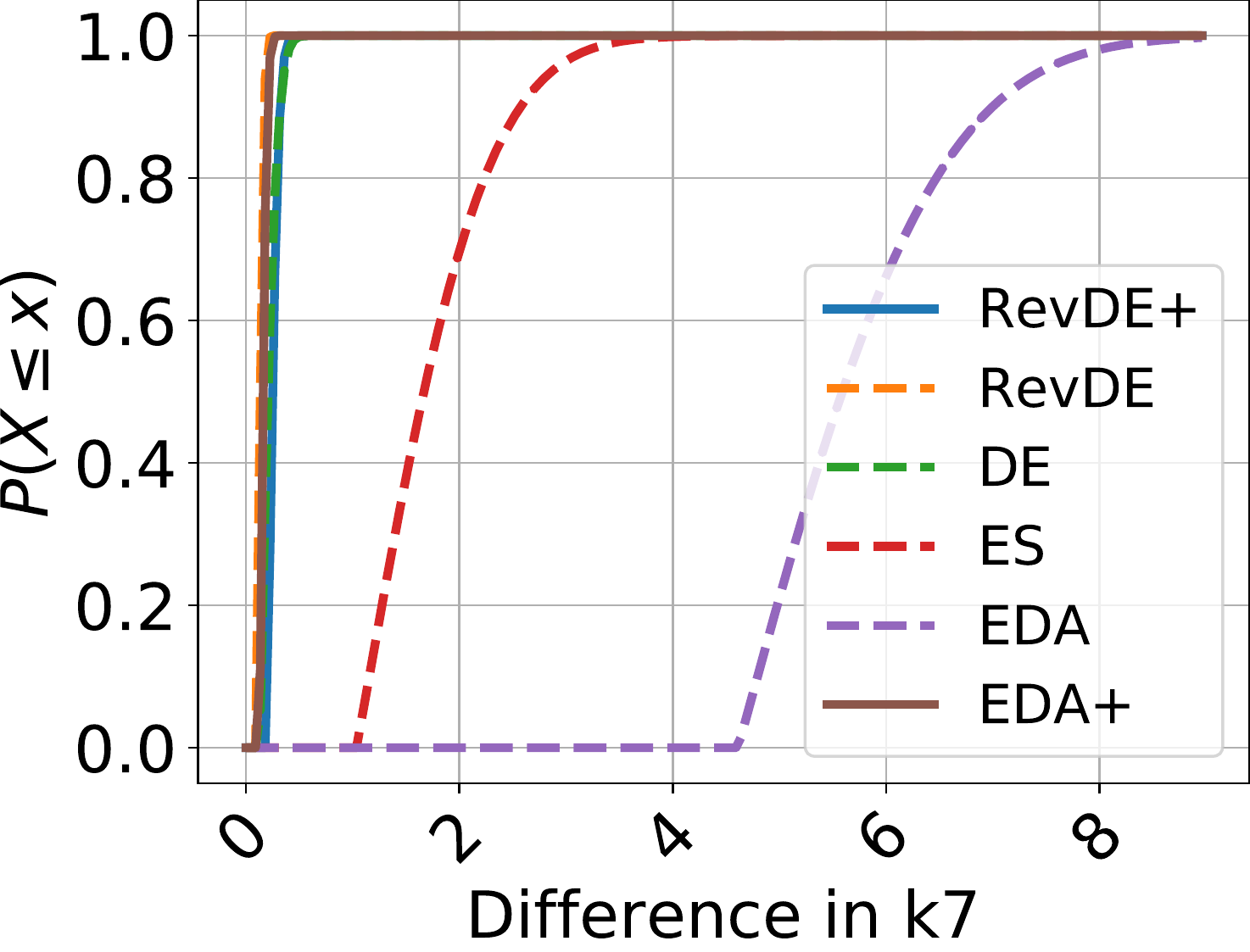} \\
        \vskip 3mm
    \includegraphics[width=110px,height=90px]{figs/parameters_difference/mutation_differences_params_k8.pdf}\quad
    \includegraphics[width=110px,height=90px]{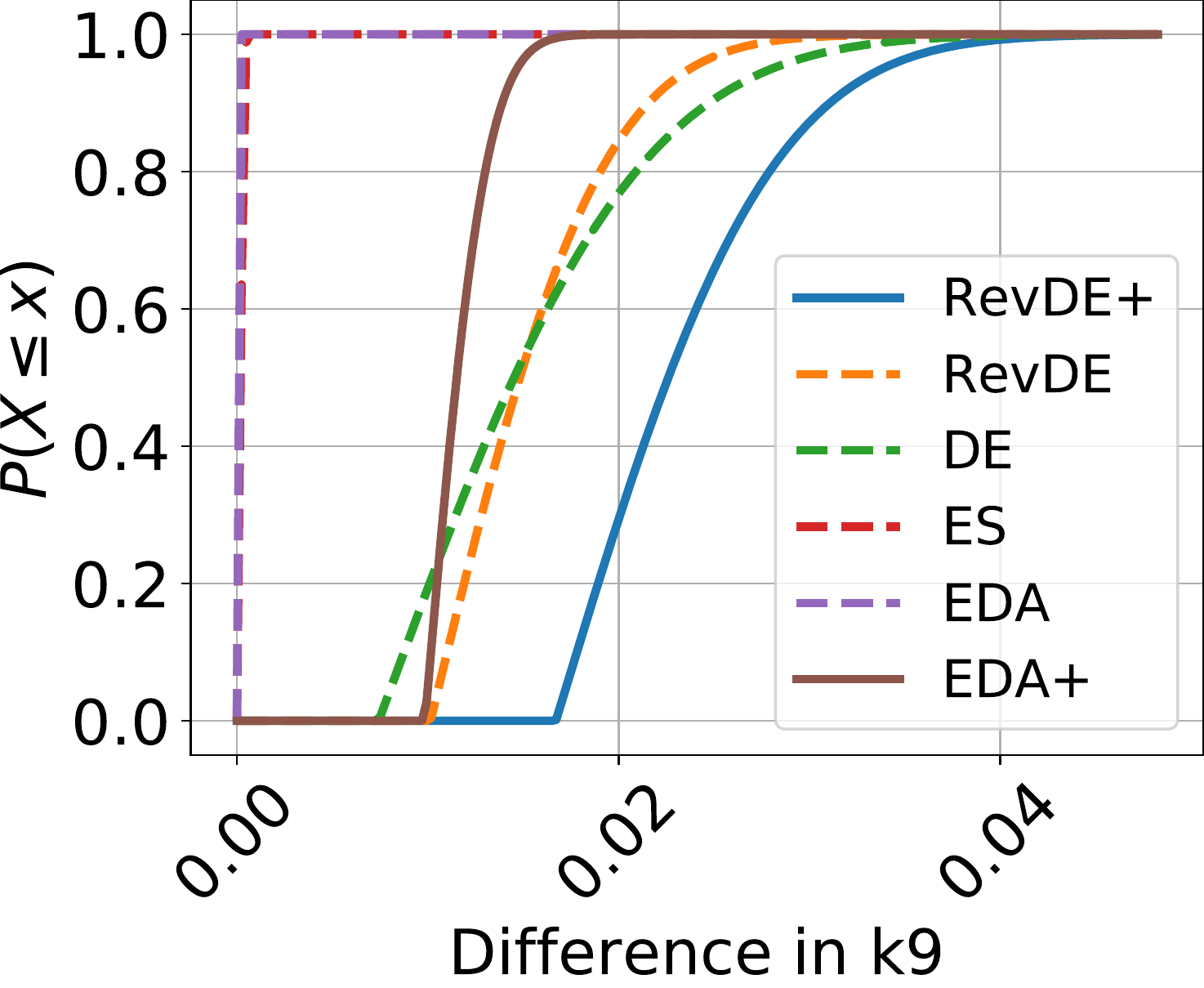}\quad
    \includegraphics[width=110px,height=90px]{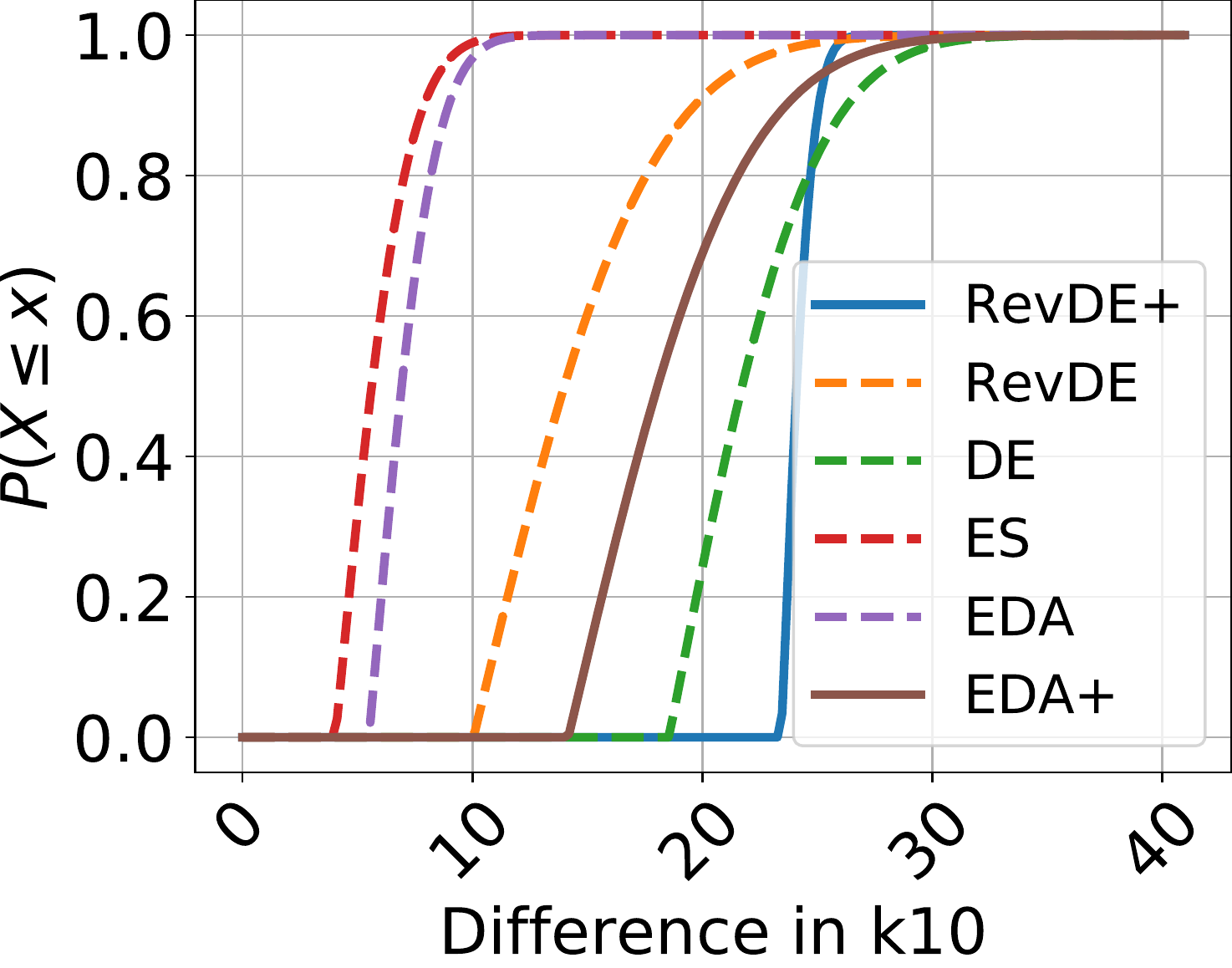}\quad
    \includegraphics[width=110px,height=90px]{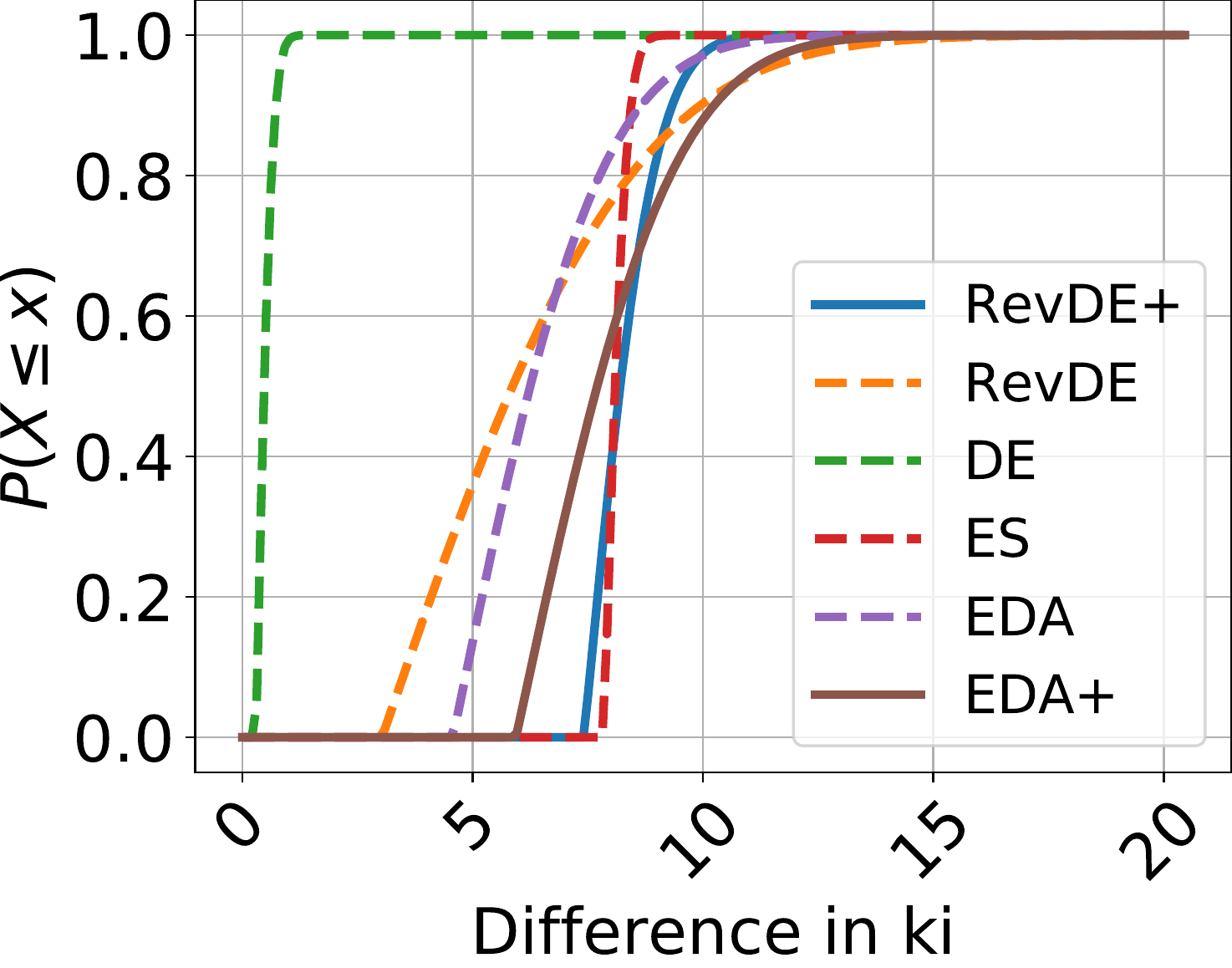} \\
        \vskip 3mm
    \includegraphics[width=110px,height=90px]{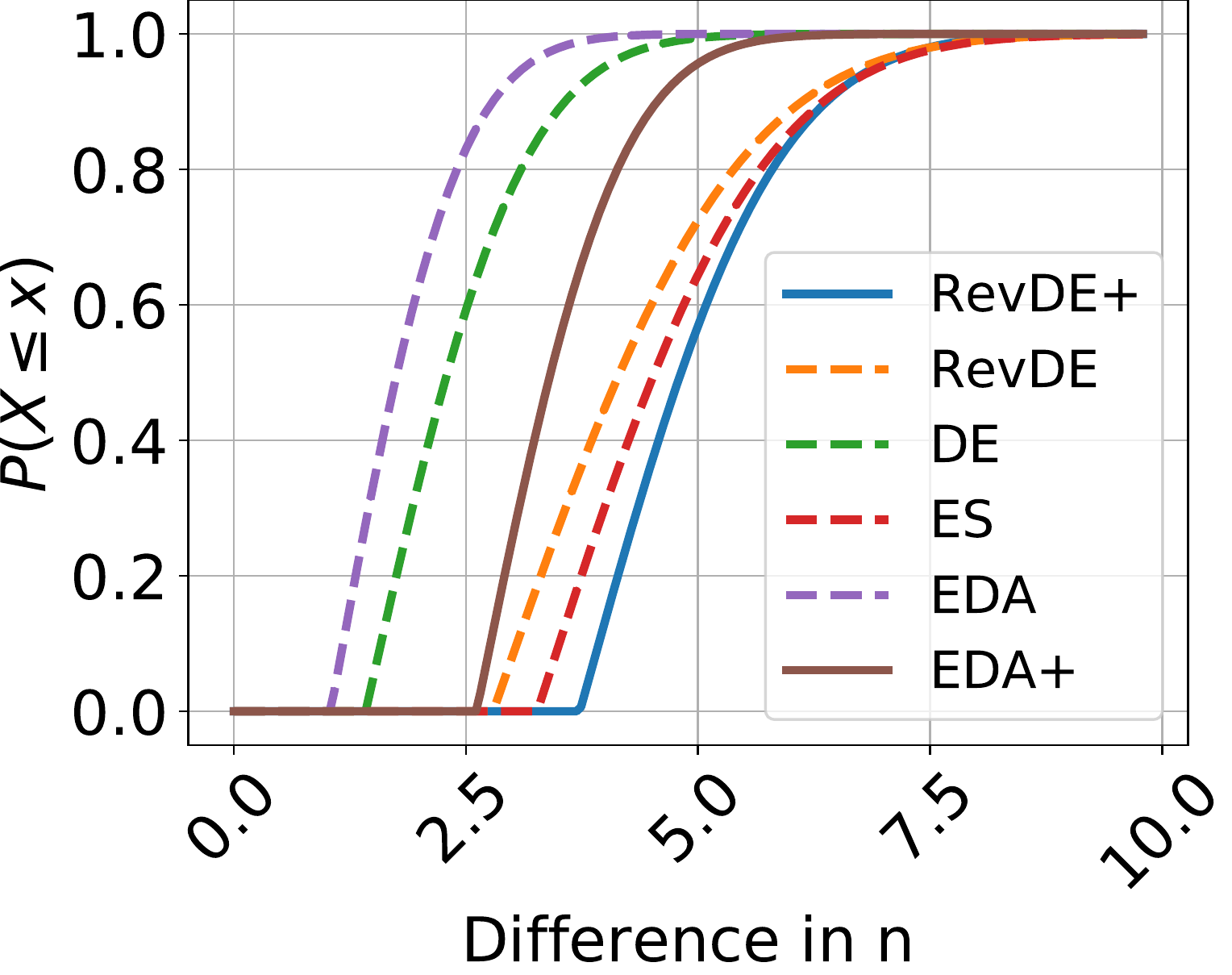}\quad
    \includegraphics[width=110px,height=90px]{figs/parameters_difference/mutation_differences_params_ntot.pdf}
    \caption{The cumulative distribution functions (cdfs) of the differences for all parameters. Ideally, a cdf of an optimization method should resemble a step-function centered at $0$. The averages and the scales are calculated over $3$ repetitions of the experiment in the Case 2.}
    \label{fig:differences_2}
\end{figure*}

\section{Appendix: The glycolysis model description}
\label{app:model}

In the considered model of the glycolysis we distinguish the following metabolites:
\begin{itemize}
    \item glycolysis (\textit{glu});
    \item fructose-1,6-bisphosphate (\textit{fru});
    \item triosephosphates (\textit{triop});
    \item triphosphoglycerate (\textit{tp});
    \item pyruvate (\textit{pyr});
    \item acetaldehyde (\textit{ac});
    \item external acetaldehyde (\textit{ace}).
\end{itemize}

Following the same assumptions as in \cite{wolf2000transduction} (\ie, a homogeneous distribution of the metabolites in the intracellular and in the extracellular solution), the systems of ordinary differential equations of the glycolysis model in \textit{Saccharomyces cerevisiae} is the following \cite{wolf}:
\begin{align}
\dot{glu} &= v_{1} - v_{2} \\
\dot{fru} &= v_{2} - v_{3} \\
\dot{triop} &= 2 v_{3} - v_{4} - v_{5} \\
\dot{tp} &= v_{5} - v_{6} \\
\dot{pyr} &= v_{6} - v_{7} \\
\dot{ac} &= v_{7} - v_{8} - v_{9}\\
\dot{ace} &= 0.1 v_{9} - v_{10} \\
\dot{atp} &= -2 v_{2} + v_{5} + v_{6} - v_{11} \\
\dot{nad} &= v_{4} - v_{5} - v_{8}
\end{align}
with the rate equations:
\begin{align}
v_{1} &= k_0\quad\\
v_{2} &= \frac{k_{1}\cdot glu\cdot at}{1 + (at/k_i)^{n}}  \\
v_{3} &= k_{2}\cdot fru \\
v_{4} &= \frac{k_{31}\cdot k_{32}\cdot triop\cdot nad A - k_{33}\cdot k_{34}\cdot tp\cdot atp\ N}{k_{33}\cdot N + k_{32}\cdot A}\\
v_{5} &= k_{4}\cdot tp\cdot A \\
v_{6} &= k_{5}\cdot pyr \\
v_{7} &= k_{6}\cdot ac\cdot nad \\
v_{8} &= k_{7}\cdot atp \\
v_{9} &= k_{8}\cdot triop\cdot nad \\
v_{10} &= k_{9}\cdot ace \\
v_{11} &= k_{7}\cdot atp
\end{align}
where:
\begin{align}
    A &= \left(a_{tot} - atp\right) \\
    N &= \left(n_{tot} - nad\right)
\end{align}

The initial conditions are the following:
\begin{align}
atp &= 2.0 \\
nad &= 0.6 \\
glu &= 5.0 \\
fru &= 5.0 \\
triop &= 0.6 \\
tp &= 0.7 \\
pyr &= 8.0 \\
ac &= 0.08 \\
ace &= 0.02 .
\end{align}

The model is schematically depicted in Figure \ref{fig:glycolysis}.

The real values of the parameters are the following \cite{wolf}:
\begin{align}
a_{tot} &= 4 \in [0, 10]\\
k_0 &= 0  \in [0, 10]\\
k_1 &= 550  \in [550, 600]\\
k_2 &= 9.8  \in [0, 10]\\
k_{31} &= 323.8  \in [300, 350]\\
k_{32} &= 76411.1  \in [76400, 76450]\\
k_{33} &= 57823.1  \in [57800, 57850]\\
k_{34} &= 23.7  \in [20, 50]\\
k_4 &= 80  \in [80, 100]\\
k_5 &= 9.7  \in [0, 10]\\
k_6 &= 2000  \in [2000, 2050]\\
k_7 &= 28.0  \in [20, 50]\\
k_8 &= 85.7  \in [80, 100]\\
k_9 &= 0  \in [0, 10]\\
k_{10} &= 375  \in [350, 400]\\
k_i &= 1  \in [0, 10]\\
n &= 4  \in [0, 10]\\
n_{tot} &= 1  \in [0, 10],
\end{align}
where we indicate the set of possible values of the parameters in the square brackets.

We note that for the sake of our experiments, we set $k_0$ to $0$ (originally: $k_0 = 50$ \cite{wolf}) in order to forbid a constant injection of $glu$, and $k_9$ to $0$ (originally: $k_9 = 80$ \cite{wolf}) in order to avoid oscillatory behavior of the system.

\section{Appendix: Additional results}
\label{app:results}

The detailed results for the timecourses of the unobserved metabolites are presented in Figures \ref{fig:obs_1} and \ref{fig:obs_2}.

The detailed results of the differences between the real parameters and found parameters are depicted in Figures \ref{fig:differences_1} and \ref{fig:differences_2}.

\end{document}